· REVIEW ·

# Deep Learning-based Software Engineering: Progress, Challenges, and Opportunities*

Xiangping CHEN[2*], Xing HU[3*], Yuan HUANG[4], He JIANG[5*], Weixing JI[6],
Yanjie JIANG[1*], Yanyan JIANG[7*], Bo LIU[6], Hui LIU[6], Xiaochen LI[5], Xiaoli LIAN[8*],
Guozhu MENG[9*], Xin PENG[10*], Hailong SUN[11*], Lin SHI[11*], Bo WANG[12*],
Chong WANG[10], Jiayi WANG[7], Tiantian WANG[13*], Jifeng XUAN[14*], Xin XIA[15],
Yibiao YANG[7*], Yixin YANG[11], Li ZHANG[8], Yuming ZHOU[7] & Lu ZHANG[1*]

1 Key Laboratory of High Confidence Software Technologies (Peking University), Ministry of Education;
School of Computer Science, Peking University, Beijing 100871, China;
2 School of Journalism and Communication, Sun Yat-sen University, Guangzhou 510275, China;
3 School of Software Technology, Zhejiang University, Hangzhou 310058, China;
4 School of Software Engineering, Sun Yat-sen University, Guangzhou 510275, China;
5 School of Software, Dalian University of Technology, Dalian 116024, China;
6 School of Computer Science and Technology, Beijing Institute of Technology, Beijing 100081, China;
7 State Key Laboratory for Novel Software Technology, Nanjing University, Nanjing 210023, China;
8 School of Computer Science and Engineering, Beihang University, Beijing 100191, China;
9 Institute of Information Engineering, Chinese Academy of Sciences, Beijing 100864, China;
10 School of Computer Science, Fudan University, Shanghai 200433, China;
11 State Key Laboratory of Complex & Critical Software Environment (CCSE);
School of Software, Beihang University, Beijing 100191, China;
12 School of Computer and Information Technology, Beijing Jiaotong University, Beijing 100044, China;
13 School of Computer Science and Technology, Harbin Institute of Technology, Harbin 150001, China;
14 School of Computer Science, Wuhan University, Wuhan 430072, China;
15 Huawei Technologies, Hangzhou, China

**Abstract**  Researchers have recently achieved significant advances in deep learning techniques, which in turn has substantially advanced other research disciplines, such as natural language processing, image processing, speech recognition, and software engineering. Various deep learning techniques have been successfully employed to facilitate software engineering tasks, including code generation, software refactoring, and fault localization. Many papers have also been presented in top conferences and journals, demonstrating the applications of deep learning techniques in resolving various software engineering tasks. However, although several surveys have provided overall pictures of the application of deep learning techniques in software engineering, they focus more on learning techniques, that is, what kind of deep learning techniques are employed and how deep models are trained or fine-tuned for software engineering tasks. We still lack surveys explaining the advances of subareas in software engineering driven by deep learning techniques, as well as challenges and opportunities in each subarea. To this end, in this paper, we present the first task-oriented survey on deep learning-based software engineering. It covers twelve major software engineering subareas significantly impacted by deep learning techniques. Such subareas spread out the through the whole lifecycle of software development and maintenance, including requirements engineering, software development, testing, maintenance, and developer collaboration. As we believe that deep learning may provide an opportunity to revolutionize the whole discipline of software engineering, providing one survey covering as many subareas as possible in software engineering can help future research push forward the frontier of deep learning-based software engineering more systematically. For each of the selected subareas, we highlight the major advances achieved by applying deep learning techniques with pointers to the available datasets in such a subarea. We also discuss the challenges and opportunities concerning each of the surveyed software engineering subareas.

Citation

* The authors are displayed in alphabetical order.
* Corresponding author (email: chenxp8@mail.sysu.edu.cn,  xinghu@zju.edu.cn,  jianghe@dlut.edu.cn,  yanjiejiang@pku.edu.cn,

# 1 Introduction

In recent years, deep learning, first proposed by Hintonetal. [1] in 2006, has achieved highly impressive advances [2]. Because of the unsupervised-layer-wise training proposed by Hinton et al. [3], the major obstacle to the training of deep neural networks has been removed. Since then, most artificial intelligence (AI) researchers have turned to deep learning, constructing deep neural networks containing dozens, thousands, and even millions of layers [4]. They have also proposed various novel structures of deep neural networks, such as convolutional neural networks (CNNs) [5], recurrent neural networks (RNNs) [6], long short-term memory networks (LSTMs) [7], bidirectional LSTM [8], and Transformer [9]. With the advances and popularity of deep learning, hardware vendors like NVIDIA release more powerful computing devices specially designed for deep learning. All of these together significantly push forward machine learning techniques and make deep learning-based AI one of the most promising techniques in the 21st century.

Given significant advances in deep learning, various deep learning techniques have been employed to fulfill software engineering tasks [10]. Although natural language processing (NLP), image and video processing, and speech processing are the major targets of current deep learning techniques, deep learning has been successfully applied to a wide range of various domains, including data mining [11], machine manufacturing [12], biomedical engineering [13] and information security [14]. Concerning software engineering, researchers have successfully exploited various deep learning techniques for various important tasks, such as code generation [15], code completion [16], code summarization [17], software refactoring [18], code search [19], fault localization [20], automated program repair [21], vulnerability detection [22], and software testing [23]. In all such tasks, deep learning techniques have been proven useful, substantially improving the state-of-the-art. One possible reason for the success of deep learning-based software engineering is the significant advances in deep learning techniques. Another possible reason is that various and massive software engineering data are publicly available for training advanced neural models. With the popularity of open-source software applications, developers share massive software requirements, source code, documents, bug reports, patches, test cases, online discussions, and logs and trace relationships among different artifacts. All such data make training specialized deep neural models for software engineering tasks feasible. To our knowledge, there have already been several surveys on deep learning for software engineering (e.g., Yanget al. [10], Watson et al. [24] and Niu etal. [25]). Although these surveys provide some overall pictures of the applications of deep learning for software engineering, there is still a lack of detailed analyses of the progress, challenges, and opportunities of deep learning techniques from the perspective of each subarea of software engineering influenced by deep learning.

In this paper, we present a detailed survey covering the applications of deep learning techniques in major software engineering subareas. We choose to provide one survey to cover the technical research of deep learning for the whole discipline of software engineering instead of several surveys for different individual subareas for the following reasons. First, software engineering has one central objective, and researchers follow a divide-and-conquer strategy to divide software engineering into different subareas. However, the advances of deep learning may provide usan opportunity to break the boundaries of research in different subareas to push forward the software engineering discipline as a whole. That is, deep learning may provide a common means to revolutionize software engineering in the future. Therefore, we believe that a survey of deep learning in all software engineering subareas may be more beneficial for software engineering researchers. Second, deep learning belongs to representation learning, and deep learning techniques for software engineering are thus highly specific to different artifacts in software development. As different subareas of software engineering typically share some common artifacts, putting different subareas into one survey would help researchers from different subareas understand the strengths and weaknesses of different deep learning techniques. For example, a deep learning technique based on source code may impact all subareas related to the comprehension, generation, and modification of codes.

One issue that raises in preparing this survey is that software engineering is a big discipline and the applications of deep learning techniques may thus touch some subareas in an uninfluential way. When facing such a subarea, where there are very few deep learning papers, we feel that these papers can hardly reflect the essence of deep learning research in that subarea. Therefore, instead of covering all subareas of software engineering, we focus on subareas where deep learning has already had significant

jyy@nju.edu.cn, lianxiaoli@buaa.edu.cn, mengguozhu@iie.ac.cn, pengxin@fudan.edu.cn, sunhl@buaa.edu.cn, shilin@buaa.edu.cn, wangbo_cs@bjtu.edu.cn, wangtiantian@hit.edu.cn, jxuan@whu.edu.cn , yangyibiao@nju.edu.cn, zhouyuming@nju.edu.cn, zhanglucs@pku.edu.cn)

**Table 1** Software Engineering Tasks Covered by the Survey

|    | Software Engineering Tasks | Number of Surveyed Papers |
|----|---------------------------|---------------------------|
| 1  | Requirements Engineering  | 28  |
| 2  | Code Generation           | 46  |
| 3  | Code Search               | 40  |
| 4  | Code Summarization        | 55  |
| 5  | Software Refactoring       | 19  |
| 6  | Code Clone Detection      | 53  |
| 7  | Software Defect Prediction | 32  |
| 8  | Bug Finding               | 114 |
| 9  | Fault Localization        | 42  |
| 10 | Program Repair            | 64  |
| 11 | Bug Report Management     | 51  |
| 12 | Developer Collaboration   | 57  |
|    | Total                     | 601 |

impacts. Fortunately, deep learning has deeply impacted software engineering, and through the subareas we selected for surveying in this paper, we can already form a general and insightful picture of deep learning for software engineering. To achieve our goal, for each of the selected subareas, we highlight major technical advances, challenges, and opportunities. As far as we know, our survey is the first task-oriented survey on deep learning-based software engineering, providing a technical overview of software engineering research driven by deep learning.

Another issue is the alignment between software engineering and deep learning. From the perspective of software engineering, research efforts are primarily divided into five phases (i.e., requirements, design, implementation, testing, and maintenance) according to the software development lifecycle, where each phase maybe further divided into different software development activities. Meanwhile, from the perspective of deep learning, research efforts are grouped by the target learning tasks, where the most distinctive characteristic of a learning task is its input and output as deep learning is typically of an end-to-end fashion. That is, using the same form of input–output pairs is typically viewed as being the same task, but there may be several different tasks for achieving the same goal. To balance the two perspectives, we grouped the research efforts according to their goals. That is, we viewed research for achieving the same goal or a set of similar goals as being of one subarea, and accordingly divided the research efforts into 12 subareas (Table 1 for the numbers of surveyed papers in each subarea): requirements engineering, code generation, code search, code summarization, software refactoring, code clone detection, software defect prediction, bug finding, fault localization, program repair, bug report management, and developer collaboration. We believe that this organization is friendly to readers from both software engineering and deep learning fields. Each subarea is of clear semantics in software engineering, and the tasks in each subarea are highly related from the perspective of deep learning. We further arranged the subareas in an order consistent with the order of the five phases of software development. An additional benefit is that this organization naturally prevents one subarea from being overcrowded. However, this organization also demonstrates the following drawbacks. From the software development lifecycle perspective, many subareas belong to software maintenance, but fewer subareas belong to other phases in our survey. We would like to emphasize that although this may roughly reflect that more research efforts have been devoted to software maintenance, it does not mean that software maintenance is of more importance than the other phases in software engineering.

In collecting papers for our survey, we focused on publications in major conferences and journals on software engineering and artificial intelligence between 2000 and 2023. Tables 2 and 3 present the conferences and journals we searched for papers on deep learning for software engineering. Notably, as we are not performing a systematic literature review, we did not follow a strict procedure for data collection (i.e., paper collection). Therefore, although Tables 2 and 3 provide the conferences and journals from where papers were surveyed, the selection of papers for our survey also underwent subjective judgment from the authors for appropriation. That is, the primary goal of our survey is to provide readers with a general picture of technical advances in deep learning-based software engineering but not to characterize a precise distribution of research efforts across different software engineering subareas. Because there are quite many researchers interested in deep learning-based software engineering, the number of papers increases quickly. Consequently, the number of papers discussed in this survey has already significantly surpassed the number of surveyed papers in recent systematic literature reviews (i.e., Yang et al. [10] and Watson et al. [24]). Further, this survey notably belongs to the theme of AI4SE (where various techniques of artificial intelligence are applied to software engineering), whereas there is also intensive research in another related theme of SE4AI (where techniques for software engineering are applied to enhance artificial intelligence systems). An important goal of our subjective judgment is to avoid the inclusion of SE4AI papers in our paper collection.

**Table 2** Conferences Covered by the Survey

| # ID | Abbreviation | Conference |
|:---:|:---:|:---:|
| 1 | AAAI | AAAI Conference on Artificial Intelligence |
| 2 | ACL | Annual Meeting of the Association for Computational Linguistics |
| 3 | APSEC | Asia-Pacific Software Engineering Conference |
| 4 | ASE | IEEE/ACM International Conference on Automated Software Engineering |
| 5 | COMPSAC | IEEE Annual Computers, Software, and Applications Conference |
| 6 | EASE | Evaluation and Assessment in Software Engineering |
| 7 | ESEC/FSE | ACM Joint European Software Engineering Conference and Symposium on the Foundations of Software Engineering |
| 8 | ICDM | IEEE International Conference on Data Mining |
| 9 | ICLR | International Conference on Learning Representations |
| 10 | ICMLA | International Conference on Machine Learning and Applications |
| 11 | ICML | International Conference on Machine Learning |
| 12 | ICPC | International Conference on Program Comprehension |
| 13 | ICSE | IEEE/ACM International Conference on Software Engineering |
| 14 | ICSME | IEEE International Conference on Software Maintenance and Evolution |
| 15 | ICSR | International Conference on Social Robotics |
| 16 | ICST | IEEE International Conference on Software Testing, Verification and Validation |
| 17 | ICTAI | IEEE International Conference on Tools with Artificial Intelligence |
| 18 | IJCAI | International Joint Conference on Artificial Intelligence |
| 19 | IJCNN | International Joint Conference on Neural Networks |
| 20 | Internetware | Asia-Pacific Symposium on Internetware |
| 21 | ISSRE | IEEE International Symposium on Software Reliability Engineering |
| 22 | ISSTA | ACM SIGSOFT International Symposium on Software Testing and Analysis |
| 23 | IWoR | International Workshop on Refactoring |
| 24 | IWSC | International Workshop on Software Clones |
| 25 | MSR | International Conference on Mining Software Repositories |
| 26 | NeurIPS | Annual Conference on Neural Information Processing Systems |
| 27 | NIPS | Advances in Neural Information Processing Systems |
| 28 | OOPSLA | Object-Oriented Programming, Systems, Languages, and Applications |
| 29 | QRS | International Conference on Software Quality, Reliability and Security |
| 30 | RE | International Conference on Requirements Engineering |
| 31 | SANER | International Conference on Software Analysis, Evolution and Reengineering |
| 32 | SEKE | International Conference on Software Engineering and Knowledge Engineering |
| 33 | S&P | IEEE Symposium on Security and Privacy |
| 34 | WCRE | Working Conference on Reverse Engineering |
| 35 | WWW | Web Conference |

**Table 3** Journals Covered by the Survey

| #ID | Abbreviation | Journal |
|-----|-----|-----|
| 1 | ASE | Automated Software Engineering |
| 2 | COLA | Journal of Computer Languages |
| 3 | CSUR | ACM Computing Surveys |
| 4 | ESE | Empirical Software Engineering |
| 5 | FCS | Frontiers in Computer Science |
| 6 | IETS | IET Software |
| 7 | IST | Information and Software Technology |
| 8 | JSS | Journal of Systems and Software |
| 9 | NCA | Neural Computing and Applications |
| 10 | PACMPL | ACM on Programming Languages |
| 11 | RE | Requirements Engineering |
| 12 | SCIS | Science China Information Sciences |
| 13 | TC | IEEE Transactions on Computers |
| 14 | TOSEM | ACM Transactions on Software Engineering and Methodology |
| 15 | TSE | IEEE Transactions on Software Engineering |
| 16 | TSUSC | IEEE Transactions on Sustainable Computing |
| 17 | – | Soft Computing |
| 18 | – | IEEE Transactions on Reliability |
| 19 | – | IEEE Access |
| 20 | – | Expert Systems with Applications |
| 21 | – | Journal of Software: Evolution and Process |
| 22 | – | Advances in Engineering Software |

## 2 Related Work

Recently, Yang et al. [10] performed a systematic literature review to summarize, classify, and analyze relevant papers in the field of software engineering that leverage deep learning techniques. They collected in total 250 relevant papers published in 32 major software engineering conferences and journals since 2006. Based on the papers, they analyzed the development trends of deep learning, provided a classification of deep learning models, and summarized the research topics tackled by these relevant papers. The major task of the systematic literature review is to figure out which and how deep learning techniques have been applied to software engineering. Their findings suggest that four categories of DNNs (CNN,LSTM, RNN, and FNN) were frequently employed by more than 20 studies. In addition, they summarized three types of DNN-based model selection strategies, i.e., characteristic-based selection, prior study-based selection based on, and using multiple feasible DNNs where the first strategy (characteristic-based selection) is by far the most popular one. Our survey differs from the systematic literature review by Yang et al. [10] in that Yang et al. focus more on deep learning techniques whereas we focus more on software engineering tasks. Yang et al. explain what deep learning techniques have been applied to software engineering whereas we analyze each software engineering task to explain how deep learning techniques could provide help for the task, and what kind of challenges could be encountered. In total, we analyze more than 500 papers that leverage deep learning techniques for software engineering, providing a more comprehensive view on this emerging research field.

Watson et al. [24] conducted another systematic literature review on the application of deep learning techniques in software engineering. They collected and analyzed 128 papers from software engineering and deep learning conferences and journals. The major task of the paper is to answer five questions, i.e., what types of software engineering tasks have been addressed by deep learning techniques, what deep learning techniques have been applied to software engineering, how requested training data are collected, how well software engineering tasks have been addressed by deep learning techniques, and what common factors influence the replicability of deep learning applied to software engineering tasks. Their findings suggest that deep learning-based techniques had been applied in a diverse set of tasks where program synthesis, code comprehension, and source code generation are the most prevalent. They also found that a number of different data preprocessing techniques have been utilized. Tokenization and neural embeddings are the two most frequently employed data pre-processing techniques. Besides that, their analysis revealed seven major types of deep learning architectures that have been employed for software engineering tasks, including Recurrent Neural Networks (RNNs), Encoder-Decoder Models, Convolutional Neural Networks (CNNs), Feed-Forward Neural Networks (FNNs), AutoEncoders, Siamese Neural Networks, and highly tailored architectures. Our survey differs from Watson et al. [24] in that Watson et al. take software engineering as a whole to discuss the advances and challenges in applying deep learning techniques to software engineering. In contrast, we discuss the advances and challenges for each software engineering task concerning how deep learning techniques could be employed to resolve the given software engineering task.

The survey conducted by Niu et al. [25] reviews pre-trained models used in software engineering. In total, they identified and analyzed 20 pre-trained models developed for software engineering tasks. They classified the models with four dimensions, i.e., the underlying network architecture, the number of input modalities, the tasks used for pretraining, and whether they are pre-trained on a single programming languages or multiple programming languages. It also investigated how the models were pretrained for different software engineering tasks. Their goal is to raise the awareness of AI technologies, including the development and use of pre-trained models and the successful applications of such models in resolving software engineering tasks. Their findings suggest that code pre-training models area promising approach to a wide variety of software engineering tasks. Our survey differs from the survey conducted by Niu et al. [25] in that the survey by Niu et al. is confined to pre-trained models whereas our survey covers all deep neural networks employed for software engineering tasks.

Zhang et al. [26] conducted a systematic survey to summarize the current state-of-the-art research in the LLM-based SE community. They summarized 30 representative LLMs of source code across three model architectures, 15 pre-training objectives across four categories, and 16 downstream tasks across five categories. They presented a detailed summarization of the recent SE studies for which LLMs are commonly utilized. Furthermore, they summarized existing attempts to empirically evaluate LLMs in SE, such as benchmarks, empirical studies. Finally, they discussed several critical aspects of optimization and applications of LLMs in SE. Our survey differs from the survey conducted by Zhang et al. [26] in that our survey focus on all deep neural networks employed for software engineering tasks. However, the survey conducted by Zhang et al. specifically concentrates on different representative LLMs.

As a conclusion, although a few surveys have been made concerning the synergy between software engineering and deep learning, we still lack a clear picture of the advances, potentials, and challenges concerning deep learning-driven attempts to various software engineering tasks. To this end, in this paper, we select the most fundamental and most challenging software engineering tasks, and for each of them we analyze the advances, potentials, and challenges of its deep learning-based solutions.

# 3 Requirements Engineering

Requirements engineering (RE) is the process of eliciting stakeholder needs and desires and developing them into an agreed-upon set of detailed requirements that can serve as a basis for all subsequent development activities. Unlike solution-oriented SE tasks (such as software design and program repair) that aim to ensure "**doing the thing right**", requirements engineering is a problem-oriented SE task that aims to ensure "**doing the right thing**", i.e., to make the problem that is being stated clear and complete, and to guarantee that the solution is correct, reasonable, and effective [27].

Recently, more and more RE researchers employ deep learning techniques to elicit, analyze, trace,

validate, and manage software requirements. In this section, we will introduce the progress of DL-related RE literature from three perspectives, i.e., RE task taxonomy, datasets, and DL models. Then, we will summarize the challenges and opportunities faced by DL-related RE literature.

## 3.1 Requirements Elicitation

The task of requirements elicitation aims to gather accurate and complete information about what the system should do, how it should behave, and what constraints and limitations it should adhere to. This process is critical in software development because the success of a project depends on clearly understanding requirements and translating them into solutions that meet stakeholder needs.

Huang et al. [28] proposed a convolution neural network (CNN)-based approach for automatically classifying sentences into different categories of intentions: feature request, aspect evaluating, problem discovery, solution proposal, information seeking, and information giving, and meaningless. They sped up the training process by integrating batch normalization. They also optimized the hyper-parameters through an automatic hyper-parameter tuning method in order to improve accuracy. Pudlitz et al. [29] presented an automated approach for extracting system states from natural language requirements using a self-trained Named-entity Recognition model with Bidirectional LSTMs and CNNs. They presented a semi-automated technique to extract system state candidates from natural language requirements for the automotive domain. The results show that their automated approach achieves a F1-score of 0.51, with only 15 minutes of manual work, while the iterative approach achieves an F1-score of 0.62 with 100 minutes. Furthermore, manual extraction took nine hours, demonstrating that machine learning approaches can be applied with a reasonable amount of efforts to identify system states for requirements analysis and verification.

Li et al. [30] proposed a technique called DEMAR based on deep multi-task learning, which can discover requirements from problem reports, and solve the limitations that requirements analysis tasks usually rely on manually coded rules or insufficient training data. Through the three steps of data augmentation, model building, and model training, their experimental results show that the multi-task learning mode of DEMAR has higher performance than the single-task mode. Meanwhile, DEMAR also outperforms the other selected existing techniques. DEMAR provides directions for exploring the application of multi-task learning to other software engineering problems. Guo et al. [31] proposed Caspar, a technique for extracting and synthesizing app question stories from app reviews. Caspar first extracts ordered events from acquired app reviews and ranks them using NLP techniques. Caspar then classifies these events into user actions or application questions, and synthesizes action-question pairs. Finally, an inference model is trained on the operation-problem pairs. For a given user action, Caspar can infer possible program problem events, thus enabling developers to understand possible problems in advance to improve program quality. They experimented with SVM, USE+SVM, and Bi-LSTM networks, respectively; and the results show that the Bi-LSTM model performs better than the other two. Mekala et al. [32] proposed a deep learning-supported artificial intelligence pipeline that can analyze and classify user feedback. This technique includes a sequence classifier. Their experimental results show that the BERT-based classifier performs the best overall and achieves good performance. Their experimental results further demonstrate that pre-trained embeddings of large corpora are a very effective way to achieve state-of-the-art accuracy in the context of low-capacity datasets. Tizard et al. [33] proposed a technique for linking forum posts, issue trackers, and product documentation to generate corresponding requirements. They scraped product forum data for VLC and Firefox, performed similarity calculations using the Universal Sentence Encoder (USE), and matched forum posts to issue trackers. Furthermore, they demonstrated that applying clustering to USE results in impressive performance on matching forum posts with product documents. Shi et al. [34] proposed a technique called FRMiner to detect feature requests with high accuracy from a large number of chat messages via deep Siamese networks. In particular, they used Spacy to perform word segmentation, lemmatization, and lowercase processing on sentences, randomly selected 400 conversations in three open-source projects for data sampling, and annotated them with the help of an inspection team. Then, they built a context-aware dialogue model using a bidirectional LSTM structure, and used a Siamese network to learn the similarity between dialogue pairs. Experimental results show that FRMiner significantly outperforms two-sentence classifiers and four traditional text classification techniques. The results confirm that FRMiner can effectively detect hidden feature requests in chat messages, thus helping obtain comprehensive requirements from a large amount of user data in an automated manner.

Panetal. [35] introduced an automated developer chat information mining technique called F2CHAT, which aims to solve the challenge of retrieving information from online chat rooms. They constructed a thread-level taxonomy of nine categories by identifying different types of messages in developer chats. Furthermore, they proposed an automatic classification technique F2CHAT,which combines hand-crafted non-textual features with deep textual features extracted by neural models. The results show that F2CHAT outperforms FRMiner and achieves high performance in cross-project verification, indicating that F2CHAT can be generalized across various projects.

## 3.2 Requirements Generation

The task of requirements generation aims to recommend requirements drafts on top of very limited information (e.g., keywords or structured models).

Most existing studies on automated requirements generation concentrate on transforming (semi-) structured models (e.g., business process models [36,37], i* framework [38,39], KAOS and Objectiver [40–42], UML models [43–45], or other representations like security goals in [46]) into specific syntactic pattern-oriented natural language requirements specifications, based on a set of pre-defined rules. Recently, Zhao et al. [47] proposed an approach called ReqGen that aims to recommend requirements drafts based on a few given keywords. Specifically, they first selected keyword-oriented knowledge from the domain ontology and injected it into the basic Unified Pre-trained Language Model (UniLM). Next, they designed a copy mechanism to ensure the occurrence of keywords in the generated statements. Finally, they proposed a requirement-syntax-constrained decoding technique to minimize the semantic and syntax distance between candidate and reference specifications. They evaluated ReqGen on two public datasets and demonstrated its superiority over six existing natural language generation approaches. Koscinski et al. [48] investigated the usage of Relational Generative Adversarial Networks (RelGAN) in automatically synthesizing security requirements specifications, and demonstrated promising results with a case study.

## 3.3 Requirements Analysis

The purpose of requirements analysis is to systematically identify, understand, and document the software requirements for a given system or software project.

Li et al. [49] proposed a new technique named RENE, which uses the LSTM-CRF model for requirements entity extraction and introduces general knowledge to reduce the need for labeled data. This technique can more efficiently extract requirements entities from textual requirements, thereby reducing labor costs. They introduced the construction and training process of RENE, and evaluated RENE on an industrial requirements dataset, showing good results. Furthermore, they explored the value of general-purpose corpora and unlabeled data, and provided an effective practical approach that can inspire how to further improve performance on specific tasks.

## 3.4 Smelly Requirements Detection

The task of smelly requirements detection is to identify the potential issues in software requirements that might threaten the success of software projects.

Casillo et al. [50] proposed to utilize a pre-trained convolutional neural network (CNN) to identify personal, and private disclosures from short texts to extract features from user stories. They constructed a user-story privacy classifier by combining the extracted features with those obtained from a privacy dictionary. Ezzini et al. [51] proposed an automated approach for handling anaphoric ambiguity in requirements, addressing both ambiguity detection and anaphora interpretation. They developed six alternative solutions based on the choices of (1) whether to use hand-crafted language features, word embeddings or a combination thereof for classification, (2) whether pre-trained language models like BERT are a viable replacement for the more traditional techniques, and (3) whether a mashup of existing (and often generic) NLP tools would be adequate for specific RE tasks. Wang et al. [52,53] proposed a deep context-wise semantic technique to resolve entity co-reference in natural-language requirements, which consists of two parts. One is a deep fine-tuned BERT context model for context representation, and the other is a Word2Vec-based entity network for entity representation. Then they proposed to utilize a multi-layer perceptron (MLP) to fuse two representations. The input of the network is a requirement contextual text and related entities, and the output is a predicted label to infer whether the two entities are co-referent. Ezzini et al. [54] proposed QAssist, a question-answer system that provides automated assistance to stakeholders during the analysis of NL requirements. The system retrieves relevant text passages from the analyzed requirements document as well as external domain knowledge sources and highlights possible answers in each retrieved text passage. When domain knowledge resources are missing, QAssist automatically constructs a domain knowledge resource by mining Wikipedia articles. QAssist is designed to detect incompleteness and other quality issues in requirements without the tedious and time-consuming manual process.

## 3.5 Requirements Classification

The purpose of requirements classification is to categorize and organize the collected requirements based on certain criteria or characteristics. By classifying requirements, the goal is to improve the understanding, management, and communication of requirements throughout the software development process.

Baker et al. [55] proposed to classify software requirements into five categories by using machine learning techniques: maintainability, operability, performance, security, and availability. They conducted experimental evaluations on two widely used software requirements datasets for convolutional neural network (CNN) and artificial neural network (ANN) evaluation. The results show that this technique is effective, reaching high precision and recall. In addition, they explored potential applications of the technique in software engineering life cycle, including automating the process of analyzing NFRs, reducing human errors and misunderstandings, and reducing potential requirements-related defects and errors in software systems. Tobias et al. [56] proposed an automated requirements classification technique called NoRBERT, which uses the transfer learning capabilities of BERT. They evaluated NoRBERT on different tasks, including binary classification, multi-class classification, and classification in terms of quality. Their experimental results show that NoRBERT performs well when dealing with natural language requirements and can effectively improve the performance of automatic classification methods. Luo et al. [57] proposed a new requirements classification technique named PRCBERT, which is based on BERT's pre-trained language model and applies flexible prompt templates to achieve accurate classification. They also conducted experiments on a large-scale demand dataset to compare it with other techniques, showing that the technique significantly improves accuracy and efficiency. Winkler et al. [58] proposed an automated approach for classifying natural language requirements by their potential verification techniques. They proposed to use a convolutional neural network for text classification, and trained it on a dataset created by industry requirements specifications. They performed 10-fold cross-validation on a dataset containing about 27,000 industrial requirements, achieving a high F1 score. AlDhafer et al. [59] used Bidirectional Gated Recurrent Neural Networks (BiGRU) to classify requirements into binary labels (functional and non-functional) and multiple labels (Operational, Performance, Security, Usability, etc).

## 3.6 Requirements Traceability

The task of requirements traceability is to build the associations between software requirements and other artifacts, which may be different types of requirements, design artifacts, source code, test cases and other artifacts. Compared to other tasks in RE, much more research uses DL techniques for requirements traceability.

To the best of our knowledge, the first DL for requirements traceability research was from Guo et al. [60] in 2017. They proposed the technique of TraceNN to build the links between software requirements and design documents. In particular, they modeled the task of traceability construction as a classification problem. They embedded the features of text sequences based on the Recurrent neural network (RNN) and used the multi-layer perceptron (MLP) to conduct the classification task. They evaluated two types of RNN models, namely long short-term memory (LSTM) and gated recurrent unit (GRU) on large-scale industrial datasets, and showed GRU achieved better mean average precision (MAP). In contrst, RNNs can only encode one side of contextual information, which will be weakened for long sequences [61]. With the wide application of BERT since 2018, different variants such as BioBERT [62] and CodeBERT [63] have been developed for various domains. Lin et al. proposed TraceBERT [64] to construct the trace links between requirements and source code based on CodeBERT, and modeled the traceability as a code search problem. They designed a three-fold procedure of pre-training, intermediate-training and fine-tuning towards their tasks. Particularly, they first pre-trained CodeBERT on source code to build TraceBERT. Then in the intermediate-training phase, they provided adequate labeled training examples to train the model to address the code search problem with the expectation that the model can learn general traceability knowledge. Finally, in the fine-tuning phase, they applied the model to the specific

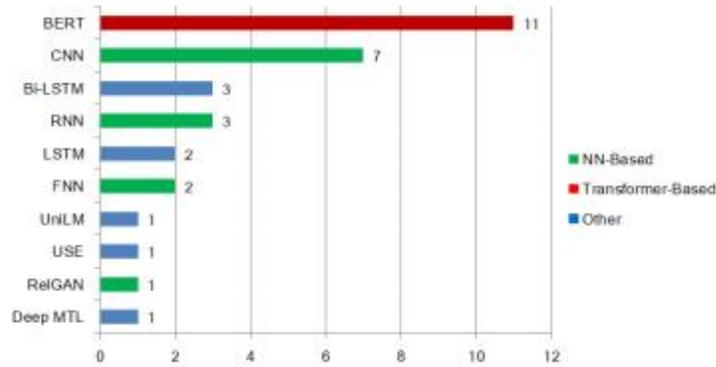

**Figure 1** The deep learning models involved in the related work

"issue (natural language)-commit (programming language) tracing" problem to improve the tracing effect. They evaluated three commonly used BERT architectures (i.e., single, twin, and Siamese) on open-source projects. Their experimental results showed that the single architecture achieves the best accuracy, while the Siamese architecture achieves similar accuracy with faster training time. Tian et al. [65] proposed a technique named DRAFT to build the traceability between new requirements and other requirements in different abstraction levels, during the system evolution process. They performed a second-phase pre-training on BERT based on the project-related corpus for the purpose of project-related knowledge transformation. Then, they designed 11 heuristic features and embedded them with requirements text. The performance of DRAFT has been evaluated with eight open-source projects. Lin et al. [66] explored the performance of information retrieval and deep learning techniques on building trace links in 14 English-Chinese projects. The involved approaches include Vector Space Model (VSM), Latent Semantic Indexing (LSI), Latent Dirichlet Allocation (LDA), and various models that combine mono- and cross-lingual word embeddings with the Generative Vector Space Model (GVSM), and adeep-learning approach based on a BERT language model. Theyed show that their TraceBERT performed best in large projects, while IR-based GVSM worked best on small projects.

From the perspective of paper distribution over different RE tasks, the most four tasks using DL techniques are requirements elicitation, requirements traceability, smelly requirements detection and requirements classification, with 9, 7, 5 and 5 papers, respectively. For the other tasks, including requirements generation and analysis, the involved papers are few. We did not find any papers related to requirements management and requirements validation.

### 3.7 DL Models

Figure 1 shows the DL models involved in our surveyed studies on requirements engineering. We can see that there are over 10 different kinds of models utilized in these DL4RE studies. Among these models, BERT is the most widely-used one, followed by CNN, and the number of papers related to other models is far smaller than these two. This may be because BERT and CNN are relatively mature and universal deep learning models, which can adapt to different RE scenarios and datasets with high performance and reproducibility. It may also be due to the active and extensive research community of BERT, which can provide rich reference materials and the latest developments, stimulating the interest and innovation of researchers.

We can also see that the numbers of papers on the transformer-based model and NN-based model are 11 and 13 respectively, indicating that they have certain competition and complementarity in requirements engineering, and have certain research value and practical significance. There is no absolute difference between good and bad, but it is necessary to select or design appropriate in-depth learning models according to the specific RE tasks.

**Usage of DL models.** Figure 2 illustrates the ways of using deep learning models in requirements engineering. Primarily, there are three types: direct-training, pre-training, and fine-tuning. Among them, 43% of the studies directly build neural networks from training. The advantage of direct training is that it can train a model from scratch according to the dataset of the target task, without being limited to the pre-trained model, and can customize the network structure and parameters towards specific domains and tasks. The disadvantage is that if the dataset of the target task is not large enough, direct training may lead to problems such as model non-convergence, overfitting, and low generalization ability, etc. Considering that most of the studies use the way of direct training, it may indicate that this way is relatively simple and effective. Another possible reason is the lack of suitable pre-trained models or domain knowledge.

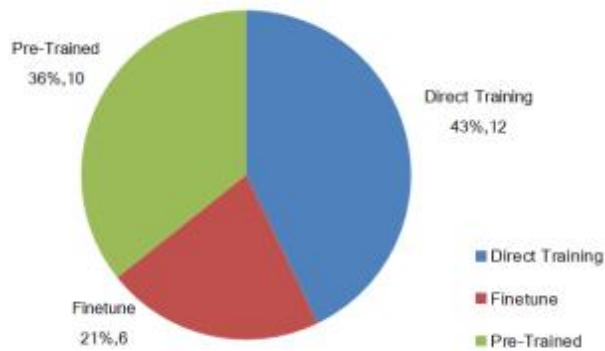

**Figure 2**  The proportion of different usage of DL models

36% of studies use pre-trained DL models. The advantage is that it can use the model parameters trained by large-scale datasets, saving time and computing resources, and improving computational efficiency and accuracy. The disadvantage is that if the dataset of the pre-trained model and that of the target task are not highly similar, the effect of the pre-trained model may not be satisfactory, because different tasks need to extract different features.

21% of studies use fine-tuned DL models. The advantage of fine-tuning is that it can adjust part or all of the parameters based on the pre-trained model according to the dataset of the target task, retaining the ability of the pre-trained model to extract general features and increasing the ability of the model to adapt to new task features. The disadvantage is that it needs to choose a suitable pre-trained model, a suitable fine-tuning level and range, a suitable learning rate, and other hyperparameters. Otherwise, it may affect the effect of fine-tuning. Pre-training and fine-tuning each account for approximately a quarter, possibly because the use of pre-training and fine-tuning has certain limitations and complexity. Pre-training needs to select appropriate pre-trained models and objective functions, and the matching degree and difference between pre-training models and RE problems or data need to be considered. Fine-tuning needs to select appropriate fine-tuning strategies and parameter settings, and consider the similarities and differences between different RE tasks or datasets.

**Performance of DL models.** Most papers use three indicators to measure the performance of deep learning models: precision, recall and F1. We find that the precision and recall of deep learning models are often above 80%, and the F1 score is often above 75%. The performance of deep learning models is related to many factors, such as the structure and parameters of the model, the training methods, and the size and quality of the dataset. Researchers often collect as many data as possible when using deep learning models to improve the expected performance of the models. However, it is not necessarily true that the larger the dataset is, the better the model performance is. The size of the selected dataset in [54] is only 387, but the average recall and precision of its model are over 90% and 84%. The dataset size used in [55] is 914, with a precision between 82% and 94%, a recall between 76% and 97%, and an F1-score between 82% and 92%. In [35], the size of the used dataset is 2959, the largest one among these three work, but its F1-score is only 62.8%, smaller than the other two. We can see that there are many factors that affect the performance of deep learning models, and the size of the dataset is just one of them. Different factors also affect and constrain each other. We need to analyze and experiment based on specific situations.

### 3.8  Datasets

Datasets are a very important part of deep learning, as they are the foundation for training deep learning models. We focus on the selected datasets for the DL4RE studies and summarize the results into a bubble chart, as shown in Figure 3. The x-axis indicates the RE tasks that the studies focused on, and the y-axis shows the size of data entries used in training and testing the involved DL models. The size of the bubble indicates the number of publications. As shown in Figure 3, we can see that 61% of the selected publications have a dataset size of less than 6,000, which may be due to the difficulty and cost of data collection. In some fields, data may be difficult to obtain and require professional equipments, personnel, or licenses.

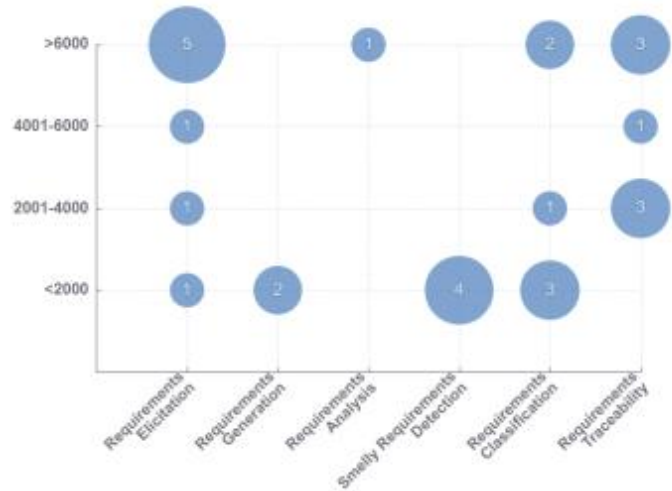

**Figure 3**   The size distribution of the datasets involved in the DL4RE studies.

We can also find that the datasets used in the tasks of requirements detection, requirement generation, and requirement classification are relatively smaller, which may be due to the phenomenon of data-hungriness in the RE community. As software requirements often embed the value that software products aim to deliver to their users, the requirements documents are typically private and confidential. Thus, it is difficult for DL4RE studies to obtain enough data for training effective DL models. Requirements generation is a text generation problem that can be improved by utilizing pre-trained language models without requiring a large amount of domain data. Besides, some techniques, such as data enhancement, transfer learning, and miniaturization networks, can be leveraged to reduce corpus size. In the meanwhile, requirements elicitation studies have larger datasets since it is a comprehensive task involving knowledge in multiple fields and multiple information sources. Moreover, many studies leverage the data in the open-source community, such as feature requests or live chats, to build their deep-learning models.

### 3.9   Challenges and Opportunities

#### 3.9.1   Challenges

- **Low transparency of public requirements.** Unlike other software development activities, the publicly accessible data for requirements typically are small in volume and lack details. For proprietary software, since requirements reflect the delivered value of the software, organizations usually consider requirements as confidential assets and are reluctant to open them. For open-source software (OSS), their requirements are scattered in massive informal online discussions, such as issue reports [34] or live community chats [30]. Although there are some widely-used requirements benchmarks, such as PROMISE and MODIS, the scale cannot be compared to that of open-source code in GitHub. Besides, many publicly accessible benchmarks depict requirements in textual and entry-oriented formats, lacking details on project-level information, such as detailed background, stakeholder preferences, and graphic presentations. Thus, it becomes difficult for RE researchers to comprehensively understand the rationale behind the textual requirements, as well as efficiently train DL models for RE.

- **High diversity in representations.** Software requirements are typically depicted in a variety of formats, including textual formats, data-entry formats, user stories, use cases, as well as (semi-) structured models. Notably, the representation within each format can also vary significantly. For instance, when it comes to modeling requirements, representations might take the form of Unified Modeling Language (UML), Systems Modeling Language (SysML), Goal-oriented models, etc.

Although guideline standards like ISO/IEC/IEEE 29148:2018 [67] and Easy Approach to Requirements Syntax (EARS) [68] are provided, their adoption in the industry is limited. It is reported that nearly half (47%) of RE practitioners are unaware of these standards [69]. Consequently, the representation of requirements often aligns with individual writing practices rather than standardized guidelines. This significant diversity in representations poses a considerable challenge for DL models in accurately interpreting requirements. Additionally, it presents a notable obstacle in training DL models too, compounding the complexity of achieving reliable performance across varied formats.

- **Hidden domain-specific knowledge and experiences.** Many requirements-related tasks (such as specification, modeling, and analysis) hinge on domain-specific knowledge and the accumulated experience of experts. For example, it is commonly accepted that high-quality requirements must embody three key characteristics: Correctness, Consistency, and Completeness — collectively known as the '3Cs'. Yet, these attributes are not clearly and formally defined in the industry, nor are there universally recognized detailed metrics to evaluate whether a set of requirements meets the 3C standard. Consequently, this aspect of the work often depends on the nuanced insights of experienced engineers. Regrettably, requirements engineers typically do not document this critical domain knowledge or their hands-on engineering experiences. This documentation is not typically required by project management protocols. Moreover, the challenge of accurately capturing the objective essence of such experiential knowledge remains unresolved. This gap means that DL models are currently difficult to absorb and apply such valuable expertise to support RE tasks.

### 3.9.2   Opportunities

- **Needs for research on piecing scattered OSS requirements into a big picture.** The influence and momentum of OSS on other SE tasks, like code generation and repair, is substantial and revolutionary. To advance AI4RE research, we appeal to construct evolving requirements for OSS. As the above discussion, OSS requirements are scattered in massive informal online discussions. Therefore, it is crucial to embark on research focused on pinpointing discrete pieces of requirements information, assembling them into coherent and intelligible requirements documents, and ensuring their evolution in tandem with changes in the code base.

- **Exploring the feasibility and usability of AI-based RE approaches in real industrial settings.** While AI-based approaches have demonstrated efficacy in a multitude of tasks, their effectiveness for RE-tasks in real-world industrial settings remains uncertain. Indeed, there was even evidence that practicing engineers are reluctant to rely on AI models for higher-level design goals [70]. This presents a significant opportunity to delve into and expand upon this uncharted area of RE research.

- **Leveraging the interaction capability of LLMs to enhance requirements activities.** High-quality requirements lay the foundation for robust software products. In essence, the success of LLMs in various SE tasks is contingent upon the quality of these underlying software requirements. Given the impressive interactive capabilities of LLMs, a promising avenue is to harness these models to engage with diverse stakeholders intelligently. Through such interactions, LLMs can be leveraged to efficiently carry out tasks like requirements discovery, negotiation, and validation, anchored by rapid prototyping.

## 4   Code Generation

In practice, a highly efficient code generation model may significantly boost developers' productivity, enabling them to complete programming tasks by simply inputting target descriptions into code generators [71]. Therefore, automated code generation can be deemed as an important objective for developers, highlighting the crucial role of the code generation task in the software engineering (SE) field. In addition, code completion is one of the most widely used code generation tasks in integrated development environments (IDEs) [72]. According to the study on the Eclipse IDE by Murphy et al. [73], code completion is one of the top ten commands used by developers. Recent years have seen increasing interest in code completion systems using deep learning techniques [74,75].

To achieve this goal, within the academic community, a range of studies [76] have put forth deep learning (DL)-based models for automatically generating code fragments from input utterances. To gain a deeper understanding of these advanced techniques, in this section, we survey different studies on DL-based code generation. In particular, we survey the following three types of research: 1) research on proposing new techniques, 2) research on empirical evaluations, and 3) research on constructing datasets. Due to the specific characteristics of code, it is necessary to invent specialized deep learning techniques to deal with code. Therefore, we can see a number of influential deep learning techniques for code stem from the area of code generation. Table 4 highlights the main technical contributions in code generation. Since code completion is one of the most important code generation applications, we also survey deep learning based code completion.

The primary contributions of the analyzed studies can be classified into nine distinct categories, highlighting the diverse range of advancements and innovations within code generation.

## 4.1   Enhancing Code Structure Information

Six studies highlight the oversight of rich structural information in many code generators [77]. To address this concern, diverse approaches have been proposed for incorporating additional code structure information into their DL-based models, aiming to enhance the overall performance. Rabinovich et al. [78] introduced abstract syntax networks, a DL-based model that incorporates supplementary structural information from abstract syntax trees (ASTs). Their model utilizes an encoder-decoder BiLSTM with hierarchical attention, aiming to generate well-formed and executable Python code fragments. Jiang et al. [79] observed that the standard Seq2Tree model translates the input natural language description into a sequence based on the pre-order traversal of an Abstract Syntax Tree (AST). However, this traversal order might not be optimal for handling multi-branch nodes in certain cases. To address this, they put forward the idea of enhancing the Seq2Tree model with a context-based Branch Selector, enabling it to dynamically determine the optimal expansion orders for multi-branch nodes.

## 4.2   Special Code Generation

Four studies employ DL techniques to generate code fragments for less common programming languages or in unusual logical forms. Yuetal. [84] introduced a novel DL model specifically designed for generating SQL code for test code scenarios. Yang et al. [86] introduced a pre-trained model to generate assembly code from NL descriptions.

## 4.3   Multi-mode based Code Generation

The studies in this category construct the code generators by taking into account multiple code artifacts in a comprehensive manner. Le et al. [90] noticed that most code generators overlook certain crucial yet potentially valuable code specifications, such as unit tests, which frequently leads to subpar performance when addressing intricate or unfamiliar coding tasks. They thus introduced a new generation procedure with a critical sampling strategy that allows a model to automatically regenerate programs based on feedback from example unit tests and NL descriptions. Wang et al. [91] integrated NL descriptions with the specific attributes of programming languages, such as token types, to construct CodeT5, which is a unified pre-trained encoder-decoder Transformer model designed for generating code fragments.

## 4.4   Compilability

Three studies concentrate on the development of DL-based code generators with the objective of generating executable code fragments across multiple programming languages. Sun et al. [92] identified a significant number of inaccuracies and non-executable SQL code generated as a result of the mismatch between question words and table contents. To mitigate this problem, they took into account the table structure and the SQL language syntax to train a DL-based SQL generator. Wang et al. and Poesia et al. [93,94] leveraged the powerful capability of large language models (LLMs) to improve the compilability of generated code.

**Table 4** Characteristics of studies within the **Design Research** category.

| Contribution | Model Type | Programming Language (PL) | Reference |
|---|---|---|---|
| Enhanced Code Structural Info | DL | Python | [78] |
| | DL | Java | [80] |
| | DL | Java, Python | [81] |
| | DL | Python, SQL | [82] |
| | Pre-trained | Python | [79] |
| Special Code Generation | DL | Conditional Statement | [83] |
| | DL | SQL (Test Code) | [84] |
| | DL | Pseudo-code | [85] |
| | Pre-trained | Assembly, Python | [86] |
| Multi-mode based | DL | Java, Python | [87,88] |
| | Pre-trained | Python | [89,90] |
| | Pre-trained | Go, Java, JavaScript, PHP, Python, Ruby | [91] |
| Compilability | DL | SQL | [92] |
| | Pre-trained | Python | [93] |
| | Pre-trained | SQL, Vega-Lite, SMCalFlow | [94] |
| Dual Learning based | DL | Java, Python | [95] |
| | Pre-trained | Java | [96] |
| | Pre-trained | Python, SQL | [97] |
| Search-based | DL | Python | [98] |
| | DL | C++ | [99] |
| | Pre-trained | Java, Python | [100] |
| Context-aware | DL | Java | [101,102] |
| | Pre-trained | Python | [103] |
| Practicality | DL | python, SQL | [104] |
| | DL | Javascript | [105] |
| Long Dependency Problem | DL | Python | [77,106,107] |

## 4.5    Dual-Learning based Code Generation

Three studies capitalize on the dual connections between code generation (CG) and code summarization (CS) to generate accurate code fragments from NL descriptions. Wei et al. [95] exploited the duality between code generation and code summarization tasks to propose a DL-based dual training framework to train the two tasks simultaneously. Ahmad etal. and Yeetal. [96,97] leveraged the inherent relationship between CG and CS and utilized pre-trained models to enhance the accuracy of the code generation task.

## 4.6    Code Generation on Top of Existing Code

Several studies have observed that generating code based on existing related code fragments can yield superior performance compared to generating code from scratch. As a result, these studies incorporate code search techniques and leverage DL to construct their code generators. For instance, Hashimoto et al. and Kulal et al. [98,99] employed DL-based retrieval models to generate Python and C++ code. Additionally, Parvez et al. [100] utilized information retrieval techniques along with pre-trained models to develop a code generator capable of generating code in multiple programming languages, such as Java and Python.

## 4.7    Context-aware Code Generation

Recognizing that the code fragments generated by many existing code generators may not be directly applicable in software, several studies have proposed to incorporate code contexts to enhance the accuracy of code generation. Guo et al. [102] introduced a context-aware encoder-decoder model with a latent variable in their code generator, enabling it to incorporate the contextual environment during code generation. Li et al. [103] proposed SKCODER, an approach for sketch-based code generation, aiming to simulate developers' code reuse behavior. SKCODER retrieves a similar code snippet based on an NL requirement, extracts relevant parts as a code sketch, and then modifies the sketch to generate the desired code.

## 4.8    Practicality

Two relevant studies aim to improve the practicality of code generators. Dong et al. [104] introduced a structure-aware neural architecture for code generation that exhibits adaptability to diverse domains and representations. Shen et al. [105] proposed a task augmentation technique that integrates domain knowledge into code generation models, making their model the first domain-specific code generation system adopted in industrial development environments.

## 4.9    Long Dependency

Many deep learning-based code generators are trained using recurrent neural networks (RNNs) such as LSTM [108], BiLSTM [109], and GRU [110]. To overcome the long-term dependency problem, three studies introduce novel techniques to tackle this challenge. Sun et al. [77] proposed a novel tree-based neural architecture and applied the attention mechanism of Transformers to alleviate the long-dependency problem. Xie et al. [107] utilized mutual distillation learning to train a code generator in order to avoid the occurrence of this problem.

## 4.10    Code Completion

Table 5 shows code completion techniques published in the premier publication venues (i.e., ASE, ICSE, and FSE) from 2020 to 2023. Since 2020, there have been six papers focusing on the code completion task, in which three of them exploit the pre-trained models or large language models. GPT-2 is widely used in completing source code. Among all programming languages, Java and Python are the most popular programming languages. Wang et al. [111] conducted an empirical study that investigated developers' perspectives on code completion. Liu et al. [112] and Izadi et al. [113] integrated the multi-task learning techniques by learning different types of information of the source code (e.g., token sequences and ASTs). Tang et al. [114] introduced the retrieval-augmented language model to conduct domain adaption and improve the performance of existing LLMs (e.g., ChatGPT and UniXcoder) on the code completion task. Their results show that retrieval techniques can be seamlessly integrated with black-box code completion models and as a plugin to further enhance the performance of LLMs.

**Table 5** Deep Learning based Code Completion Studies

| Year | Venue | Literature | Language | Model |
|------|-------|------------|----------|-------|
| 2020 | ASE | [112] | Java, TypeScript | Multi-task learning, Pre-trained models |
| 2022 | ICSE | [113] | Python | Multi-task learning, GPT-2 |
| 2023 | ASE | [114] | Java, Python | Retrieval-augmented language model |
| 2023 | FSE | [115] | Java, Python | GPT-2, CodeT5 |
| 2023 | FSE | [111] | - | Empirical Study |
| 2023 | ICSE | [116] | Java | Transformer |

## 4.11 Empirical Studies

Four studies [117–120] undertake empirical investigations to examine the characteristics of existing code generators. Dahal et al. [117] leveraged text-to-tree, linearized tree-to-tree, and structured tree-to-tree code generation models to perform an empirical analysis of the significance of input forms for code generation. They found that using a structure-aware model improves the performance of models on both two datasets. Norouzi et al. [118] examined whether a generic transformer-based Seq2Seq model can achieve competitive performance with the minimal design of code-generation-specific inductive bias. They observed that it is possible for a transformer-based Seq2Seq model with minimal specific prior knowledge to achieve results that are superior to or on par with state-of-the-art models specifically tailored for code generation. Mastropaolo et al. [119] presented an empirical study to investigate whether different but semantically equivalent NL descriptions yield the same code fragments. The experimental results demonstrate that modifying the description leads to generating different code in approximately 46% of cases. Furthermore, differences in semantically equivalent descriptions can have an impact on the correctness of the generated code (±28%). Xu et al. [120] conducted a comprehensive user study on code generation and retrieval within an integrated development environment (IDE), developing an experimental harness and framework for analysis. They noticed that developers raise concerns about the potential side effects of code generators on their workflow, encompassing aspects such as time efficiency, code correctness, and code quality.

## 4.12 Datasets

There are four studies that construct available benchmark datasets for the code generation task. Table 6 provides detailed information about the studies within the dataset construction category. Specifically, Iyer et al. [101] emphasized the significance of code contexts in the code generation task. To achieve accurate code fragment generation based on corresponding code contents and natural language descriptions, they developed CONCODE, a dataset including over 100,000 examples comprising Java classes sourced from online code repositories. Liang et al. [121] introduced a novel code generation task: to generate a program in a base imperative language with an embedded declarative language, given a natural language comment. To support this task, they created a dataset (i.e., Lyra) consisting of 2,000 carefully annotated database manipulation programs extracted from real-world projects. Each program is associated with both a Chinese comment and an English comment. To accurately assess the performance of code generation, Hendrycks et al. [122] introduced APPS, a benchmark specifically tailored for code generation in more restricted settings compared with the prior benchmarks. Their benchmark comprises 10,000 problems, spanning from simple online solutions to substantial algorithmic challenges. Lu et al. [123] developed a comprehensive dataset known as CodeXGLUE. This dataset covers a diverse set of 10 tasks across 14 different datasets, encompassing eight programming languages, and it serves as a platform for evaluating and comparing models in the field of code generation.

## 4.13 Challenges and Opportunities

In this section, we highlight some challenges and opportunities for future research in deep learning techniques for code generation:

**Table 6** The detailed information of studies within the **Dataset Construction** category.

| PL | Dataset Name | Reference |
|---|---|---|
| Java | CONCODE | [101] |
| Python | Lyra | [121] |
| Python | APPS | [122] |
| Python, Java, PHP, JavaScript, Ruby, Go, C/C++ | CodeXGLUE | [123] |

### 4.13.1 Challenges

• **Unsafe Code Generation.** Deep learning techniques, especially recently proposed large language models (LLMs), are (pre-)trained on massive code bases and then applied to code generation. The code bases may include vulnerable code snippets that lead to generation of unsafe code. Thus, how to generate functionally correct and safe code is challenging.

• **Benchmarks.** Existing benchmarks for code generation mainly include hand-written programming problems and their corresponding solutions (such as HumanEval). However, there is a huge difference between these human-written benchmarks and real projects. In addition, the human-written benchmarks are time-consuming and strongly dependent on experts' knowledge. Thus, constructing a benchmark from real projects automatically is important for code generation.

• **Hallucination of LLMs.** Recently, many LLMs have been exploited for code generation, such as Copilot. Existing studies have shown that LLMs, such as ChatGPT, often generate fabricated or inaccurate responses, which are commonly referred to as the hallucination phenomena [124]. Hallucination makes LLMs not always reliable in generating code snippets. It is crucial to combine the capabilities of ChatGPT with human expertise to ensure the quality and reliability of the generated code.

### 4.13.2 Opportunities

• **Knowledge-Augmented Code Generation.** Existing studies have shown that recently proposed LLMs can generate code effectively. To better adapt LLMs to generate code for a specific domain, knowledge-augmented code generation is helpful. It thus is important to integrate different information, such as project information and similar code snippets, to boost existing LLMs.

• **Managing datasets as software.** Datasets (including the training data and benchmarks) are important in training and evaluating a code generation model. As more and more datasets have been proposed in recent years, we need to better manage the datasets for code generation models. Nowadays, the datasets are also evolved and hosted in collaborating platforms, such as GitHub and Hugging Face. Similar to software, we should improve dataset productivity, quality, and security.

## 5 Code Search

Code search is the process of finding relevant code snippets from online or local code repositories based on query statements, typically expressed in natural language or code itself.

### 5.1 Natural Language based Code Search

#### 5.1.1 Information Retrieval

Early code search engines primarily relied on Information Retrieval (IR) techniques to match query keywords with code snippets. These techniques assess the relevance of queries and code based on their textual similarity [125–128]. Subsequent approaches enhance code search by delving into the structural aspects of source code. They consider diverse relationships between code entities and sought matches between relevant APIs. A notable trend is the representation of code as directed graphs, effectively transforming the search task into a graph exploration problem. For instance, McMillan et al. [129] introduced Portfolio, a tool that pinpoints potential methods containing query keywords within an API call graph. The tool then ranks the results using a combination of PageRank scores to evaluate node importance and Spreading Activation Network (SAN) scores to gauge query relevance. In a similar vein, Li et al. [130] presented RACS, a technique founded on relations. It parses natural language queries into action graphs and code into relation invocation graphs. This enables a structural alignment between the two graph representations, thereby elevating the accuracy of matches. However, these approaches encounter limitations stemming from the pronounced disparities between programming languages and natural languages. Consequently, the comprehension of semantics remains challenging for IR-based approaches.

### 5.1.2 Deep Learning

To establish more robust semantic connections between natural language queries and code, researchers have increasingly turned to deep learning models to tackle code search tasks. The fundamental approach involves encoding both the query statement and code into separate vector representations, then assessing the semantic correlation between the two through vector similarity analysis. Gu et al. [19] innovatively employed perceptrons and Recurrent Neural Networks (RNNs) to embed various code elements such as method names, API call sequences, and code sequences into a shared high-dimensional space. This lays the foundation for DeepCS,a code search tool. Sachdevetal. [131] introduced NCS (Neural Code Search), tailored for extensive code repositories. NCS combines word embeddings with TF-IDF to generate vector representations for code snippets and query statements. It then gauges relevance through vector distances, simulating the significance of code snippets in relation to queries.

As deep learning progresses, subsequent research integrates more complex representations and more sophisticated models for code vectorization. Ling and Zou [132] introduced a novel source code search technique employing graph embedding. It involves creating a code graph from a software project's source code, representing code elements using graph embedding, and then utilizing this structure to answer natural language queries by returning relevant subgraphs composed of code elements and their relationships. Gu et al. [133] proposed CRaDLe, a novel approach for code retrieval based on statement-level semantic dependency learning. CRaDLe distills code representations by merging dependency and semantic information at the statement level, ultimately learning unified vector representations for code-description pairs to model their matching relationship. Wanetal. [134] introduced MMAN,a Multi-Modal Attention Network designed for semantic source code retrieval. They created a holistic multi-modal representation by utilizing LSTM for sequential tokens, Tree-LSTM for code's AST, and GGNN for its CFG, followed by a multi-modal attention fusion layer that combines and assigns weights to different components for an integrated hybrid representation. Ling et al. [135] introduced an end-to-end deep graph matching and searching (DGMS) model for semantic code retrieval. They represented query texts and code snippets as unified graph-structured data, and used the DGMS model to retrieve the most relevant code snippet by capturing structural information through graph neural networks and fine-grained similarity through cross-attention based semantic matching operations. Liu et al. [136] presented GraphSearchNet, a neural network framework that improves source code search accuracy by simultaneously learning from source code and natural language queries. They introduced bidirectional GGNN (BiGGNN) to create graphs for code and queries, capturing local structural details, and enhanced BiGGNN using a multi-head attention module to incorporate global dependencies for enhanced learning capacity. Li et al. [137] introduced CodeRetriever, which obtains function-level code semantic representations via extensive code-text contrastive pre-training. This involves unimodal contrastive learning that uses function names and documentation to build code pairs, and bimodal contrastive learning that utilizes code comments and documentation for code-text pairs, both contributing to effective pre-training using a vast code corpus. Jiang et al. [138] introduced ROSF, a technique that enhances code snippet recommendations by combining information retrieval and supervised learning. The approach involves two stages: generating a candidate set using information retrieval and then re-ranking the candidates based on probability values predicted by a trained model, resulting in improved code snippet recommendations for developers.

In recent years, significant progress has been made in the realm of large pre-trained models based on the Transformer architecture, driving advancements across numerous NLP tasks. This progress leads to the emergence of code understanding pre-training models leveraging transformers, fostering the growth of code intelligence. For instance, Feng et al. [63] introduced CodeBERT, a pioneering large-scale pre-trained model that integrates natural language and programming language understanding across multiple programming languages. CodeBERT harnesses Masked Language Modeling (MLM) to capture the semantic relationship between natural language and code. Researchers have explored the incorporation of multiple modal representations of source code into the Transformer paradigm to gain a comprehensive understanding. Guo et al. [139] developed the GraphCodeBERT model, seamlessly combining the variable sequence from data flow graphs with the code token sequence. This model undergoes training via MLM, Edge Prediction (EP), and Node Alignment (NA) tasks to encompass both code structures and data dependencies. Similarly, Guo et al. [140] introduced UniXcoder, which fuses serialized Abstract Syntax Trees (ASTs) with comment text sequences. By utilizing MLM, Unidirectional Language Modeling (ULM), DeNoiSing (DNS), Multi-modal Contrastive Learning (MCL), and Cross-Modal Generation (CMG), this model enriches its comprehension of code syntax and semantics. Some researchers further leveraged contrastive learning to enhance model performance. Shi et al. [141] introduced CrossCS,a technique that improves code search through cross-modal contrastive learning. They devised a novel objective considering both inter- and intra-modality similarity, used data augmentation for semantic consistency, and boosted pre-trained models by ranking code snippets with weighted similarity scores based on retrieval and classification scores. Bui et al. [142] presented Corder, a self-supervised contrastive learning framework for source code models. It aims to reduce the need for labeled data in code retrieval and summarization tasks by training the model to differentiate between similar and dissimilar code snippets using contrastive learning and semantic-preserving transformations. Additionally, Shi et al. [143] introduced CoCoSoDa, which employs contrastive learning for code search, incorporating soft data augmentation and negative samples. They also applied multimodal contrastive learning to enhance code-query pair representations.

### 5.1.3 Query Expansion and Refinement

Significant differences in expression and vocabulary between natural languages and code are key factors contributing to the mismatch between high-level intents implied in natural languages and low-level code implementations [19], impacting the accuracy of code search. Improving the query statement or the candidate code has been proved to be an essential approach for enhancing code search effectiveness. Bajracharya et al. [144] introduced Sourcerer, an open-source code search engine that extracts detailed structural information from code and stores it in a relational model. This information facilitates the implementation of CodeRank and supports search forms beyond traditional keyword-based searches. Lu et al. [145] introduced an approach that extends queries using synonyms from WordNet, which involves extracting natural language phrases from source code identifiers, matching expanded queries with these phrases, and sorting the search results. Fei et al. [146] introduced CodeHow, a code search technique capable of recognizing potential APIs referenced in a user query. After identifying relevant APIs, CodeHow expands the query with these APIs and performed code retrieval using the extended boolean model, incorporating both text similarity and potential APIs for improved search. Mohammad et al. [147] used context-awareness and data analysis to apply appropriate term weighting in query reformulation, thereby enhancing code search. Hill etal. [148] presented a search technique based on method signature analysis, involving the rewriting of code method names and subsequent matching of the altered method names with queries to facilitate the search process. Additionally, Liu et al. [149] introduced the NQE model, which predicts keywords related to the query keywords in the corpus based on natural language queries. This technique expands query statements and subsequently improves code search effectiveness. Alongside utilizing identifiers in source code, researchers explores leveraging search logs from platforms like stack overflow to enhance code search. Cao et al. [150] analyzed large-scale search logs from stack overflow to identify patterns in query reformulation. They constructed a corpus encompassing both original queries and their reconstructed versions, and then trained a model using this corpus. The trained model can generate a list of candidate reconstructed queries when provided with a user query, offering improved search options. Li etal. [151] introduced a generation-augmented query expansion framework that utilizes code generation models to enhance code retrieval. Instead of relying solely on documentation queries, the approach involves augmenting queries with generated code snippets from the code generation model, drawing inspiration from the human retrieval process of sketching an answer before searching.

## 5.2 Code-to-Code Search

In addition to searching for code based on natural language input, code snippets are also utilized as input for code search, divided into searching within the same programming language and across different programming languages. A notable work for searching within the same language is Aroma proposed by Luan et al. [152]. Aroma takes incomplete code snippets as input and searches for similar complete code snippets from pre-indexed open-source projects. Compared to searching within the same language, cross-language code search is more challenging due to syntactic and semantic differences across languages. Mathew et al. [153] introduced the COSAL approach, which performs non-dominated sorting based on similarities between code snippets, including AST structures and input-output behaviors, to facilitate code search within the same language and across languages. Additionally, cross-language code search is used for code translation, such as converting Java code into Python code with the same functionality. Perez et al. [154] employed LSTM networks to model clone similarity between cross-language code snippets based on ASTs, and Nguyen et al. [155] utilized the API2Vec model, inspired by Word2Vec, to embed APIs into high-dimensional vectors for cross-language code translation. Chen et al. [156] introduced BigPT, a technique for interactive cross-language retrieval from Big Code, involving a predictive transformation model based on auto-encoders to aid program translation using retrieved code representations. Users are able to further refine the retrieval results to improve the process.

**Table 7** Natural Language based Code Search Datasets

| Dataset | Language | Size | Source | Release Year |
|---|---|---|---|---|
| StaQC [133] | Python, SQL | 267k | SO | 2018 |
| CoNaLa [135] | Python, Java | 2.8k | SO | 2018 |
| FB-Java [141] | Java | 287 | SO, GitHub | 2019 |
| CodeSearchNet [142] | Python, Java, Ruby, Go, PHP, JavaScript | 2M | GitHub | 2019 |
| SO-DS [143] | Python | 2.2k | SO, GitHub | 2020 |
| CosBench [151] | Java | 52 | SO, GitHub | 2020 |
| CodeXGLUE [128] | Python | 281k | Bing, GiHub | 2020 |
| CoSQA [157] | Python | 20k | Bing, GitHub | 2021 |
| XCodeEval [158] | C#, C++, C, D, Go, Haskell, Java, Javascript, Kotlin, Ocaml, Pascal, Perl, PHP, Python, Ruby, Rust, Scala | 11k | Codeforces | 2023 |

### 5.3 Datasets

The following datasets in Table 7 are commonly used for natural language based code search.

The StaQC dataset [133] is tailored for predicting the suitability of code snippets in addressing specific queries. Comprising (question, code) pairs, it was curated by filtering Python and SQL Stack Overflow posts tagged with "how-to" questions, resulting in 147,546 Python pairs and 119,519 SQL pairs.

The CoNaLa dataset [135] consists of 2,379 training and 500 test examples, manually annotated with natural language intents and corresponding Python snippets.

The FB-Java dataset [141] comprises 287 natural language queries and relevant code snippet answers from Stack Overflow threads tagged with "java" or "android". Additionally, it includes code snippet examples from the search corpus, sourced from public repositories on GitHub, that correctly answer the corresponding queries.

The CodeSearchNet corpus [142] is an extensive collection of approximately 6 million functions automatically gathered from open-source code spanning six programming languages (Go, Java, JavaScript, PHP, Python, and Ruby). It includes 2 million functions with query-like natural language descriptions obtained via scraping and preprocessing associated function documentation. Furthermore, it contains 99 natural language queries with around 4,000 expert relevance annotations of likely results from the CodeSearchNet Corpus.

The SO-DS corpus [143] consists of code snippets mined from Stack Overflow posts with the most upvoted posts labeled with "python" and tags related to data science libraries such as "tensorflow," "matplotlib," and "beautifulsoup." The ground truth is collected by creating queries from duplicate Stack Overflow posts, resulting in 2,225 annotated queries.

The CosBench corpus [151] comprises 475,783 Java files and 4,199,769 code snippets (Java methods) extracted from the top-1000 popular Java projects on GitHub. It includes 52 queries with ground truth code sbippets indicating three types of intentions: bug resolution, code reuse, and API learning, chosen from Stack Overflow.

The CodeXGLUE [128] serves as a benchmark dataset and an open challenge for code intelligence, encompassing various code intelligence tasks. For NL-based code search, it includes two sub-datasets: AdvTest and WebQueryTest. AdvTest, constructed from the CodeSearchNet corpus [142], uses the first paragraph of documentation as a query for the corresponding function. WebQueryTest is a testing set of Python code questions answered with 1,046 query-code pairs and expert annotations.

The CoSQA dataset [157] is based on real user queries collected from Microsoft's Bing search engine logs. It encompasses 20,604 labels for pairs of natural language queries and code snippets, each annotated by at least 3 human annotators.

The xCodeEval [158] is recognized as one of the most extensive executable multilingual multitask benchmarks, encompassing seven code-related tasks that span across 17 programming languages. Derived from a pool of 25 million openly available samples from codeforces.com, a platform hosting competitive programming contests, this dataset comprises 7,514 distinct problems. In terms of code retrieval, xCodeE-val introduces a novel and more demanding task, specifically centered on matching a natural language problem description to the most relevant and accurate code within a candidate pool containing similar solutions. To facilitate this, all submitted code snippets and their associated test cases are aggregated for each programming language, creating a retrieval corpus and a suite of test cases. The primary objective is to evaluate the correctness of these code snippets against the provided test cases. In this context, the natural language problems serve as queries, and the correct solutions, verified by successful execution outcomes (PASSED), are considered as the ground truth.

For code-to-code search, existing datasets designed for code clone detection, featuring clusters of se-mantically equivalent implementations, can be utilized. One such dataset is the BigCloneBench provided within CodeXGLUE [128]. In these datasets, a group typically comprises variations of the same imple-mentation. In practice, one implementation within a group can serve as a query, while the remaining implementations within the cluster act as the ground truth. Additionally, xCodeEval [158] offers a dataset specifically tailored for code-to-code tasks. This dataset consists of 9,508 queries, created from correct submitted solutions to the same natural language problems, adding diversity to the evaluation of code-to-code search capabilities.

## 5.4  Challenges and Opportunities

### 5.4.1  Challenges

- Quality Assurance of Search Results. Ensuring the quality of code search results goes beyond merely matching search intent. Factors such as correctness, security, and timeliness must be considered to guarantee the reliability and suitability of the returned code snippets.

- Long Tail Issues. Addressing the challenges posed by less common, long-tail issues is essential. Code search systems need to effectively handle diverse and infrequent queries, ensuring comprehensive coverage across a spectrum of coding scenarios.

- Result Interpretability. Achieving interpretability in code search results is also a challenge. Only little research effort (e.g., [159]) on this direction. It involves presenting search outcomes in a clear and understandable manner, aiding developers in comprehending the context and relevance of the returned code snippets.

- Integrating Retrieved Code into Development Context. Effectively utilizing code search results poses a significant challenge. Typically, developers expend considerable effort in adapting and integrating a retrieved code snippet into the current context of their code development.

- Ambiguity in Search Intents. The inherent ambiguity in certain search intents poses a challenge. Code search systems need to navigate and interpret vague or imprecise queries to provide relevant and accurate results.

- Dataset Quality Issues. Existing datasets used for training and evaluating code search models, such as CodeSearchNet [142], often consider method comments as search queries, which fails to align well with real-world code search scenarios [160]. While this can lead to strong model performance during evaluation, it may not be effectively transferred into practical use.

• Efficiency vs. Effectiveness. Learning-based code search has shown promising search accuracy, but it often demands substantial computing and storage resources. Although certain approaches [161, 162] have been proposed to address efficiency concerns in learning-based code search, there is still significant room for improving the trade-off between efficiency and effectiveness.

### 5.4.2 Opportunities

• Incorporating Richer Input Information. Expanding the scope of input information beyond the query itself presents an opportunity. Incorporating details from IDE development contexts, historical patterns, and personal preferences can enhance the accuracy and relevance of code search results.

• Enhanced Code Search with LLMs. Leveraging advancements in LLMs offers an opportunity to augment code search capabilities. Focusing on result interpretability, code auto-adaptation, and harnessing the advanced natural language and code semantic understanding abilities of LLMs can elevate the efficiency of code search systems.

• Improved Intent Clarification Techniques. Advancing techniques for Intent clarification in code search represents an opportunity. Developing techniques to better understand and refine user search queries contributes to a more streamlined and effective code search experience.

• High-quality Dataset Construction. This presents an opportunity for the community to enhance the quality of current datasets [160] or create more reliable ones for code search evaluation.

## 6 Code Summarization

Code summarization, also known as code comment generation, is a process that aims to enhance the understanding and documentation of source code by automatically generating concise and informative summaries for software artifacts [63,163–165]. It helps address the challenge of comprehending large and complex code repositories by providing developers with high-level descriptions that capture the code's essential functionality and usage patterns.

Code summarization has been a hot research topic in software engineering in recent years. Initially, researchers explore template-based methods like SWUM [166] and Stereotypes [167] for generating code comments automatically; Meanwhile, information retrieval-based techniques such as VSM [168] and LSI [169] are also applied for code summarization. However, with the rapid advancements in deep neural networks within machine learning, deep learning-based approaches have gained momentum and become predominant in code summarization research. Typically, researchers leverage deep learning models to capture implicit relationships between relevant information within source code and natural language descriptions. These approaches have significantly contributed to the development of code summarization, facilitating more effective comprehension and documentation of software products.

Deep learning-based approaches for comment generation primarily mimic the Neural Machine Translation (NMT) [170] models in natural language processing. However, compared to translation tasks in natural language, source code typically has a much greater length than comments and contains rich structural information. Most deep learning-based research takes the source code token sequence as the input to the model, while some studies also consider other information sources, such as Abstract Syntax Trees (ASTs), API, and so on. We divide these studies into five categories based on different sources of information: techniques that utilize source code sequences as the model input, techniques that employ Abstract Syntax Tree (AST) sequences as the model input, techniques that use tree structures as the model input, techniques that utilize graph structures as the model input, and techniques that consider other sources of information. These techniques generate comments for code snippets (class level, function level or line level), as well as comments for code commits (i.e., commit messages).

## 6.1  Using source code sequences as model input

A source code sequence refers to a simple stitching of a code snippet into a sequence with a token as a basic unit. Using source code sequences as model input is simple and convenient and could preserve the most original semantic information of the code.

Iyer et al. [17] proposed CODE-NN, the first deep learning model in code summarization, which uses LSTM network structures and attention mechanisms to generate natural language descriptions of C# code and SQL code. Allamanis et al. [171] applied convolutional neural networks (CNNs) with attention mechanisms to an encoder that helps detect long-range topical attention features and local time-invariant features of code sequences. Ahmad etal. [172] first used the Transformer model for source code summarization, innovatively adding an attention layer to the encoder for replicating rare tokens of the source code. Wang et al. [173] proposed Fret, which combines Transformer and BERT to bridge the gap between source code and natural language descriptions and alleviate the problem of long dependencies. Zhang et al. [174] tried to fuse two techniques: deep learning and information retrieval. Specifically, they proposed Rencos, which first trains an encoder-decoder model based on a training corpus. Subsequently, two code segments are selected from the training corpus according to the syntax and semantic similarity. Finally, the input code segment and the retrieved two similar code segments are encoded and decoded to generate comments. LeClair et al. [175] explored the orthogonality of different code summarization techniques and proposed an integration module to exploit this for better overall performance. Gong et al. [176] proposed SCRIPT, which first obtains the structural relative position matrix between tokens by parsing the AST of the source code, and then encodes this matrix during the computation of the self-attention score after the source code sequence is input into the encoder.

Some research considers both comment generation and code generation tasks. Chen et al. [177] focused on both code retrieval and comment generation tasks, and they proposed a framework, BVAE, which allows a bidirectional mapping between code and natural language descriptions. The approach attempts to construct two VAEs (variational autoencoders), where C-VAE mainly models code and L-VAE primarily models the natural language descriptions in comments. The technique jointly trains these two VAEs to learn the semantic vectors of code and natural language representation. Similarly, Wei et al. [95] considered code summarization and code generation as dyadic tasks, as there is a correlation between the two tasks. They proposed a dual framework to train both tasks simultaneously. They exploited the pairwise nature and the duality between probability and attention weights. Then they designed corresponding regularization terms to constrain this duality. Clement et al. [89] focused on the Python language and proposed PYMT-5. They also focused on dual tasks: code generation from signatures and documentation generation from method code.

Some researchers focused on comment generation for commits. Jiang et al. [178,179] used NMT to generate concise summaries of commits while designing a filter to ensure that the model is trained only on higher-quality commit messages. Jiang et al. [180] preprocessed code changes into more concise inputs, explicitly using a code semantic analysis approach for the dataset, which can significantly improve the performance of the NMT algorithm. Liu et al. [181] used a modified sequence-to-sequence model to automatically generate PR descriptions based on submission information and source code comments added in pull requests (PRs). Bansal et al. [182] proposed a project-level encoder to generate vector representations of selected code snippets in software projects to improve existing code summarization models. Xie et al. [183] considered method names as refined versions of code summaries. Their approach first uses the prediction of method names as an auxiliary training task and then feeds the generated and manually written method names into the encoder separately. Finally, the outputs are fused into the decoder.

## 6.2  Using AST sequences as model input

An Abstract Syntax Tree (AST) is a hierarchical representation of the syntactic structure of a program or code snippet. An AST represents the structure of the code by breaking it down into its constituent parts and organizing them in a tree-like format. ASTs are commonly used in computer science and programming language theory to analyze and manipulate code. Each node in an AST corresponds to a syntactic element of the code, such as a statement, expression, or declaration.

Hu et al. [184] proposed Deepcom, a technique to preserve the structural information of the code intact by parsing the source code into an AST. The authors designed a new traversal strategy, SBT (Structure-based Traversal), which solves the problem that the source code cannot be effectively restored from the AST sequence. Subsequently, they proposed Hybrid-DeepCom [185] based on DeepCom, which mainly improves DeepCom in three aspects. First, it uses a combination of code information and AST sequence information. Second, the OOV problem is mitigated by subdividing the identifier into multiple words based on the camel naming convention. Finally, Hybrid-DeepCom uses beam search to generate code comments. Huang et al. [186] proposed a statement-level AST traversal approach that preserves both code text information and AST structure information, and achieves good results in code snippet-oriented comment generation tasks. Tang et al. [187] proposed AST-Trans, which exploits two node relationships in ASTs: ancestor-descendant and sibling relationships. The authors applied the attention of a tree structure to assign weights to related nodes dynamically. Liu et al. [188] proposed ATOM, which explicitly incorporates an AST of code changes and utilizes a hybrid sorting module to prioritize the most accurately retrieved and generated messages based on a single code change.

Some approaches receive both AST and source code sequences as input. Wan et al. [189] combined an LSTM that receives code sequences and an LSTM that receives ASTs to extract a hybrid vector representation (named Hybrid-DRL) of the target code synthetically. It further uses a reinforcement learning framework (i.e., actor-critic network) to obtain better performance. LeClair et al. [190] proposed ast-attendgru, which also considers two representations of the code: a word-based text sequence and an AST-based tree structure. It processes each data source as a separate input and later merges the vectors produced by the attention layer. Xu et al. [191] proposed CoDiSum to extract AST structures and code semantics from source code changes and then jointly model these two sources of information to learn the representation of code changes better. Li et al. [192] designed a new semantic parser, SeCNN, using two CNN components that receive source code and AST, respectively, and proposed a new AST traversal technique ISBT to encode structural information more sufficiently. Specifically, they used the serial number of the AST via pre-order traversal to replace the brackets in the SBT sequence. Gao et al. [193] proposed M2TS, which uses cross-modal fusion further to combine AST features with the missing semantic information and highlight the key features of each module. Zhou et al. [192] proposed GSCS, which uses a graphical attention network to process AST sequences and a multi-head attention mechanism to learn features of nodes in different representation subspaces.

### 6.3 Using tree structure as model input

Unlike the approaches discussed in the previous sub-section, which transforms the AST into a sequence, approaches discussed in this sub-section retain the tree structure of the AST directly as input.

Liang et al. [193] proposed Code-RNN based on tree-LSTM, which is for the case where a node has multiple children, thus overcoming the restriction of converting ASTs into binary trees. For decoding, code-GRU is used. Wang et al. [194] built a tree structure based on code indentation, where the nodes of the tree are statements in the code, and statements with the same indentation are sibling nodes. They then fed this tree structure into a tree-transformer-based encoder. Lin et al. [195] partitioned the AST into several subtrees according to the control flow graph of the method, and then fed the subtrees into a tree LSTM for pre-training to obtain their vector representations. They used these representations in the subsequent comment generation task. Similarly, Shi et al. [196] used user-defined rules to split the AST tree hierarchically. The model learns the representation of each subtree using a tree-based neural model, i.e., RvNN. The difference is that RvNN finally combines the representations of all subtrees by reconstructing the split AST to capture the structural and semantic information of the whole tree.

### 6.4 Using graph structure as model input

A graph is a versatile and powerful data structure that captures complex relationships and interconnections among entities. Some approaches treat the source tokens as graph vertices and represent the relationships between tokens by edges.

Fernandes et al. [197] added graph information to sequence encoding. Source code is modeled as a graph structure, which helps infer long-distance relationships in weakly structured data (e.g., text). LeClair et al. [198] used a graph neural network (GNN) based encoder to model the graph form of an AST and an RNN-based encoder to model the code sequences. Liu et al. [199] constructed a code property graph (CPG) based on an AST while augmenting it with CPGs of ASTs of the retrieved similar code snippets. Then, the CPGs are input into a graph neural network for training. Liu et al. [200] proposed a graph convolutional neural network (GCN) based on a hierarchical attention mechanism for encoding graphs parsed from code sequences. The code encoded by the sequence encoder is further combined with the document text information for decoding. Cheng et al. [201] designed three encoders that receive source code sequences, code structure information, and code context information. A bipartite graph is used to represent the structure information evolved from the AST, with the addition of a keyword guidance module.

**Table 8** Datasets for code summarization

| Dataset | Literature | Language | Size |
|---|---|---|---|
| TL-CodeSum | [205] | Java | 87,136 |
| Deepcom | [185] | Java | 588,108 |
| Funcom | [190] | Java | 2.1 million |
| CodeSearchNet | [157] | Go, Java, JavaScript, PHP, Python, and Ruby | 2 million |
| code-docstring-corpus | [214] | Python | 150,370 |
| SCGen | [215] | Java | 600,243 |
| commitMessage | [179] | Java | 2,027,734 |

Guo et al. [202] proposed CODESCRIBE, which models a code snippet's hierarchical syntactic structure (i.e., AST) by introducing new triadic positions. Then, they used a graph neural network and Transformer to preserve the structural and semantic information of the code, respectively. Ma et al. [203] proposed MMF3, which uses a graph convolutional network to encode AST graph embeddings while fusing the sequence of source code features to determine the matching relationship between each token in the code sequence and each leaf node in the AST by comparing the position order. Wang et al. [204] proposed GypSum to introduce specific edges associated with the control flow of code snippets into the AST for building graphs, and designed two encoders for learning from the graph and source code sequences.

## 6.5 Considering other sources of information

Other sources of information include APIs, control flow graphs, unified modeling languages, and byte-code,etc. Hu etal. [205] argued that APIs called within code may provide certain information, and they proposed TL-CodeSum, which first trains the mapping relationship between APIs and code comments and subsequently migrates the learned knowledge to the code summarization task. Shahbazi et al. [206] generated comments using API documentation, code snippets, and abstract syntax trees. They showed that API documentation is an external knowledge source, and the performance improvement is negligible. Gao et al. [207] proposed GT-SimNet, a code semantic modeling approach based on local application programming interface (API) dependency graphs (local ADG). This approach is accomplished by computing the correlation coefficients between dependency graphs and AST nodes.

Zhou et al. [208] proposed ContextCC to obtain ASTs by parsing code to find methods and their associated dependencies (i.e., contextual information) and then generate code comments by combining the filtered contextual information. Wang et al. [209] constructed a type-augmented abstract syntax tree (Type-augmented AST) and extracted control flow graphs (CFGs) as an alternative syntax-level representation of the code, with a hierarchical attention network to encode this data. Wang et al. [210] introduced class names and associated Unified Modelling Languages (UMLs) for method comment generation, where the UMLs are fed into the graph neural network as graph forms. Son et al. [211] found that Program Dependency Graphs (PDGs) can represent the structure of code snippets more effectively than ASTs, proposed an enhancement module (PBM) that encodes PDGs as graph embeddings, and designed a framework for implementing PBMs with existing models. Zhang et al. [212] proposed Re_Trans to enhance structural information by adding data flow and control flow edges to the AST and using GCN to encode the entire AST.

Huanget al. [213] explored the feasibility of using bytecode as a source of information to generate code comments. They used pre-order traversal to serialize the bytecode control flow graph, and combined it with a bytecode token sequence as model input to achieve automatic code summarization in a scenario without available source code.

## 6.6 Datasets

The following datasets in Table 8 are commonly used for automatic code summarization.

TL-CodeSum [205] comprises 69,708 method-comment pairs obtained by crawling Java projects developed between 2015 and 2016, each having a minimum of 20 stars on GitHub. The average lengths of Java methods, API sequences, and comments are 99.94, 4.39, and 8.86, respectively.

Deepcom [185] is collected from GitHub's Java repositories created from 2015 to 2016 considering only those having more than 10 stars to filter out low quality repositories. It uses the first sentences of the Javadoc as the target comments and excludes the setter, getter, constructor and test methods. After the preprocessing, there are 588,108 method-comment pairs in total.

Funcom [190] constitutes a compilation of 2.1 million method-summary pairs derived from the Sourcerer repository. After the removal of auto-generated code and exact duplicates, the dataset is partitioned into training, validation, and test sets by project.

CodeSearchNet [157] is a large well-formatted dataset collected from open source libraries hosted on GitHub. It contains 2 milllion code-summary pairs and about another 4 million functions without an associated documentation, spanning six programming languages (i.e., Go, Java, JavaScript, PHP, Python, and Ruby).

code-docstring-corpus [214] is a dataset downloaded from repositories on GitHub, retaining Python 2.7 code. The dataset contains 150,370 code-comment pairs. The vocabulary size of code and comment is 50,400 and 31,350, respectively.

SCGen [215] is a dataset of Java code snippets constructed from 959 Java projects of GitHub. Data cleaning is performed to filter out invalid data according to templates, such as comments in setter and getter methods or comments generated by the template predefined in the IDE comment plugin. Using a comment scope detection approach, 600,243 code snippet-comment pairs are collected.

commitMessage [179] contains 967 commits from the exsiting work and all the commits from the top 1,000 popular Java projects in Github. The rollback commits, merge commits and the commits with messages that are empty or have non-English letters are filtered. In the end, there are 2,027,734 commits in the dataset.

## 6.7 Challenges and Opportunities

### 6.7.1 Challenges

- High-quality Dataset. The code summarization techniques based on deep learning need a high-quality dataset to improve their performance. Although several datasets have been published in this area, the data is selected from the perspective of the project popularity. As a result, there may be duplicate data and machine-generated comments in the dataset. Developing practical techniques to identify high-quality comments that really reflect the code intent is challenging.

- Evaluation Metrics. Many studies employ the BLEU metric to evaluate the performance of the code summarization models. However, there are many variations of BLEU, which results in different ways of calculation, such as BLEU-L and BLEU-C. On one hand, due to the different emphasis of each variation of BLEU, the performance of the same model shows significant differences under different BLEU metrics. On the other hand, it is possible that the BLEU score does not accurately reflect the actual effect of the generated comment because the BLEU score is based on the repetition of tokes in two sentences. Two sentences with the same semantics but different words have a low BLEU score, which is unreasonable.

- Adaptation Ability : Code summarization needs to be adaptable to various programming languages, each with its own syntax and semantics. Developing a universal summarization model that performs well across diverse languages is a significant challenge.

### 6.7.2 Opportunities

At present, most code summarization models cannot be directly applied to the production practice, and are still in the experimental prototype. It comes down to the fact that the model is not powerful enough to apply. There is still a lot of room for improvement.

- Utilizing more implementation information: When the information at the source code level (e.g., ASTs, tokens, APIs, CFGs) is almost mined, a feasible way maybe to mine more useful information from outside the source code (e.g., bytecode, API documents, design documents) to characterize the internal patterns of the code, and further improve the performance of code comment generation models.

- Considering richer information in the comments: Most code summarization datasets take the first sentences of comment or commit message as the code summary because the first sentences are considered to describe the functionalities of Java methods according to Javadoc guidance. With the development of deep learning models, other information in the summary (e.g., the intent or rationale of the code) may be extracted and used for summary generation.

# 7  Software Refactoring

Software refactoring is to improve software quality by changing its internal structures whereas its external behaviors are kept intact [216]. Ever since Opdyke proposed the concept of software refactoring in 1992, researchers have tried to automate software refactorings aiming to reduce the cost of software refactoring and to improve the safety of refactorings. Thanks to such hard work, automatic or semi-automatic software refactoring has been provided as a default feature in all mainstream IDEs, such as Eclipse, IntelliJ IDEA, and Visual Studio, thus significantly increasing the popularity of software refactorings.

Various techniques have been exploited for software refactorings [217–220] whereas recently deep learning techniques have become the main force in this field [221,222]. Traditional software refactoring heavily depends on static code analysis, code metrics, and expert-defined heuristics to identify code smells [217,218](i.e., what should be refactored) and to recommend refactoring opportunities. However, it is challenging to formalize complex refactorings with human-defined simple heuristics, making traditional heuristics-based refactoring less accurate. In contrast, deep learning techniques with complex networks and numerous weights, have the potential to learn complex refactorings [219]. Consequently, various deep learning techniques have been recently employed for software refactoring.

However, applying deep learning techniques to software refactorings is nontrivial, encountering a sequence of challenges. The first challenge is to collect a large number of high-quality items requested for training. Since deep models often contain a large number of parameters, they usually request a large number of labeled items as training data. However, we lack such large-scale high-quality datasets in the field of software refactoring. The second challenge is to figure out how deep models (deep learning techniques) could be adapted for different categories of software refactorings. There are various deep learning techniques (e.g., CNN, LSTM, and GNN), which are originally designed for tasks (e.g., natural language processing or image processing) other than software refactoring. Consequently, such techniques should be substantially adapted for this specific task.

## 7.1  Detection of Code Smells

Code smell detection is often taken as the first step in software refactoring because code smells often indicate the problems of source code as well as their solutions (i.e., refactorings). Traditional approaches usually distinguish software entities associated with code smells from smell-free code by code metrics, taking the task of code smell detection as a binary classification problem. Since deep learning techniques have proved effective in classification tasks [223,224], it is reasonable to investigate deep learning-based detection of code smells.

To the best of our knowledge, the automated approach to detecting feature-envy smells proposed by Liu et al. [219] is the first attempt at deep learning-based code smell detection. They exploited traditional code metrics, e.g., coupling between code entities, by a CNN, and exploited the identifiers of code entities (i.e., names of the to-be-tested method and names of its enclosing class as well as its potential target class) by another CNN. The outputs of the two CNNs are fed into a dense layer (Fully-Connected Network, FCN) and its output generates how likely the method should be moved from its enclosing class to the given target class. Their evaluation results suggest that trained with automatically generated data, the deep learning-based approach is more accurate than traditional approaches that do not leverage deep learning techniques. Based on the success, they [222] expanded the deep learning-based detection to additional categories of code smells (i.e., long methods, large classes, and misplaced classes), and their evaluation results suggest that deep learning techniques have the potential to improve the state of the art in code smell detection.

Barbez et al. [225] applied CNN to capture the evolution of God classes. For a class to be tested, the proposed approach (called CAME) extracts the latest n versions of this class, and for each version, it extracts the selected code metrics (e.g., the complexity of the class). As a result, it expresses the evolution of the class as a matrix $X_{n,m}$ where n is the number of versions and m is the number of involved code metrics. This matrix is fed into a CNN whose output is forwarded to a MLP (Multilayer Perceptron). The MLP will make the final prediction, i.e., whether the class under test is a God class. Notably, although this approach depends on deep neural networks, it leverages only 71 real-world God classes for training and testing.

Yu et al. [226] employed a Graph Neural Network (GNN) to detect feature-envy smells. They represented methods as nodes in the graph and the calling relationships among methods as edges. They leveraged a GNN technique to extract features and vectorize the nodes, and finally classified the nodes (methods) as smelly methods or smell-free ones. Their evaluation results suggest that their GNN-based approach is more accurate than the CNN and FCN-based approach proposed by Liu et al. [222] for detecting feature-envy smells.

Zarina et al. [220] proposed a hybrid approach to identifying feature-envy smells by leveraging both deep learning techniques and traditional machine learning techniques (i.e., SVM). The approach first represents methods and classes as vectors with Code2Vec [227]. Code2Vec parses a method into an AST, and represents each path between two AST leaves as a vector. Based on the resulting vectors that represent patches within the AST, Code2Vec represents the whole method as a vector. A class is presented as a vector that equals the average of the vectors of methods within this class. The vector of the to-be-tested method and the vector of its potential target class are fed into SVM-based binary classifier to predict whether the method should be moved to the given target class. Note that, the deep learning model employed by this approach (i.e., Code2Vec) is unsupervised. Consequently, it does not request a large number of labeled training data, which is a significant advance of the hybrid approach. Similarly, Di et al. [228] also leveraged Code2Vec (or Code2Seq) to turn a method (i.e., AST) into a vector, and employed graph embedding techniques to represent its dependency with other methods. With the resulting embeddings, they also leveraged a traditional machine learning technique (Naive Bayes) to make predictions.

Code smell detection, if taken as a binary classification, often encounters serious class imbalances because software entities associated with code smells (noted as positive items) are often significantly fewer than smell-free entities (noted as negative items). Fuzzy sampling, proposed by Yedida and Menzies [229], is a novel technique to handle class imbalance. It adds points concentrically outwards from points (items) of the less popular class. The oversampling thus may push the decision boundary away from these points if the newly added points belong to the same class. As a result, the classifier trained with additional items may learn better to identify similar items belonging to the less popular class. Yedida and Menzies [230] validated whether this novel technique can boost deep learning-based code smell detection. Their results demonstrate that fuzzy sampling boosts the deep learning-based approaches (proposed by Liu et al. [222]) to detect feature envy, long methods, large class, and misplaced classes by fuzzy sampling. That is to say, the results suggest that fuzzy sampling does improve the state of the art in deep learning-based code smell detection.

## 7.2　Recommendation of Refactoring Opportunities

Although we may suggest where and which refactorings should be applied by automated detection of code smells as introduced in the preceding section, we may also need to recommend refactoring opportunities (and even detailed refactoring solutions) directly without code smell detection. For example, Liu et al. [231] proposed an automated approach to recommending renaming opportunities based on renaming refactorings that developers recently conducted. In this approach, they do not detect any specific code smells but recommend refactorings similar to what has been conducted recently. The approach proposed by Liang et al. [232] is similar to the approach by Liu et al. [231] in that both approaches recommend renaming opportunities according to the evolution history of the source code. The key difference is that Liang et al. [232] employed deep learning techniques whereas Liu et al. [231] depended on heuristics and static source code analysis. Notably, Liang et al.'s approach recommends renaming opportunities on methods only. For a given method, it leverages BERT [233] and textCNN [234] to vectorize the method. After that, it employs an MLP classifier to predict whether the method should be renamed. The predicted method is renamed only if at least one of its closely related entities has been renamed recently. Liu et al. [235] employed unsupervised deep learning techniques to identify and recommend renaming opportunities. First, for each method in the given corpus, it vectorizes method names and method bodies by CNN and Paragraph Vector [236], respectively. For a given method whose method name is mn, and whose method body is md, it retrieves the top k most similar method names from the corpus and top k method names whose corresponding method bodies are the most similar to md. If the two sets of method names are highly similar, method name mn is consistent with the corresponding method body md. Otherwise, they are inconsistent, and thus it selects a method name from the latter set with a set of heuristics and recommends replacing mn with it.

Since useful refactorings should be frequently applied by various developers on various software applications, it is likely that we may infer such refactorings from the rich evolution histories of software applications without knowing exactly what the refactorings are in advance. Consequently, by applying advanced learning or mining techniques to evolution histories, we may learn (discover) some less-known refactorings, and can even learn where and how such refactorings could be applied. For example, Tufano et al. [237] proposed a deep learning-based technique to learn from pull requests and to infer how code is changed. Among the most frequent changes learned by this approach, refactorings are dominating. After training the deep neural model with various pull requests, the technique applies the resulting model to predict expected changes on a given application. Most of the predicted changes are refactorings, and thus the prediction could be viewed as an automated recommendation of refactorings.

Nyamawe et al. [238] suggested that refactoring activities, as well as other software development activities, should be traced to software requirements as well as their changes. Based on this assumption, they proposed a novel technique to recommend refactorings based on feature requests. It first associates feature requests with code smells and refactorings by mining software evolution histories. For a new feature request, it employs various machine learning-based models to predict the required refactorings based on the feature request as well as code smells associated with related source code.

AlOmar etal. [239] proposed the first just-in-time recommendation approach for extract-method refactorings based on copy-and-paste actions. When developers copy and paste a piece of source code, the approach determines whether the copied fragment of source code should be extracted as a new method. It leverages a large number of code metrics and uses a CNN to classify code fragments based on the code metrics.

Chi et al. [240] proposed a novel and more reliable approach called ValExtractor to conduct extract-local-variable refactorings. The primary challenge in automating extract-local-variables refactorings is the efficient identification of side effects and potential exceptions between extracted expressions and their contexts without resorting to time-consuming dynamic program execution. ValExtractor addresses this challenge by utilizing lightweight static source code analysis to validate the side effects of the selected expressions. It also identifies occurrences of the selected expression that can be extracted together without introducing program semantics or potential exceptions.

Besides the generic approaches that could be applied to various software applications, deep learning-based refactoring recommendation has also been proposed for some special domains, e.g., microservices. Desai et al. [241] proposed a deep learning-based approach to recommending refactoring opportunities, i.e., extracting some classes from a monolith application as micro-services. The approach represents classes as nodes and invocation among them as edges. It also identifies entry points (i.e., APIs of web applications) and represents such information as attributes of the classes. It then employs a graph neural network to cluster the nodes (i.e., classes), aiming to minimize the effect of outlier nodes. The resulting outlier nodes are finally recommended to be extracted as micro-services.

We conclude that both supervised and unsupervised deep learning techniques have been applied to recommending generic and domain-specific refactoring opportunities. However, we also notice that existing approaches support only a limited number of refactoring categories, and thus it is potentially fruitful to recommend more categories of refactoring opportunities by deep learning techniques in the future. We also notice that some latest advances in deep learning techniques, like large language models (e.g., GPT) have not yet been fully exploited in automated software refactoring approaches.

## 7.3 Datasets

Lacking of large-scale and high-quality training datasets is one of the biggest obstacles to deep learning-based software refactoring. Notably, most existing datasets for software refactorings are built manually [242] or built by mining refactoring histories [243]. For example, to validate the automated move-method refactorings, JMove [218] requested developers to manually check the suggested refactoring opportunities, and thus such manually confirmed items could serve as training or testing dataset for future research in this line. However, such a manually constructed dataset is often too small for sufficient training of deep neural networks. It is also challenging to enlarge such datasets because the manual identification of refactoring opportunities is time-consuming and error-prone. Another way to construct datasets of software refactorings is to leverage automated refactoring miners to discover actually conducted refactorings recorded in open-accessed version control systems. A few approaches, e.g., RefactoringMiner [244], RefDiff [245], and Ref-Finder [246], have been proposed for such purpose. Although such mining tools could identify a large number of refactorings from real-world applications, they often result in false positives, making the resulting dataset unsuitable for model training.

To this end, Liu et al. [219] proposed a novel approach to generating large-scale training data for the training of deep learning-based refactoring models. In the paper, the authors focused on a single category of code smells (i.e., feature envy) and its corresponding refactoring (i.e., move-method refactoring). To create positive items (i.e., methods associated with feature envy smells), they randomly moved methods across classes (with precondition checking of Eclipse move method refactoring), and took the moved methods as positive items because they had better be moved back to the original place (i.e., their enclosing classes before the movement). Methods that could be moved (i.e., satisfying the precondition of move method refactorings) but have not been moved could be taken as negative items, i.e., methods not associated with feature envy smells. By applying this novel approach to high-quality open-source applications, they generated huge data sets of refactorings (code smells) where the ratio of positive (negative) items could be accurately controlled as well. Based on the generated data, they trained a CNN-based deep neural model to detect feature envy smells and to suggest solutions (i.e., where the associated methods should be moved). Their evaluation results suggest that the resulting model significantly outperforms the state-of-the-art approaches. Later, Liu et al. [222] successively expanded this approach to more categories of code smells, i.e., long methods, large classes, and misplaced classes. Long methods are created by automated inline refactorings that merge multiple methods into a single one, large classes are created by merging multiple classes whereas misplaced classes are created by moving classes across packages. Their evaluation results suggest that employing such automatically generated large datasets to train deep neural networks could significantly improve the state of the art in code smell detection and automated recommendation of refactoring opportunities [222]. Currently, this automated data generation has been employed by almost all deep learning-based refactoring approaches that request labeled training items [230,247].

Although the quantity of automatically generated training items is satisfying, their quality may still be questionable. Because the code smells (i.e., positive items) are automatically generated, they could be significantly different from code smells introduced unconsciously by developers. As a result, deep neural models trained with such generated artificial data may learn only how to identify artificial smells instead of real-world code smells. To this end, Liu et al. [243] aimed to improve the precision of refactoring miners, and thus their discovered real-world code smells and refactorings could be taken directly as high-quality training data. The key to their approach is to leverage a sequence of heuristics and machine learning-based classifiers to exclude false positives. Notably, they employed a traditional machine learning technique (i.e., decision trees) instead of deep learning techniques to exclude false positives. The major reason for the selection is that traditional machine learning techniques may work well with small (but high-quality) training data whereas deep learning ones often request much larger dataset that they were unable to provide. Their evaluation results suggest that by filtering out false positives with their approach, the precision of the employed refactoring miner (RefactoringMiner [244]) is able to reach a high level comparable to human experts in discovering move-method refactorings. Compared to the artificial feature-envy methods automatically generated by previous approach [219], such real-world feature-envy methods discovered by the proposed approach could significantly improve the performance of deep learning-based model in detecting feature-envy smells and in recommending move-method opportunities.

Most of the current refactoring detection approaches often result in non-negligible false positives and false negatives. To solve this problem, Liu et al. [248] proposed a novel refactoring detection approach (called ReExtractor). The rationale of ReExtractor is that an entity matching algorithm takes full advantage of the qualified names, the implementations, and the references of software entities. Compared against the state of the art, it improves the accuracy of entity matching between two successive versions and thus substantially reduces false positives and false negatives in refactoring detecting.

Based on the preceding analysis, we conclude that large-scale and high-quality training data are critical for deep learning-based refactoring, and data collection remains an open question that deserves further investigation.

## 7.4 Challenges and Opportunities

Based on the preceding analysis of deep learning-based software refactoring, we present here a list of potential challenges and opportunities for future research in deep learning-based software refactorings.

### 7.4.1 Challenges

- Large-scale high-quality dataset. It remains challenging to collect large-scale and high-quality refactoring data to train deep learning models. Although generating refactorings to be automatically reversed as suggested by Liu et al. [219] should result in large-scale refactoring data, the quality and representativeness are in question. In contrast, discovering refactoring histories in open-source applications may result in high-quality real-world refactoring data, such data are often small and lack diversity.

- Generalization across different paradigms. Most of the code bases are written in various programming languages, each with its own syntax and semantics. Developing deep learning models that generalize well across different languages is a considerable challenge. Currently, most of the studies in deep learning based refactoring use applications written in Java. Thus, there is little or slow adoption in other programming languages. Another challenge concerning the generalization is to make the approaches applicable to all categories of code smells. The identification of code smells is very crucial in the process of software refactoring. It has been proven challenging to have a general deep learning model to detect code smells for refactoring as different models behave differently for specific smells.

- Generic classification and feature engineering. It is very challenging to design general classifier which may be used for the process of software engineering as different features may be needed for different refactoring processes.

- Complexity of Code Patterns. Code bases often contain complex patterns and structures. Capturing and representing these patterns effectively for training deep learning models can be difficult, especially when dealing with large and diverse code bases.

- Context Sensitivity. Refactoring decisions are often context-dependent, considering the broader system architecture, design patterns, and usage scenarios. Deep learning models might struggle to capture and understand such context-sensitive information.

- Interpretability. Deep learning models are known for their "black-box" nature, making it challenging to understand the rationale behind their refactoring recommendations. This lack of interpretability can be a significant hurdle for developers who need to trust and adopt these suggestions.

### 7.4.2 Opportunities

- Large language model-based refactoring. Large language models have great potential in smell detection and refactoring suggestions. Up to date, various deep learning techniques have been proposed to detect code smells, and/or to suggest solutions to identified code smells. However, to the best of my knowledge, large language models have not yet been applied to such tasks. Since large language models are good at understanding and generating source code, it is likely that they can be employed to format source code and to identify abnormal parts of source code (i.e., code smells). They may also be employed to discover (and measure) the relationship (like coupling, similarity, cohesion, and dependency) among software entities. Such a relationship, currently measured by statistic code metrics only, is critical for the detection of code smells as well as suggestion of refactoring solutions.

- Automated discovery of new refactorings. Up to date, the academic community focus on well-known code smells and refactorings. All such smells and refactorings are coined by human experts. However, with the development in programming languages and new paradigms, new categories of code smells as well as new categories of refactorings are emerging. One possible way to discover such new smells and new refactorings is to mine the evolution history of open-source applications. With advanced deep learning techniques, it is potentially feasible.

# 8 Code Clone Detection

Code clones are code fragments that have the same or similar syntax and semantics. The wide presence of code clones in open-source and industrial software systems makes clone detection fundamental in many software engineering tasks, e.g., software refactoring, evolution analysis, quality management, defect prediction, bug/vulnerability detection, code recommendation, plagiarism detection, copyright protection, and program comprehension.

The recent research in code clones has attracted the use deep learning techniques. As shown in Figure 4, most of the research focuses on source code clone detection and source code representation for clone detection. A little of research focuses on the binary clone detection, cross language clone detection, clone evaluation and validation.

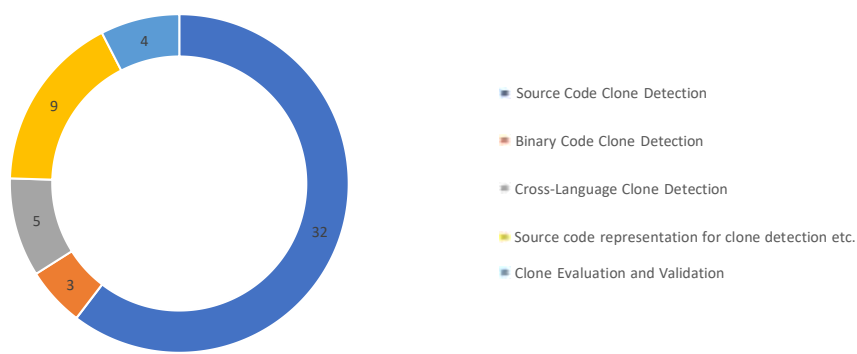

**Figure 4**  Code clone research tasks where deep learning has been applied.

## 8.1  Source code clone detection

Source code clones are typically classified as follows based on their degree of similarity.

1. Type1: Duplicate code fragments, except for differences in white space, comments, and layout.

2. Type2: Syntactically identical code fragments, except for differences in variable names, literal values, white space, formatting, and comments.

3. Type3: Syntactically similar code fragments with statements added, modified, or deleted.

4. Type4: Syntactically different code fragments implementing the same functionality.

As the boundary between Type3 and Type4 clones is often ambiguous, in benchmarks like Big-CloneBench [249] researchers further divide these two clone types into three categories: strongly Type3, moderately Type3, and Weakly Type3/Type4. Each type is harder to detect than the former one. Weak Type3 and Type4 clones are usually called semantic clones.

The research efforts on source code clone detection are shown in Table 9. As can be seen from the table, most of recent research tries to leverage deep neural networks to effectively capture complex semantic information in code fragments, so as to improve the effectiveness of semantic clone detection.

The code clone detection process begins by modeling the semantic of the source code. To achieve this goal, diverse program representations such as tokens, Abstract Syntax Trees (ASTs), Control Flow Graph (CFGs), Data Flow Graphs (DFGs), Program Dependency Graphs (PDGs) are being used to learn program features.

Various deep learning models, such as CodeBERT, GraphCodeBERT, Graph Neural Network (GNN), Graph Attention Network, Convolutional Neural Network (CNN), Graph Convolutional Network (GCN),

ASTNN, Tree-Based Convolutional Neural Network (TBCNN), Recursive Autoencoders (RAE), Recurrent Neural Network (RNN), Recursive Neural Network(RvNN), LSTM, have been used.

These studies have achieved higher recall and better precision than the best classical approaches.

**Table 9** Source Code Clone Detection

| Year | Venue | Literature | Language | Code Representation | Deep Learning Models | Clone Type | Benchmark | Baseline |
|------|-------|-----------|----------|--------------------|--------------------|-----------|-----------|----------|
| 2022 | ICSME | SSCD [250] | Java, C/C++ | token | CodeBERT,GraphBERT | Type 3 and Type 4 | BigCloneBench,CompanyC, CompanyC++ | SourcererCC [251] |
| 2022 | TSE | [252] | Java | token | CodeBERT | All | BigCloneBench,SemanticCloneBench etc. | NA |
| 2022 | IWC | Hemon [253] | Java | PDG | GNN,LSTM | ALL | GoogleCodeJam, SeSaMe,BigCloneBench | TROCD [254] |
| 2022 | ICSR | SEEC [255] | Java,C,C++ | Semantic graph | GMN | Type 4 | OJClone, BigCloneBench | TBCCD [254],ASTNN [256] |
| 2022 | ADES | CCDD-DL [257] | Java | AST,CFG,FCG | DNN | Type 4 | BigCloneBench,opensource projects | LV-CCD [256], FCCA [258], and TBCNN [260] |
| 2021 | QRS | CAEM [261] | C | Event Embedding Tree | GAT | Type1,Type4 | OJClone | CCLearner [262],Deckard [263],CloneWork [264],SourcerCC [251] |
| 2021 | APSEC | [265] | Java,C | BC-CNN | BC-CNN | Type3 and Type4 | BigCloneBench, OJClone | Deckard [263], GLC [266],CDLH [267],Deepsim [265],ASTNN [256] etc. |
| 2021 | Applied Sciences | [260] | Java | AST | TBCNN | All | BigCloneBench | OJClone,BigCloneBench |
| 2020 | IEICE T INF SYST | [268] | Java | AST | CNN,Siamese | ALL | BigCloneBench | NiCad [270] |
| 2020 | ASE | SCDetector [271] | Java,C / C++ | CFG | GRU,Siamese | All | BigCloneBench | RtvNN [269],ASTNN [256], Deckard [263],SourcererCC [251] |
| 2020 | APSEC | Bra-RAE [272] | Java | AST | recursive autoencoders (RAE) ,Siamese | All | OJClone,BigCloneBench | DeepSim [246], CDLH [267],weighted RAE [273] |
| 2020 | DSA | [274] | Java | PDG | Bi-RNN,GCN | Type4 | opensource projects | Nicad [270] |
| 2020 | T-R | FCCA [258] | Java | token,AST,CFG | RNN | All | BigCloneBench | Deckard [263], CDLH [267], DeepSim [265], DLC [266],SourcerCC [251],TBCCDoteod-Yu2019 ect. |
| 2020 | SANER | FA-AST-GMN [275] | Java,C/C++ | FA-AST | GNN | All | GCJ,BigCloneBench | Deckard [263], RtvNN [269],CDLH [267], ASTNN [256] |
| 2020 | ISSTA | [276] | C++ | AST | DNN | Type3 | OJClone | Deckard [263], DLC [266],CDLH [267], ASTNN [256], DeepSim [265] |
| 2020 | IEEE Access | [277] | Java | AST | CNN | Type1,Type2 | BigCloneBench | SourcerCC [251],NiCad [270],Deckard [263],CCLearner [262],Oreo [278] |
| 2020 | Complexity | [279] | Java,C/C++ | AST | BiLSTM | All | OJClone,BigCloneBench | RAEm-Whote24,CDLH [267], ASTNN [256] |
| 2019 | IEEE Access | weighted RAE [273] | Java | AST | RAE | All | BigCloneBench | Oreo [278], DeepSim [265], CCLearner [262],CDLH [267], Nicad [270] |
| | | | | | | | | Deckard [263] SourcerCC [251],CloneWorks [264] |
| 2019 | AAAI | ACD [280] | Java,C | AST | LSTM | All | OJClone,BigCloneBench | Deckard [263], SourcerCC [251],CDLH [267] |
| 2019 | SANER | [281] | Java | AST | RvNN,Siamese Network | Type4 | OJClone,BigCloneBench | CDLH [267],Deckard [263],SourcerCC [251],DLC [266] |
| 2019 | ICPC | TBCCD [254] | Java,C | AST,token | PACE | All | OJClone,BigCloneBench | CDLH [267],Deckard [263],SourcerCC [251],DLC [266] |
| 2019 | ISSTA | Go-Clone [283] | Golang | LSFG | CNN | NA | from Github | NA |
| 2019 | TII | [283] | C | AST | GCN | NA | open source projects | VUDDY [284], LSTM |
| 2018 | ESEC/FSE | Oreo [278] | Java | metrics,token | DNN with Siamese architecture | All | opensource projects | Deckard [263], SourcerCC [251],CloneWorks [264] |
| 2018 | ESEC/FSE | DeepSim [265] | Java | Semantic features matrix | Feed-forward nerual network | Type3,Type4 | GCJ ,BigCloneBench | Deckard [263], RtvNN [269], etc. |
| 2018 | ICMLA | CCDL2 [285] | Java | BDG,PDG,AST | CNN | All | BigCloneBench | NA |
| 2017 | IJCAI | CDDU [286] | Java | AST | LSTM,word2vec | All | OJClone,BigCloneBench | CDLH [267],Deckard [263] ,SourcerCC [251] ,RtvNN [269] etc. |
| 2017 | IJCAI | CDLH [267] | Java,C | AST | LSTM | All | OJClone,BigCloneBench | Deckard [263], SourcerCC [251] ,RtvNN [269] etc. |
| 2017 | ICSME | CCLearner [262] | Java | token | DNN | Type1,Type2,Type3 | BigCloneBench | Deckard [263], NiCad [270] |
| 2016 | ASE | RtvNN [269] | Java | AST | RvNN | All | open source projects | NA |
| 2016 | ICMLA | [287] | Java | metrics | MLP | All | BigCloneBench | Deckard [263], SourcerCC [251],Nicad [270] ,CCFiner,IClone |

## 8.2 Code representation learning for clone detection

The studies in Table 10 focus on learning source code representation, so as to automatically capture the syntactic and semantic information from source code. Then the embedding is applied to code clone detection, code classification tasks etc. Code similarities can be learned from diverse representations of the code, such as identifiers, tokens, ASTs, CFGs, DFGs, and bytecode.

Siowetal. [288] performed an empirical study on code representation. They found that the graph-based representation is superior to the other selected techniques across these tasks. Different tasks require task-specific semantics to achieve their highest performance; however, combining various program semantics from different dimensions such as control dependency, data dependency can still produce promising results. Tufano et al. [289] demonstrated that combined models relying on multiple representations can be effective in code clone detection and classification.

**Table 10** Source Code Representation Learning for Clone Detection

| Year | Venue | Literature | Language | Code Representation | Deep Learning Models | Clone Type | Bechmark | Baseline |
|------|-------|-----------|----------|--------------------|--------------------|-----------|----------|----------|
| 2022 | ICSE | [290] | Java,C/C++ | token | CodeBERT,GraphCodeBERT | NA | OJClone,BigCloneBench | CDLH [267],FA-AST-GMN [275],TBCCDoteod-Yu2019 ect. |
| 2022 | SANER | [288] | C | features,token,AST,CFG | BiLSTM,LSTM,Transformer,Tree-LSTM,Code2Vec,GAT,GCN,GGNN | NA | OJClone,BigCloneBench | NA |
| 2022 | EMSE | [291] | Java | AST | code2vec | NA | opensource projects | NA |
| 2021 | ICSE | InferCode [292] | Java,C/C++ | AST | TBCNN | All | OJClone,BigCloneBench | code2vec,code2seq,Deckard [263],SourcererCC [251],RtvNN [266] |
| 2021 | ICONIP | HDCR [293] | C,Java | SG, EDFG | T-GCN,E-GAT | All | GCJ,BigCloneBench | ASTNN [256],FA-AST,FCCA |
| 2019 | ICSME | ASTNN [256] | Java,C | AST | ASTNN | All | OJClone,BigCloneBench | TBCNN [294] ,CDLH [267] ,RAE [246] etc. |
| 2019 | ICCF | TBCAA [295] | Java,C | AST | Tree-based Convolution | All | OJClone,BigCloneBench | CDLH [267],Deckard [263] SourcerCC [251],DLC [266] |
| 2019 | ICSME | TECCD [296] | Java | AST | Sentence2Vec | Type3 | BigCloneBench,opensource projects | CCLeaener [262],Nicad [270],CCAligner [297] |
| 2018 | ICSE | [289] | Java | identifier,ASTs, CFGs, and Bytecode | RNN | All | Qualitas.class Corpus [298] | NA |

## 8.3 Cross-language code clone detection

The above work on source code clone detection focuses on clones in a single programming language. However, software systems are increasingly developed on a multi-language platform on which similar functionalities are implemented across different programming languages [154,299].

The main challenge of cross-language code clone detection is how to reduce the feature gap between different programming languages.

The studies on cross-language code clone detection are shown in Table 11. These studies focus on extracting syntactic and semantic features of different programming languages. Nafi et al. [300] used a Siamese architecture to learn the metric features. Perez et al. [154] used an unsupervised learning approach for learning token-level vector representations and an LSTM-based neural network to predict clones. Bui et al. [301] proposed a Bi-NN framework to learn the semantic features of two different programming languages. Wang et al. [302] proposed a Unified Abstract Syntax Tree neural network. Yahya et al. [299] used AST embeddings from InferCode [292] as input of the Siamese architecture.

**Table 11** Cross-Language Code Clone Detection

| Year | Venue | Literature | Language | Code Representation | Deep Learning Models | Benchmark | Baseline |
|------|-------|-----------|----------|--------------------|--------------------|-----------|----------|
| 2023 | Computers | CLCD-I [299] | Java,Python | AST | Siamese architecture [303] | coe from programming competition | LSTM |
| 2022 | ICPC | UAST [302] | C,C++,Java,Python, JavaScript | token | Bi-LSTM,GCN | JC, dataset collected from Ieecode | InferCode [292] |
| 2019 | SANER | [301] | Java,C++ | AST | Bi-NN | OJClone,opensource projects | TBCNN [294] etc. |
| 2019 | MSR | [154] | Java,Python | token,AST | tree-based skip-gram,LSTM | code from programming competition | sequential input model |
| 2019 | ASE | CLCDSA [300] | Java,Python, C# | Metrics | DNN | code from programming competition | LICCA [304],CLCMiner [305], [154] |

## 8.4 Binary code clone detection

Binary code clone detection can be used in the context of cross-platforms as well as legacy applications that are already deployed in several critical domains.

Research on binary code clone detection is shown in Table 12.

Xue et al. [306] combined program slicing and a deep learning based binary code clone modeling framework to identify pointer-related binary code clones. Xu et al. [307] proposed a neural network-based approach to compute the embedding based on the control flow graph of each binary function, and then to measure the distance between the embeddings for two binary functions. Marastoni et al. [308] tackled the problem of binary code similarity using deep learning applied to binary code visualization techniques. They found that it is important to further investigate how to build a suitable mapping from executables to images.

**Table 12** Binary Code Clone Detection

| Year | Venue | Literature | Code Representation | Deep Learning Models | Benchmark | Baseline |
|------|-------|-----------|---------------------|---------------------|-----------|----------|
| 2018 | MASES | [308] | visualization graph | CNN | GoogleCodeJam etc. | Shallow Neural Net |
| 2018 | FEAST | Clone-Slicer [306] | CFG,DDG | RNN | SPEC2006 | CloneHunter [309] |
| 2017 | CCS | Gemini [307] | ACFG | Structure2vec | dataset from [310] etc. | Genius [310] |

## 8.5 Clone evaluation and validation

Mostaeen et al. [311] proposed a machine learning based approach for predicting the user code clone validation patterns. The proposed method works on top of any code clone detection tools for classifying the reported clones as per user preferences. The automatic validation process can accelerate the overall process of code clone management.

Saini et al. [312] presented a semi-automated approach to facilitating precision studies of clone detection tools. The approach merges automatic mechanisms of clone classification with manual validation of clone pairs, so as to reduce the number of clone pairs that need human validation during precision experiments.

Liu et al. [313] proposed an evaluation methodology that can systematically measure the cross-functionality generalizability of neural clone detection. They also conducted an empirical study and the results indicate that the studied neural clone detectors cannot generalize well as expected. They found that the performance loss on unseen functionalities can be reduced by addressing the out-of-vocabulary problem and increasing training data diversity.

Yuetal. [314] presented an experimental study to show that BigCloneBench typically includes semantic clone pairs that use the same identifier names, which however are not used in non-semantic-clone pairs. To alleviate these issues, they abstracted a subset of the identifier names (including type, variable, and method names) in BigCloneBench to result in AbsBigCloneBench and used AbsBigCloneBench to better assess the effectiveness of deep learning models on the task of detecting semantic clones.

Jens and Ragkhitwetsagul [315] performed a manual investigation on BigCloneBench. They demonstrated that the way BigCloneBench being constructed makes it problematic to use BigCloneBench as the ground truth for learning code similarity. BigCloneBench fails to label all clone pairs. Moreover, only a small set of true negatives has been created and, for most of the possible pairs in the dataset, the ground truth is unknown. This leads to a strong impact on the validity of the ground truth for Weakly Type3 and Type4 clone pairs, threatening the validity of results for evaluations in the Weakly Type3 and Type4 category and approaches to learning code similarity.

**Table 13** Source Code Clone Detection

| Dataset | Literature | Language |
|---------|-----------|----------|
| BigCloneBench | [249] | Java |
| OJClone | [294] | C |
| GoogleCodeJam (GCJ) | [268] | Java |
| SemanticCloneBench | [316] | Java, C, C#, Python |
| SeSaMe | [317] | Java |
| JC | [301] | Java, C++ |
| Leetcode | [302] | C, C++, Java, Python, JavaScript |

## 8.6 Datasets

We summarize datasets used in clone detection in Table 13.

BigCloneBench is a benchmark of inter-project clones from IJaDataset [249], a big Java source code repository. It has about 8M labeled clone pairs, as well as 260,000 false clone pairs, covering 43 functionalities. BigCloneBench divides Type3 and Type4 clones into four categories: Very-Strongly Type3, Strongly Type3, Moderately Type3, and Weakly Type3/Type4.

OJClone is generated based on OJ dataset [294] covering 104 functionalities. Each functionality is a programming question with 500 verified solutions written in C, submitted by students. Two solutions to the same question can be considered as a clone pair.

GoogleCodeJam (GCJ) is a benchmark similar to OJClone. It contains 1,669 solutions written in Java for 12 functionalities collected from GoogleCodeJam.

SemanticCloneBench is a benchmark of semantic clone pairs [316]. It consists of four thousand clone pairs, each for four programming languages (i.e., Java, C, C#, and Python). The clone pairs are collected from the StackOverflow answers. The method pairs to the same questions on Stack Overflow are considered as semantic clones.

The SeSaMe dataset consists of 857 semantically similar method pairs mined from 11 open-source Java repositories [317].

The JC dataset includes 10 categories of programs crawled by Bui etal. [301] from GitHub. It contains 5822 Java files and 7019 C++ files. The code files for each category implement the same function.

The Leetcode dataset contains 50 categories of programs from Leetcode, each of which contains 400 semantically similar solutions of five different programming languages, with a total of 20000 files [302].

## 8.7 Challenges and Opportunities

This section presents the challenges and opportunities for further work on code clone detection.

### 8.7.1 Challenges

- Challenge to build comprehensive learning-oriented code clone detection datasets. Most of the existing code clone datasets are of a limited scale due to the effort required in manually constructing the benchmarks. BigCloneBench is a large-scale dataset and becomes a standard to evaluate and compare the performance of clone detection tools. Many researchers also use it to train deep learning models. However, as pointed by Jens and Ragkhitwetsagul, the incomplete ground truth and the bias and imbalance of the ground truth will have a strong impact on deep learning approaches for code similarity that learned from Big-CloneBench's ground truth [315]. Besides, there still lack standard datasets for cross-language clone detection. Also, there is no clear taxonomy about cross-language clones. Therefore, it is necessary to build large, cross-language, learning oriented clone datasets.

- Challenge to cross-functionality generalizability of deep learning-based clone detection. Most of the deep learning based code clone detection approaches are proposed for detecting semantic clones, and they have achieved impressive results based on the tested benchmarks. However, according to the research by Liu et al., these studies are limited in detecting clones whose functionalities have never been previously observed in the training dataset [313]. Further research on deep learning algorithms for improving the cross-functionality generalizability is required. Moreover cross-functionality generalizability should also be considered in detection result evaluation.

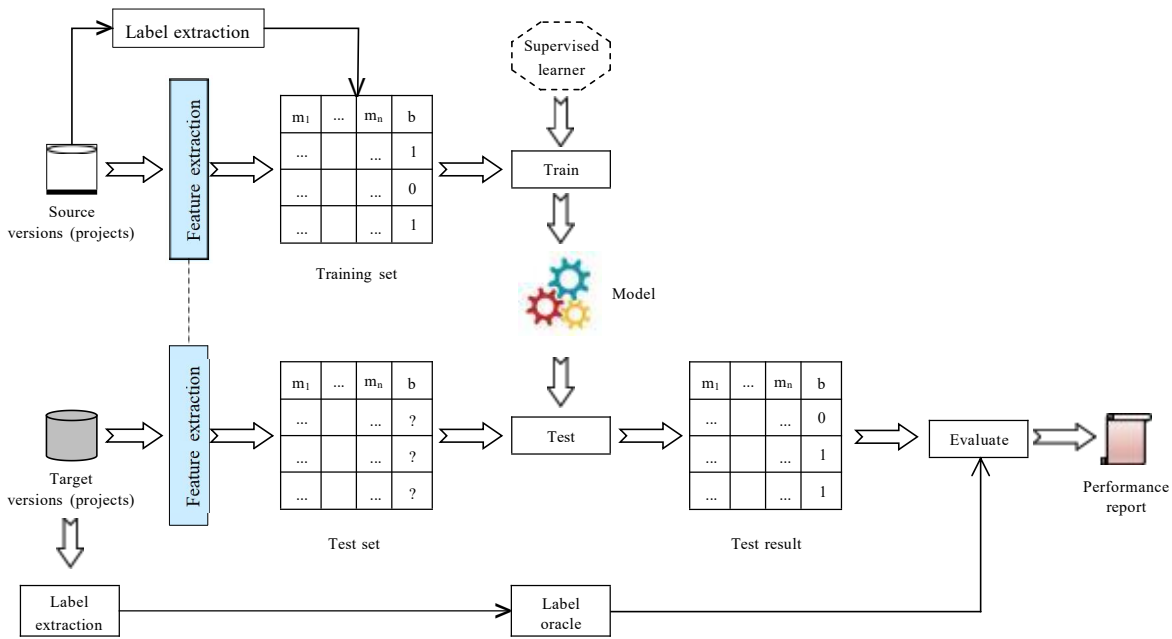

**Figure 5**  The overall structure of supervised defect prediction

- Challenge to selecting suitable source code representation. Current deep learning based clone detection approaches use different representations of source code like tokens, AST, CFG, PDG. These represents vary in clone detection scalability, efficiency and effectiveness. For example, approaches representing code fragments as sequences of tokens may be more efficient, but they may not be generalized to source code having out-of-vocabulary tokens, as vocabulary of tokens is unlimited. Approaches representing code fragments as PDGs help in semantic analysis. However the low efficiency may prevent them from being used on a large code base. Further research on source code representation is still needed to deeply analyze the scalability, efficiency and effectiveness.

### 8.7.2  Opportunities

- Opportunity to explore deep learning in more clone related tasks. A software may contain a lot of clones but not all clones need to be manipulated. History of clone management helps extract useful features that can help take various decisions of clone related tasks. For example, deep learning based recommendation can help automatically identify important clones for refactoring. Deep learning based methods can also help predict the code clone quality, detect clone related bugs, and reduce the maintenance cost caused by harmful or risky clones. However, only a few existing studies focus on these tasks. Research can be further conducted in these areas.

- Opportunity to explore large language model in detecting cross-language code clones. In recent years, large language models have experienced rapid development. By pre-training on a large amount of corpus, the model is able to "remember" knowledge from the corpus, including syntax, semantics, etc., thus typically being able to understand the syntax and semantics of different programming languages. This lays a good foundation for detecting cross-language code clones.

## 9  Software Defect Prediction

Software Defect Prediction (SDP) aims to forecast potential defect locations in a software project, predicting which modules (such as files, classes, and functions) may contain defects. This predictive information is crucial for the software quality assurance process. On one hand, it prioritizes modules awaiting inspection or testing, facilitating early identification of defective modules in the software project under test. On the other hand, it guides testing personnel to allocate code review or testing resources more efficiently and sensibly for each module. Allocating more review resources to modules with a higher likelihood of defects can help quality assurance personnel discover as many software defects as possible within the given budget.

As depicted in Fig. 5, in the supervised defect prediction scenario, a learner is utilized to establish the connections between features and labels using the training set. Subsequently, the learned model is applied to the test set to predict defect-proneness. For each instance in the test set, the corresponding features are used to calculate the probability of being defect-prone. If the probability surpasses a predefined threshold (typically set at 0.5), the instance is classified as "buggy"; otherwise, it is labeled as "clean". Traditionally, numerous manually crafted features have been employed, such as size metrics, complexity metrics, cohesion metrics, and coupling metrics. However, these conventional hand-crafted features primarily rely on syntactic information. Consequently, they fail to capture semantic information, resulting in limited predictive capability for defect-proneness. To address this limitation, a variety of deep learning techniques have been adopted to generate powerful semantic features for defect prediction.

Table 14 provides a summary of notable studies that employ deep learning techniques in defect prediction. The table includes information on the publication year and the type of defect prediction addressed in the first and second columns, respectively. To simplify the presentation, "WPDP" represents Within-Project Defect Prediction, while "CPDP" represents Cross-Project Defect Prediction. The third column presents the granularity at which defect prediction is conducted, spanning from file-level to statement-level. Note that GDRC (Graph of Defect Region Candidates) corresponds to an area of the code file. The fourth column specifies the utilized deep learning technique. The fifth to ninth columns outline the inputs provided to the deep learning model for extracting powerful semantic features. An entry marked with "•" or non-blank signifies that the corresponding information is used as input, while a blank entry indicates its absence. The last column provides the reference for each study.

From Table 14, we can see that numerous studies have emerged, employing a variety of deep learning techniques for defect prediction. Among these techniques, CNN, RNN (including LSTM as a variant), and GCN stand out as the most prominent ones, each offering distinct capabilities that researchers find advantageous for defect prediction. CNN proves highly effective in capturing localized patterns and spatial dependencies within the data. RNN excels at capturing sequential and long-term dependencies, which are particularly valuable in understanding the temporal aspects of defect occurrences. GCN demonstrates remarkable proficiency in capturing the structural information and intricate inter-dependencies within code, attributes that are crucial for accurate defect prediction. Overall, the adoption of deep learning techniques has significantly advanced defect prediction research, offering valuable insights into defect detection.

Overall, in the defect prediction community, significant attention has been dedicated to utilizing deep learning in order to generate expressive features and enhance the effectiveness of defect prediction. Through the application of deep learning, remarkable progress has been made in automatically generating features that are both highly informative and discriminative. This transition from manual feature engineering to automated feature extraction using deep learning signifies a fundamental change in approaches to predicting defect-proneness in software systems. The objective is to construct models capable of comprehending the fundamental traits and complexities of the code, thereby enabling more dependable and accurate defect predictions.

## 9.1  Using manually crafted features as the input

In the literature, numerous manually crafted features have been proposed for defect prediction. However, these features often focus on specific characteristics of a module, and their high-level semantic relationships are not adequately captured, limiting their defect prediction capability. To address this limitation, Yang et al. [318] utilized DBN to generate expressive features from a set of initial change features. This approach results in more powerful change-level defect prediction models. Similarly, Tong et al. employed SDAEs to extract expressive features from traditional software metrics [323]. Following their work, a range of deep learning techniques have been explored for this purpose, including deep forest [326], DNN [327,339], Layered RNN [328], DAE-CNN [332], Bi-LSTM [334], MDA [344], SSDAE [345], and TCN [347].

**Table 14** An overview of prominent defect prediction models utilizing deep learning techniques

| Year | Type | Granularity | Deep learning technique | Input to deep learning | | | | | Reference |
|------|------|-------------|-------------------------|------------------------|-----|---------|------------------------|-------|-----------|
| | | | | Hand-crafted features | AST | CFG/CPG | Raw code (token) | Other | |
| 2015 | WPDP | Change | DBN | • | | | | | Yang et al. [318] |
| 2017 | WPDP | File | CNN | | | CFG | | | Phan et al. [319] |
| | WPDP | File | CNN | | • | | | | Li et al. [320] |
| 2018 | WPDP/CPDP | File | CNN | | | | • | Comments | Huo et al. [321] |
| | WPDP | File | RNN | • | | | | Change history | Liu et al. [322] |
| | WPDP | File/Function | SDAEs | • | | | | | Tong et al. [323] |
| 2019 | CPDP | File | CNN | | • | | | | Qiu et al. [324] |
| | WPDP | Change | CNN | | | | | commit log + change | Hoang et al. [325] |
| | WPDP | File/Function | Deep Forest | • | | | | | Zhou et al. [326] |
| | WPDP | File | DNN | • | | | | | Xu et al. [327] |
| | WPDP | File | Layered RNN | • | | | | | Turabieh et al. [328] |
| | WPDP/CPDP | File | LSTM | | • | | | | Dam et al. [329] |
| | CPDP | File | Bi-LSTM | | • | | | | Li et al. [330] |
| 2020 | WPDP/CPDP | File | CNN | | | | • (image) | | Chen et al. [331] |
| | WPDP/CPDP | Change | DAE-CNN | • | | | | | Zhu et al. [332] |
| | WPDP/CPDP | File/Change | DBN | | • | | | | Wang et al. [333] |
| | WPDP | File | Bi-LSTM | • | | | | | Deng et al. [334] |
| | WPDP/CPDP | File | Bi-LSTM +Attention | | • | | | | Shi et al. [335] |
| | WPDP | Statement | LSTM | • | | | • | | Majd et al. [336] |
| | WPDP | File | LSTM | • | | | | Change history | Wen et al. [337] |
| 2021 | WPDP/CPDP | File | CNN | | • | | | | Shi et al. [338] |
| | CPDP | File | DNN | • | | | | | Xu et al. [339] |
| | WPDP | File | GCN | | • | | | | Xu et al. [340] |
| | WPDP | File | GCN | • | | | | CDN | Zeng et al. [341] |
| | CPDP | GDRC | GCN | | | CPG | | | Xu et al. [342] |
| | WPDP | File | LSTM | | • | | | | Wang et al. [343] |
| | CPDP | File | MDA | • | | | | | Zou et al. [344] |
| | WPDP | File/Function | SSDAE | | | | | | Zhang et al. [345] |
| 2022 | WPDP/CPDP | File | Bi-LSTM | | | | • | | Uddin et al. [346] |
| | WPDP | Change | TCN | • | | | | | Ardimento et al. [347] |
| 2023 | WPDP | Code line | HAN | | | | • | | Pornprasitetal. [348] |
| | WPDP | File | CNN | | • | | | | Qiu et al. [349] |

## 9.2 Using raw source code as the input

Traditional manually crafted features in source code analysis tend to be syntax-based, disregarding the valuable semantic information embedded in the code. To overcome this limitation, Chen et al. introduced a novel technique that visualizes the source code of programs as images and then employs CNN to extract image features [331]. These extracted features are utilized for building more effective defect prediction models. Uddin et al. leveraged Bi-LSTM to capture expressive features from the embedded token vectors obtained through BERT applied to the source code [346]. This technique allows effectively harnessing the semantic information within the code for defect prediction. Furthermore, Pornprasit et al. [348] adopted HAN to learn semantic features from the surrounding tokens and lines of code. This approach enables prediction of defective lines more accurately by considering the context and relationships between different parts of the code.

## 9.3 Using abstract syntax trees as the input

One significant drawback of using raw source code as input is the disregard for crucial structural information when extracting expressive features. To tackle this issue, Li et al. introduced a novel approach, utilizing a program's AST (Abstract Syntax Tree) as input to generate semantic features [320]. In their approach, Li et al. initially extracted token vectors from the program's AST and then transformed them into numerical vectors through mapping and word embedding. Subsequently, they fed these numerical vectors into a CNN, allowing the model to learn both semantic and structural features for defect prediction. Following this breakthrough, several other deep learning techniques have been explored for the same purpose, including LSTM [329,343], Bi-LSTM [330,335], DBN [333], and GCN [340]. These approaches aim to enhance defect prediction by considering the inherent structural characteristics of the source code, leading to more effective and accurate models.

## 9.4 Using graphical representations as the input

Phan etal. emphasized the importance of utilizing precise graphs that represent program flows to capture the complex semantics of programs accurately [319]. They argued that tree structures like ASTs might fall short in this regard. To address this issue, Phan et al. adopted CFGs (Control Flow Graphs) and applied GCN to extract expressive features. By leveraging CFGs, they were able to capture intricate program dependencies and interactions more effectively. The GCN allows the generation of expressive features that incorporate both local and global information from the program's control flow. As a result, their approach improves the accuracy of defect prediction models by considering the intricate semantics of the code. In a related study, Xu et al. proposed an alternative approach using CPGs (Code Property Graphs) as input to GCN [342]. This approach aims to extract even more expressive features for defect prediction, further enhancing the capabilities of the model in predicting software defects.

## 9.5 Using hybrid-source information as the input

In addition to source code, other information such as comments, commit logs, and change history contains valuable insights for defect prediction. Recognizing this, researchers have proposed innovative approaches to leverage this diverse information for more accurate defect prediction models. Huo et al. introduced a technique that jointly learns semantic features from both source code and comments for defect prediction [321]. Likewise, Liu et al. utilized the historical version sequence of manually crafted features from continuous software versions as input for RNN in defect prediction [322]. Subsequent studies follow a similar path, combining different sources of information. Some explore the joint use of change and commit logs [325], while others investigate the combination of manually crafted features and CDN (Class Dependency Networks) [341]. Integrating these diverse data sources provides richer and more expressive semantic features, leading to further improvements in defect prediction capabilities.

## 9.6 Datasets

The field frequently relies on datasets like NASA, PROMISE, AEEEM, and Relink, valued for their extensive coverage of Java and C, the primary programming languages [326,345]. These datasets predominantly focus on file-level granularity and typically encompass fewer than 1,000 instances within a project.

- **Programming language**. The majority of datasets comprise open-source projects, prominently featuring programming languages such as Java, C, and C++ [326]. Typically, these studies encompass around 10 projects, with over 80% utilizing Apache projects.

- **Prediction granularity**. The datasets exhibit varying levels of prediction granularity, spanning from file-level, change-level, function-level, to statement-/line-level. File-level granularity is the most prevalent (nearly 80%), whereas statement-/line-level granularity receives less emphasis [336, 348].

- **Training dataset size**. In the majority of studies, within-version defect prediction is the norm, leading to training sets typically comprising fewer than 1,000 instances. Notably, only a handful of change-level defect prediction studies leverage notably larger datasets, with instance counts reaching tens of thousands [332].

In summary, existing studies primarily concentrate on projects involving a small number of programming languages, mainly focusing on coarse-grained defect prediction. In particular, the training set sizes for deep learning are often limited.

## 9.7  Challenges and Opportunities

While numerous studies have put forward different defect prediction methodologies, several significant challenges hinder the development of accurate and cost-effective approaches for defect prediction. These challenges encompass issues like model interpretability, thorough assessment, and the replication of experiments. Simultaneously, there are promising opportunities on the horizon that could potentially address these challenges in future studies on defect prediction.

### 9.7.1  Challenges

The application of deep learning to defect prediction presents the following main challenges that need to be addressed:

- **Limited interpretability of control- and data-flow in relation to defect-proneness**. Practitioners highly value understanding the root causes within code that contribute to defect-proneness, particularly concerning control- and data-flow dynamics. Such understanding is pivotal in comprehending the occurrence of defects and subsequently addressing them. However, the opacity of deep learning models presents a challenge, as they are often perceived as black boxes, impeding the ability to interpret the rationale behind their predictions. This lack of transparency restricts the interpretability of the models, posing a barrier to the adoption in defect prediction, an area where interpretability holds critical importance.

- **Insufficient assessment of the effectiveness in comparison to traditional models**. Many studies evaluating new deep learning models tend to solely compare them against earlier deep learning approaches, often neglecting comprehensive comparisons with established traditional models. This oversight results in a significant gap in understanding the true advancements brought about by deep learning techniques in defect prediction. Consequently, it becomes challenging to gauge the tangible progress and assess how deeply these adopted deep learning methodologies truly drive the field forward compared to well-established traditional models.

- **Substantial challenges in replicating the reported findings externally**. The absence of standardized practices for distributing replicated packages, encompassing datasets and their corresponding scripts, poses a significant challenge. In practice, it is well known that even minor variations in the re-implementation of prior models can lead to substantial differences in defect prediction performance. This lack of transparency and accessibility hinders external validation of insights derived from past studies. As a result, ensuring the robustness and applicability of reported conclusions becomes progressively more challenging.

Future studies should aim to address the aforementioned challenges to accurately gauge the progress made in the field of defect prediction. Failing to do so will hamper our understanding of the extent to which advancements have been made. Furthermore, there is a risk of unintentionally drawing misleading conclusions regarding the advancement of the state-of-the-art, potentially leading to missed opportunities for further progress in defect prediction.

### 9.7.2 Opportunities

There are several valuable opportunities available to address the challenges associated with deep learning-based defect prediction.

- **Development of interpretable techniques**. One promising opportunity lies in the development of novel techniques that enhance the interpretability of deep learning models in defect prediction. Researchers can explore methods such as feature attribution, saliency mapping, and attention mechanisms to gain insights into the factors contributing to defect-proneness. These techniques would enable better understanding of model behavior and facilitate the identification of critical features and patterns related to defects.

- **Comparative evaluation frameworks**. A key opportunity is the establishment of comprehensive evaluation frameworks for deep learning-based defect prediction. This involves designing rigorous comparative studies that systematically compare the performance of deep learning models with traditional approaches on diverse datasets. By incorporating various evaluation metrics and statistical analysis, researchers can provide robust evidence on the effectiveness and advantages of deep learning models.

- **Replication and external validation studies**. Another important opportunity is to conduct replication and external validation studies to verify the reported findings in deep learning-based defect prediction. Collaboration among researchers, sharing of datasets and code, and adopting open science practices can facilitate the replication of experiments across different research groups. External validation studies conducted on independent datasets can provide further validation of the generalizability and reliability of the reported results.

By embracing these opportunities, researchers can address the challenges faced in deep learning-based defect prediction and pave the way for advancements in model interpretability, performance evaluation, and the overall reliability of defect prediction systems.

## 10  Bug Finding

The term "bug" originates from a literal bug (i.e., moth) stuck in the relay of the Mark II computer, and has been used to denote any defect that violates its specification: unexpected crashes, incorrect results, and information leakages, etc. Bug finding is a process that involves examining software artifacts (source code repositories, documentation, and existing test inputs, etc.) and generating a list of potential bugs with explanations.

The process of bug finding goes beyond mere "wild guesses". Unlike defect prediction (Section 9), which identifies loose correlations between software metrics and bugs, bug finding techniques should provide concrete bug certificates. Certificates can aid developers in accurately identifying the actual presence and root cause of a bug. Such certificates could be a hint, like a bug anti-pattern, or a specific test case that triggers a program error. This section discusses the three mainstream approaches to bug finding:

1. Static analysis. Bugs reside within source code, and bugs can be identified by "reading" the source code and its associated artifacts, without the need to execute the program in a real environment. Static bug does not require resources (such as hardware platform, computational power, environment, and dependencies) to bootstrap the software, making it accessible to any party involved in a software's life cycle. From Lint tools [350] to bug finders [351,352], static analysis serves as a fundamental gate-keeping procedure in the software development process.

2. Dynamic testing. More reliable certificates provide undeniable evidence of a bug's existence, with test cases [353] being the most commonly used such certificates. Developers create unit and system test cases, and software is also extensively validated using machine-generated test cases. Passing test cases enhance the confidence that the software functions under various circumstances.

3. Formal verification. In theory, one can exhaustively test a (finite-state) system to prove its bug-freedom. To make this procedure practical, one can leverage path-based abstraction and employ a constraint solver to manage the search space [354,355]. Alternatively, one can provide a machine-checkable proof to a proof assistant. There have been successful reports on the verification of compilers [356], operating system kernels [357], and other software systems [358].

The three mainstream approaches have undergone decades of evolution, particularly the classical (formal) algorithmic bug-finding techniques. These techniques rely on algorithms, logical procedures that can be mechanically implemented over simple axioms, to identify bugs [359]. With the advent of deep learning, probabilistic techniques based on machine learning have gained popularity, because learned models can effectively digest inputs from the chaotic and complex human world.

The algorithmic and probabilistic paradigms complement each other in the deep-learning era. This is because software, which projects the requirements of the probabilistic human world onto the algorithmic computing world, intersects both of these realms [360,361]. To provide a bug certificate, one must possess not only a thorough understanding of the requirements but also the ability to engage in complex logical reasoning.

Therefore, we argue that significant research efforts should be invested to answer how to leverage learned (albeit imperfect) domain-knowledge to facilitate an effective algorithmic bug-finding procedure. This approach builds upon the success of AlphaGo [362], harnessing the potential of AI-in-the-(algorithmic)-loop", akin to having "experienced human experts" consistently offering insights into software artifacts and intermediate results whenever decisions are required in static checking, dynamic testing, and formal verification. On the other hand, it would also be equally interesting to explore whether bug-finding can be driven by an autonomous agent, like the chain of thought [363,364], by automatically exploiting the existing algorithmic bug-finding techniques.

In addition to the classical papers that shed light on the fundamentals of bug finding, this section also comprehensively surveys recent papers by conducting Google Scholar search using the combinations of the following two sets of keywords:

- $S_1$: Testing, validation, verification, fuzzing, program analysis, static analysis, dynamic analysis, symbolic execution, formal methods;

- $S_2$: Neural network (NN),deep learning (DL),machine learning (ML),reinforcement learning (RL), large language model (LLM), transformer.

We collect recent (2018–2023) papers and narrow our selection to papers from top-tier conferences and journals in the fields of software engineering, programming languages, and computer systems. Additionally, we include widely-cited papers with 30 or more citations. In total, we select 72 papers for this section (and more in the case study on vulnerability detection in Section 10.4). These papers are categorized and discussed as follows.

## 10.1  Static Bug-finding: Program Analysis

### 10.1.1  Code proofreading

Even if we do not check against any specification on application behavior, we expect that source code reads like natural language texts, following the famous quote from student's first programming class [365]:

> Programs are meant to be read by humans and only incidentally for computers to execute;

Readability implies that the code is easier to maintain, and researchers do find that programs resemble natural-language texts to some extent [366]. This "naturalness" of software serves as the foundation of code proofreading (by either a human or a checker of probabilistic distribution) by identifying anomalies that correlate to bugs.

Such "proof reading" can be algorithmic, i.e., checking against formal best-practice specifications [367,368] or lint rules [350]. Adding a bit of probability to linters yields interesting chemical reactions: any "odd" (infrequent) pattern may suggest a bug [369]. This "may belive" approach paves a way for harnessing the power of both formal templates and code-level counter-coincidental couplings, leading to the family of bug-finding techniques that mine frequent items and consider minor items as bugs, e.g., CP-miner [370].

Today's deep learning models, which are fine-tuned over billion lines of code (e.g., llama2-code), are far more capable than finding these counter-coincidental couplings between lexical tokens. Large language models exhibit strong usefulness in finding (and even fixing) the "unnatural" bugs, even for semantics bugs that violate application specifications. For example, GitHub Copilot is an AI pair programmer which facilitates the finding and repairing of bugs through its context-sensitive recommendations [371]. BUGLAB is a self-supervised approach that trains bug detectors by co-training a bug selector, which produces bugs that are elusive to detect [372].

Seemingly surprising, the intuition behind these papers is straightforward: a small code snippet is "almost completely determined" by its surrounding context. Based on the fact that most of the bugs are not "new" and there are similar bug samples in the training data, a language model can well predict the existence of them–code proofreading is very useful as a gate-keeping procedure for improving software quality. On the other hand, we still cannot fully rely on it, either algorithmically or probabilistically, to find bugs. Probability distributions like ChatGPT are always happy to provide proofreading results and explanations on the bug, but many times "talks nonsense with a straight face". Algorithmic lint tools produce an excessive number of false alarms. These fundamental limitations can be alleviated, but is not expected to be resolved in the near future [360,373].

### 10.1.2 Semantic analysis

To catch subtle semantic bugs that appear to be correct and escape proofreading, it is necessary to understand precisely what a piece of program can and cannot do (i.e., conducting a semantic analysis) [374]. Observing that each program statement's behavior has been rigorously defined as a part of the language specification, it would be theoretically possible to predict a program's all possible behaviors by an algorithmic procedure (see Section 10.3; however, the general problem is undecidable [375]). On the other hand, while today's language models can somehow find semantic bugs, their probabilistic nature make them less predictable (and thus less reliable) than algorithms.

Classical semantic analysis strives to over-approximate the programs semantics to obtain sound predictions of program outcomes. In practice, static analysis techniques have to trade off between usefulness, thoroughness, and engineering efforts. For decades, the data-flow and abstract interpretation framework dominates the field, and numerous properties about programs can be automatically proved under rigorous logical reasoning upon the formal operational semantics [376], and scaling to millions of lines of code [352].

Still, static analyzers are limited to finding bugs of limited "general" types of semantic bugs like pointer errors and data races, which are neutral to the application logic. It seems still a long way for today's analyzers to automatically prove arbitrary developer's assertions about program states. Certificates from state analysis are also imperfect for bug-finding. Static analyzers have to draw indefinite conclusions like "two variables may have the same value" (but actually not) to achieve scalability, resulting in false bug certificates.

The paths for algorithmic and probabilistic static analysis have long been diverged, and the progress to unite them is still exploratory. Probabilistic distributions are useful in providing heuristic hints, e.g., control of the degree of analysis sensitivity to meet resource constraints [377]. [378] proposes a technique to learn effective state-selection heuristics from data, in order to keep track of a small number of beneficial program states in Infer [352]. LLift [379] complements semantic analysis by providing speculate conclusions on undecided (timeout) symbolic-execution paths. Static analyses can also be complemented by learned features over reduced programs [380], or leveraging anomaly detection techniques to learn a balance between precision and scalability [381].

Machine learning can also be incorporated with specific static analysis techniques for better effectiveness and precision. [382,383] address the challenge of manually developing cost-effective analysis heuristics for pointer analysis using an algorithm for heuristic learning. LAIT [384] utilizes an iterative learning algorithm that develops a neural policy to identify and eliminate redundant constraints throughout the sequence in order to produce a faster and more scalable numerical analysis.

Today's deep learning models are fundamentally limited in predicting the execution results of even a small code snippet [385], and we believe that static analysis is a field to be revolutionized by deep learning. The potential lies in the naturalness of software, which "informally" reflects program semantics, like the following piece of code, which is difficult to be fully (rigorously) analyzed. On the other hand, a language model can speculate variable's types or even possible values at a specific program point [386,387], even if the static analyzer cannot prove it. These results may further benefit analyses on other parts of the program, which may be useful in developing efficient analyses. Unfortunately, there still lacks a framework that can simultaneously exploit the power of rigorous and long logical reasoning and the power of understanding the human-world semantics of programs from LLMs [388,389].

Another promising direction is approximating program semantics via deep-learning models. For example, neural program smoothing [390,391] approximates programs as differentiable function of inputs and accordingly generates new test inputs using gradient descent. [392] constructs a benchmark suite with 28 open-source projects and proposes PreFuzz, which guides fuzzing by a resource-efficient edge selection mechanism and a probabilistic byte selection mechanism. However, the efficacy of neural program smoothing remains an area for further exploration according to extensive evaluation [393].

Enabling AI-in-the-loop for static analysis brings another challenge: machine learning models can be as large as billions of parameters, and it is impractical for static analyzers that frequently query the model. Today's AI-aided static analysis is still limited to manual feature engineering, in which simple classifiers like gradient boosting [378] are used. Deep learning model inference is considerably more expensive, and the problem remains open.

## 10.2  Dynamic Bug-finding: Software Testing

### 10.2.1   Test oracle

Before one can test a program, a fundamental question should be answered first: what do we mean by a program to be "correct"? Some correctness criteria are obvious: a program should not crash, should be race-free, memory-safe, assertion-pass, and, most importantly, functional. A somehow desperate fact is that as long as software reflects a physical-world procedure (i.e., requirements), being functional becomes a myth1), and this is referred to as the "specification crisis" [394].

In the context of testing, such a specification, which decides whether each test case passes, is called a test oracle. General programming errors, e.g., semantic errors discussed in Section 10.1, can be a part of a test oracle and can be effectively checked at runtime. Sanitizers [395,396] are famous for being effective in bug-finding for practical systems. However, finding a generic test oracle that decides whether a program is functional remains as an open problem.

Despite that the research community has a strong focus on modeling and lightweight formal methods [397], we argue that the test oracle problem may be one of the first problem to be effectively solved by deep learning. A key observation is that, while programs are generally hard for neural networks to understand, an execution trace really looks like a (maybelong) story consisting of events. The implication of this observation is twofold:

1. Deep learning models can mimic a human developers that write testcases or directly act as a human test operator, and simultaneously generate expected program behavior (test oracle) [398], following that a test case describes a "natural" procedure in the human world, e.g., generating natural faults for mutation testing [399].

2. Deep learning models can digest software's execution traces. Traces can be projected to the human domain, and a language model can find "common sense violation" cases [400].

Test oracles for unit testing is a promising direction to be solved by machine learning [401,402]. AthenaTest [403] is a method designed to generate unit test cases through a sequence-to-sequence learning model, trained first on a large unsupervised Java corpus for denoising and fine-tuned on real-world methods and developer-written test cases. Atlas [404] is a Neural Machine Translation (NMT) based approach designed to predict a meaningful assert statement to assess the correctness of the focal method. [405] leverages a transformer model, first pre-trained on an English textual corpus and then semi-supervised trained on a substantial source code corpus, culminating in finetuning for generating assert statements in unit tests. CallMeMaybe [406] is a technique employing natural language processing (NLP) to analyze Javadoc comments for identifying temporal constraints, thereby aiding test case generators in executing method call sequences that adhere to these constraints. TOGA [407] is a unified transformer-based neural method designed to deduce exceptional and assertion test oracles from the context of the focal method, effectively addressing units with unclear, absent documentation, or even missing implementations. ChatUniTest [408] is a ChatGPT-based automated unit test generation tool which creates adaptive focal contexts from projects, uses them in prompts for ChatGPT, and then validates and repairs the generated tests using rule-based and ChatGPT-based approaches. A3Test [409] is a DL-based approach that addresses the limitations of AthenaTest [403] by incorporating assertion knowledge with a mechanism to verify naming consistency and test signatures. CodaMosa [398] leverages an LLM to generate unit tests (oracles) to reach a designated program part. To generate test oracles for games, Neatest [410] navigates through a program's statements, constructing neural networks that manipulate the program for dependable statement coverage, essentially learning to strategically explore various code segments. One can also model the test generation problem as a completion problem [116].

---

1) For example, more types of genders are recognized by the society over time, but software systems often fall behind.

Existing studies show that the quality of LLM-generated test cases (including oracles) still have significant room for improvement [411–414]. [415] points out that unit tests generated by ChatGPT still suffer from correctness issues, including diverse compilation errors and execution failures.

The challenge of implementing a "neural test oracle" is that program traces are too verbose for today's transformer architecture to digest. Neural networks fall short on finding long-range logical dependencies in the trace. The probabilistic nature of deep learning models also implies that they are imperfect test oracles. Nevertheless, we can always generate more test cases, and deep neural networks are not likely (though still possible) to make false-negative predictions on all test cases that trigger the same bug.

### 10.2.2 Test input generation

Even if the specification crisis is resolved, bug-finding is still a challenge. The number of test inputs is astronomically high, and we do not have the resources to examine all of them. This issue, which we refer to as the "search-space curse," is akin to finding needles in a haystack. Consequently, we only afford sampling useful test inputs that exercise diverse program behaviors (in hope of revealing bugs). A natural idea is to decompose the input space over its structure:

1. Connecting the search space, based on the observation that useful (generally available, e.g., regression tests [416,417]) inputs can be slightly mutated (modified) to obtain another useful input, which may manifest different program outcomes. Then, the input space can be decomposed into a graph, where vertices (inputs) are connected by mutation operators. The exploration procedure can be guided, e.g., by leaning towards test inputs that cover new code [418–420]. Machine learning is also useful in creating mutants [391,399,421].

2. Partitioning the search space, and only sample representative test input(s) are selected in the equivalent classes. To alleviate the search-space curse, symbolic execution essentially merges all test inputs that share the same control-flow path. Then, for every path, we only care about whether it is reachable or not. The search is postponed, and a clever constraint solver may quickly draw a conclusion.

We see solid progress in improving how we decompose (and explore) the input space. Learned probability distributions over test inputs, traces, and execution logs are undoubtedly useful in whenever decisions should be made, e.g., in ranking the mutants [422,423] or providing useful "golden" seeds [424,425].

Deep learning-based generative models, particularly LLMs, exhibit exceptional performance in code generation tasks and are thus frequently employed in generating seeds for testing compilers or deep learning libraries. Deng et al. presented TitanFuzz as the first technique to directly leveraging LLMs to generate seeds for fuzzing DL libraries [424]. FuzzGPT [425] follows up the work by using LLMs to synthesize unusual programs for DL fuzzing. WhiteFox [426] uses LLMs to produce test programs with source-code information to fuzz compilers. Fuzz4All [427] is the first universal fuzzer capable of targeting a wide range of input languages and their various features. It capitalizes on LLMs for input generation and mutation, generating diverse and realistic inputs for any language. COMFORT [428] is another compiler fuzzing framework designed to identify bugs in JavaScript engines, utilizing advancements in deep learning-based language models for automatic JS test code generation. DeepSmith [429] uses generative models to generate tens of thousands of realistic programs, thereby accelerating the fuzzing process of compilers.

Machine-learning can also provide heuristic decisions for accelerating fuzzing. Specific work includes RegFuzz, a directed fuzzing approach that employs a linear regression model to predict seed effectiveness, thereby allocating more energy and fuzzing opportunities to efficient seeds [430]. HATAFL [431] utilizes pre-trained LLMs to construct grammars for protocol message types and assist in mutating messages for protocol fuzzing. SEAMFUZZ [421] learns effective mutation strategies by capturing the characteristics of individual seed inputs. [422] proposes a reinforcement learning-based hierarchical seed scheduling strategy for greybox fuzzing. RLF [432] models the fuzzing process of smart contracts as a Markov decision process and uses a specially designed reward system that considers both vulnerability and code coverage. [433] leverages the Monte Carlo Tree Search (MCTS) to drive DL model generation, thus improves the quality of DL inference engines. [434] employs machine learning models and meta-heuristic search algorithms to strategically guide the fuzzing of actuators, aiming to maneuver a Cyber-Physical System (CPS) into various unsafe physical states. NeuFuzz [423] utilizes deep neural networks for intelligent seed selection in graybox fuzzing which learns vulnerability patterns in program paths. Some studies also utilize machine learning models to integrate fuzzing with symbolic execution. For example, [435] predicts the timing for switching between concrete and symbolic executions. [436] trains a neural network-based fuzzing policy on the dataset generated by symbolic execution, enabling the application of the learned policy to fuzz new programs. JOpFuzzer [437] learns the relationships between code features and optimization choices to direct seed mutation for JIT compiler fuzzing.

The challenge here is similar to incorporating deep learning within static analysis: excessive amount of expensive queries may outweigh simply exercising more test cases.

Machine learning models generally perform better for domain-specific test input generation. For example, WebExplor [438] leverages a curiosity-driven reinforcement learning to generate high-quality action sequences (testcases) for web testing. FIGCPS [439] adopts deep reinforcement learning to interact with the Cyber-Physical Systems (CPS) under test and effectively searches for failure-inducing input guided by rewards. Mobile applications provide a natural human interface, which can be effectively understood by machine-learning models. QTypist [440] utilizes a pre-trained LLM for generating text inputs based on the context of a mobile application's GUI. [441] proposes Deep GUI, which enhances black-box testing by utilizing deep learning to generate intelligent GUI inputs. AdaT [442] is a lightweight image-based approach that uses a deep learning model to dynamically adjust inter-event times in automated GUI testing based on the GUI's rendering state. Badge [443] is an approach for automated UI testing which uses a hierarchical multi-armed bandit model to prioritize UI events based on their exploration value and exploration diversity. Q-testing [444] employs a curiosity-driven strategy to focus on unexplored functionalities and uses a neural network as a state comparison module to efficiently differentiate between functional scenarios. [445] introduces Avgust, a system that automates the creation of usage-based tests for mobile apps by using neural models for image understanding.

Machine learning models are also capable of understanding "stories"–API call sequences. APICAD [446] and NLPtoREST [447] are tools that enhance REST API testing by applying NLP techniques from API documents and specifications. [448] describes an adaptive REST API testing technique that employs reinforcement learning to prioritize API operations and parameters, using dynamic analysis of request and response data and a sampling-based strategy to efficiently process API feedback.

## 10.3 Proving Bug-freedom: Formal Verification

### 10.3.1 Searching for needles in the haystack

To the extreme end of testing, one can theoretically test over all possible inputs to verify that a program is bug-free (or to find all bugs). Exhaustive enumeration is the ultimate victim of the search-space curse. While we cannot leverage the small explanation hypothesis in verification (we cannot leave any corner case unchecked), the idea of search space decomposition still applies[2]. For example, one can separately verify each program path, in which each verification is essentially a smaller search problem that can be solved by a constraint solver. Search spaces may have their own structures and pruning opportunities [449], which can be accelerated by machine learning [450].

On the other hand, path-based verification is not a silver bullet. Loops, even nested with simple control flow, post significant scalability challenges to a symbolic verifier. For example, a loop-based popcount implementation, which sequentially checks each bit of a 32-bit integer and increments a counter when the bit is set, consists of $2^{32}$ distinct program paths, and off-the-shelf symbolic execution engines fail to verify it. Sometimes we may rewrite the above function into one formula that can be recognized by a constraint solver [451], e.g.,

$$popcount(x) = \sum_{0 \leqslant i < 32} \text{AND}(\text{SHR}(x, i), 1),$$

---

2) Decomposition is not limited to input spaces. One can also decompose the search space consisting of program states for model checking.

to "offload" the $2^{32}$ paths to the constraint solver. Generally, we do not have this luck for most of the practical cases, and symbolic program verification is still limited to small programs. The complexity issues raised by control flows, pointers, memory allocation, libraries, and environments, are all challenges to verify practical programs [452].

Dynamic symbolic execution is a path-based program verification technique, and many learning-based techniques have emerged to ease the search-space curse in symbolic execution. Most studies focus on employing machine learning techniques to devise an optimized search strategy, thereby reducing the time and space overhead of path enumeration. Learch [453] utilizes a machine learning model to predict the potential of a program state, specifically its capability to maximize code coverage within a given time budget. [454−456] dynamically adapt search heuristics through a learning algorithm that develops new heuristics based on knowledge garnered during previous testing. There are also techniques to prune the search space. Homi [457] identifies promising states by a learning algorithm that continuously updates the probabilistic pruning strategy based on data accumulated during the testing process. Others optimize symbolic execution from different perspectives, including the prediction and optimization of path constraints [458,459], as well as the fine-tuning of search parameters [460], and transformation of target code [461].

How can probabilistic techniques be useful in software verification, i.e., an exhaustive search over the input space? The use of machine learning model must be sound, i.e., the checking results remain correct even under prediction errors. This problem remains open today.

### 10.3.2 Providing a checkable proof

Proving bug-freedom does not really require an exhaustive enumeration. Programs are rigorous mathematical objects: programs can be regarded as a function taking an input and produces an execution. One would always provide such a program with a logical proof that asserts all produced executions satisfying the specification, like we prove the correctness of any algorithm, e.g., bubble sort indeed gives a sorted array after n − 1 iterations. The validity of such proofs can be checked by a proof assistant like Coq [462] or Isabelle/HOL [463,464], to provide a certificate that the proof is correct [356,357].

Fully automatic theorem proving is hard, even for short mathematical proofs. We could also search for the proof, carrying the search-space curse, and exploiting deep neural networks for heuristics. In contrast to programs that implement a human-world requirement, mechanical proofs are quite "unnatural", and understanding a proof usually requires a careful examination of the proof stack, while code in mainstream programming languages reads much more like a natural language text. This implies that training deep learning models for creating proofs is considerably more challenging [465,466], and learning to accelerate search for proof tactics is still in preliminary stage [467].

We argue that proofs for verifying software systems have a considerably different structure compared with proofs for mathematical theorems (the focus of today's research [468]), and the research community may have a paradigm shift: the core of a proof is invariants, which "summarizes" what happens in the intermediates of program execution, to form inductive hypotheses for machine-checkable proofs, and we identify strong patterns for program invariants. They can be done by humans, and we see opportunities that human work can be replaced by deep learning, e.g., machine-learning models can rank generated invariants [469].

An interesting observation is that we are trying to prove that a program satisfies a specification regardless of the input space size and the program execution length. However, the input space can be huge (or even infinite), and the program execution can be lengthy! Both the program and the proof seem much more concise compared with the set of all possible program execution traces, and proof checking can be done reasonably efficient. This phenomenon, which connects to the small explanation hypothesis, suggests that the program's execution space (inputs and their corresponding traces) follows a somehow simple structure that can be described algorithmically, and we might avoid a costly exhaustive enumeration. We speculate that practical software implementations are of the same magnitude of the minimum specification-satisfying implementation, like the "Kolmogorov complexity" of software. The implications of this phenomenon also remain open.

## 10.4 Case Study: Vulnerability Detection

Following the above framework, this section further describes how deep learning techniques can improve the effectiveness and efficiency of finding a specific kind of bugs−security vulnerabilities, which can be exploited by attackers to gain unauthorized access, perform malicious activities, or steal sensitive data.

### 10.4.1 Static Vulnerability Detection

In modern times, the dominant model for software development revolves around library-based programming. The primary objective is to enhance development efficiency, minimize program complexity, and streamline operations such as development and maintenance. Program documentation plays a crucial role in providing a natural language description of the program, aiding users in comprehending and utilizing it effectively. Within a code base, an API serves as an interface that enables users to access its various functions. These APIs subject to certain security constraints, such as manually releasing function return pointers, among others. These security constraints, known as security protocols, are documented by the code base developers within the program documentation. By documenting these protocols, developers offer users of the code base a valuable point of reference and guidance. During a call to the API, the developers must comply with the constraints of API calls. Otherwise, API misuse can occur, leading to serious software security issues, such as NULL pointer dereference, pointer use after free, and logical bugs, etc.

In recent years, many researchers have used text analysis to find various security problems automatically, including access control configuration errors, wrong access requests, and logic flaws, etc. For example, application developers provide privacy policies and notify users, but users cannot tell whether the application's natural behavior is consistent with their privacy policies. In response to this problem, Zimmeck et al. [470] proposed a systematic solution to automate the analysis of privacy policies to detect inconsistencies between them and application-specific behavior. Tools such as WHYPER [471] and AUTOCOG [472] examine whether Android applications correctly describe usage permissions in application descriptions. Similarly, Liu et al. [473] used text categorization and rule-based analysis to test for consistency between standard EU data protection regulations and applicable privacy policies.

The approaches above operate under the assumption that the documentation is accurate and do not contain errors. Consequently, if the code contains defects related to an API that lacks documentation, these defects cannot be detected. Cindy et al. [474] examined 52 filesystems and discovered discrepancies between the error codes returned by functions and those recorded in the documentation. This investigation revealed over 1,700 undocumented error codes. Tan et al. [475] utilized a series of rule templates and a pre-trained decision tree to filter out comments from code that described API usage specifications. The program was then analyzed with user-provided function names (e.g., lock/unlock function names) to identify inconsistencies between the comments and the code. TCOMMENT [476] focuses on parameter values in Java comments and verifies the consistency of exceptions thrown under those values with the actual types of exceptions thrown by the code. Wen et al. [477] performed data mining on 1,500 software code submissions and manually analyzed 500 to classify inconsistencies between code and comments. They also discussed the degree to which a code submission necessitates concurrent modification of comments, guiding for identifying and resolving inconsistencies between code and comments. Pandita et al. [478] employed machine learning models to filter out sentences in documents that describe API usage timing. They subsequently employed traditional natural language processing techniques to transform these sentences into first-order logical expressions. They further identified code defects that deviate from these specifications by constructing a semantic diagram of the document statements and inferring API usage timing specifications. Ren et al. [479] extracted an API Declaration Graph from semi-structured API declarations and derived usage specifications from the natural language descriptions within API documents. They then used this information to generate a knowledge graph encompassing API usage constraints, facilitating the detection of API misuse. Lv et al. [480] introduced Advance, the first comprehensive API misuse detection tool, employing document analysis and natural language processing techniques.

Several studies summarize security protocols from many code usage examples for vulnerability detection. One intuitive way to do this is to automate the analysis of a large amount of code, then take a majority vote and use the most frequent code used as a reference for API usage. For example, Apisan [481] extracts usage patterns from a large amount of code through parameter semantics, causality and then extracts API usage references from usage patterns based on a majority vote. Thus, Apisan no longer needs to define defect patterns manually. Apex [482] finds criteria for the API return value range on this branch based on fewer observations of code branch statements that handle errors and then infers the API's Error Specification based on the principle that most people are right and diagnosing defects for handling code snippets that do not follow the error definition for API return values. Like APEx, Ares [483] uses heuristic rules to identify error-handling blocks of code. A majority vote on the entry criteria for these blocks and a range merge results in an API error definition and diagnosing a defect by checking the return value of the API against a check that violates the error definition. Apisan, Apex, and Ares all rely on the majority vote, but the majority vote is only sometimes right, which leads to the fallacy of the inferred specification itself.

Deep learning-based large language models (LLMs), represented by the Transformer structure, are being applied to vulnerability detection tasks, mainly for static code analysis. Given a piece of code snippet, LLMs are asked through a question-answering dialogue whether it detects any vulnerabilities in the code and provides an explanation. However, LLMs still cannot handle various types of sensitive detection (including flow-sensitive, domain-sensitive, context-sensitive, etc). Therefore, it is necessary to fine-tune the large model or introduce additional knowledge through prompts to guide LLMs to gradually correct its analysis results. For specific field program vulnerability detection, such as smart contracts and shell scripts, due to their short length and low complexity, LLMs usually have a more accurate performance.

## 10.4.2  Dynamic Vulnerability Detection

In the field of vulnerability detection, fuzzy testing is an efficient dynamic detection technique. It explores and detects vulnerabilities in programs by continuously constructing unexpected abnormal data and providing them to the target program for execution while monitoring program execution anomalies [484]. During security testing, a large amount of data can be produced and further employed, such as test cases, execution traces, system states, software implementation specifications, and vulnerability descriptions. This information can be analyzed with deep learning techniques to improve fuzzy testing. For example, natural language processing can be used to understand text descriptions related to vulnerabilities, which can assist in generating test cases [485]. Through the strong fitting ability of deep learning models, mappings between program inputs and states can be accurately established, which can guide test case mutation [390]. Models trained and generated from existing testcases can automatically learn some input specifications to facilitate input generation [486]. At the same time, with the trained model, guidance and reasoning can be performed at a relatively low cost, which helps use deep learning techniques in real-time during fuzzy testing.

As for the objectives, the application of deep learning in fuzzy testing can be divided into two categories: reducing human processing overhead and increasing decision intelligence. The former includes reducing preparation work before testing, such as input model inference and mutation operation customization; the latter includes tasks such as seed file scheduling, mutation operation scheduling, and test case filtering.

Researchers have proposed to use deep learning algorithms to learn a generation model from existing test cases for input model inference or to enhance existing input models by automatically understanding auxiliary information such as input specifications using machine learning algorithms [487,488]. On the other hand, researchers have used machine learning techniques to customize mutation operations for different programs. Angora [489] first uses taint analysis technology to obtain input positions that affect specific branches in a program. By converting branch conditions into input functions, Angora uses gradient descent to mutate corresponding input positions to generate test cases covering specified branches. This adaptive gradient-based mutation operation does not require manual setting and outperforms randomly using existing mutation operations in experiments.

Subsequently, researchers [390–392] further propose to use a single function to fit the input to the corresponding branch coverage of a program, followed by selecting input positions affecting specified branches based on the gradient information of the function and performing targeted mutations accordingly. According to the Universal Approximation Theorem [490], neural networks have strong function fitting capabilities and can approximate any function. Additionally, they have good generalization ability and easy calculation of gradients. Therefore, researchers propose to use neural networks as a function to fit program behaviors, with the principle of larger gradients indicating greater impact on the corresponding edge as the basis for the automatic selection of mutation positions and adaptive mutation based on the size and direction of gradients. These techniques alleviate the cost of expert-designed mutation operations to some extent while also having adaptability for different programs.

Due to the inherent randomness of mutation operations, generated test cases may not meet specific test input generation standards for fuzzy testing. The main cost of a mutation-based fuzzy testing tool lies in the execution of test cases [491,492]. If deep learning can be used to filter inputs before execution, it can reduce unnecessary running costs of target programs and improve the efficiency of fuzzy testing. Deep learning-based directed fuzzy testing, such as Neufuzz [390] and Fuzzguard [493], provides a novel approach to filtering redundant test cases. This approach collects a large number of test samples and uses whether they are reachable on a sensitive path as the classification standard to train deep learning models to classify and predict the reachability of future test samples. However, one limitation of this approach is whether the model can accurately understand the code logic. To overcome this limitation, we need to combine the semantics of the code itself so that the model can correctly understand the logic of the code and make accurate judgments about test cases.

## 10.5   Datasets

Considering the naturalness and complexity of modern software systems, it is not likely that anyone can train neural network models from scratch. Therefore, a mainstream approach to bug-finding is embracing pre-trained models for static analysis, dynamic testing, and formal verification [398,424]. To train domain-specific machine-learning models, e.g., for GUI trace understanding or heurstic decision-making, datasets are needed [390,404]. Alternatively, one may randomly select a subset of program execution results to serve as training datasets [453]. A unified, large-scale dataset has not been identified, hence it is not explored in this discussion.

## 10.6   Challenges and Opportunities

### 10.6.1   Challenges

In pursuing effective bug-finding techniques, the challenges for either static analysis, dynamic testing, or formal verification, all point to the specification crisis and the search-space curse. Interestingly, both the crisis and the curse arise from the formal aspect of programs and the algorithmic nature of the bug-finding techniques.

Machine learning, particularly deep learning techniques, serves as a bridge between the algorithmic realm and the human realm. In short term, even if today's deep learning models are still superficial and fall short on long chain of logical inference, they are extremely good at digesting software artifacts. Whenever there is a need for heuristics, deep learning models have potential to perform significantly better than hand-crafted heuristics. The effectiveness of heuristics is multiplicative: among a huge number of decisions, even small improvements may result in magnitudes of significant efficiency improvements.

- To the probabilistic end, while deep learning models are generally replacing human beings in conducting simple, fast jobs like writing unit tests, the context-length of today's transformers makes it fundamentally limited in understanding large-scale systems—we still need an effective mechanism to simplify or reduce large systems such that neural networks can handle various analysis tasks on these systems.

- To the algorithmic end, we dream one day, a compiler[3] is sufficiently powerful to automatically generate a proof for arbitrary assertions, even in natural language, or provide a counter-example. All programs that compile will automatically be "provably correct" to some extent. However, providing a prove, particularly for large-scale programs, are far beyond the capability of today's verifiers (model checkers) and automatic theorem provers.

### 10.6.2   Opportunities

Looking back at the academia's main theme of bug finding in the past decades, we see the thrive of both fully automatic bug-finding algorithms and end-to-end models. This is partly because for both parts we have available benchmarks for the push-button, reproducible evaluation. To go even further, perhaps we have overlooked the fact that we have developers, Q/A teams, who are also "probabilistic" and whose performance varies day by day in the loop of software development. We could be more open to semi-automated techniques that invoke humans, exploit humans, and tolerate biases and errors. Such "humans" will eventually be replaced by a deep learning model on the availability of data.

---

3) The term "compiler" may no longer be appropriate at that time. Better to call it a "terminator" that kills programmers.

- To the probabilistic end, it remains open and interesting whether there is an effective chain-of-thought to draw useful conclusions on software with the ability to understand both the software artifacts and analysis results from algorithmic tools.

- To the algorithmic end, we may hit a balance in the middle: in case the compiler is not powerful enough to prove an assertion, compilation will fail and the program should improve the code to make the program easier to verify. The Rust programming language is an early-stage attempt following this pathway [494].

# 11 Fault Localization

Fault localization (FL) in software engineering is the process of identifying the specific code elements (i.e., statements or functions) in a program where defects occur [495,496]. Currently, due to the requirement of oracles, FL techniques mainly focus on functional bugs, which could be found by correctness specifications such as unit tests.

Traditionally, FL techniques involve manual debugging techniques [497], such asprint statements, code inspection, and step-wise program execution. While these techniques have been widely used, they can be time-consuming, error-prone, and inefficient, particularly in large and complex software systems. To overcome these limitations, researchers from the software engineering community have explored automated FL techniques. These techniques aim to leverage various information sources, such as program execution traces, test cases, and code coverage information, to identify the locations in the code base that are most likely responsible for the observed failures.

Automated FL techniques can be broadly categorized into Heuristic FL and Statistical FL approaches. Heuristic FL techniques rely on pre-defined heuristic rules to locate the bugs that share similar buggy behavior. For example, some utilize dynamic dependencies (i.e., program slices) [498], some utilize stack traces [499], and some mutate the crisis values during test execution [500]. On the other hand, statistical FL techniques [501–503] build statistical models of buggy programs to analyze the relationships between code elements and failures. Most statistical FL techniques utilize program spectra [504–506], a kind of coverage information collected from test execution, to learn a model or a formula and then use it to rank the code elements. Some use other information sources, such as mutation analysis [507,508]. The former family is called spectrum-based FL (SBFL), while the latter is called mutation-based FL (MBFL).

Recently, with the rapid development of deep learning, deep-learning-based fault localization (DLFL) techniques have shown the potential to automate and improve the accuracy of fault localization. By utilizing neural networks and sophisticated learning algorithms, these approaches can effectively identify fault-prone regions of code, prioritize debugging efforts, and accelerate the resolution of software defects. In the following, we divide deep learning-based fault localization into two categories: techniques for directly enhancing fault localization and techniques for augmenting input data for fault localization.

## 11.1 Fault Localization Approaches

Before the rise of deep learning, researchers had already attempted to establish a connection between the coverage information of test executions and the test results, in order to predict the location of faulty code. As early as 2011, Wong et al. proposed back-propagation neural network models for defect localization [509]. Subsequently, Wong et al. made improvements by using a more complex Radial Basis Function network [510].

To the best of our knowledge, Zheng et al. [511] and Zhou et al. [512] first proposed to adopt deep learning approaches into FL, in 2016 and 2017, respectively. They used a simple full-connection deep neural network (DNN) that is trained against the same input of the SBFL. The primary evaluation results show the potential of DNN, which significantly outperforms the DStar approach, which was recognized as one of the most effective SBFL techniques at that time.

DeepFL, the milestone of DLFL, gained huge attention in the field of software engineering [513]. To address the limitations of traditional fault localization techniques, DeepFL utilizes various deep learning architectures, such as MLP, and RNN, to capture different aspects of the software system and learn intricate patterns and correlations. By training on labeled data consisting of program execution traces, testcases, and associated fault information, DeepFL is able to make accurate predictions on the likelihood of specific code locations being responsible for failures. The complex DNN models enable DeepFL to handle multiple kinds of information (including SBFL features, MBFL features, and code-complexity features) and complex run-time traces.

Zhou et al. proposed CNN-FL [514], which utilizes a tailored CNN model to process data for FL. First, CNNs are capable of effectively learning local features of the code, leading to more accurate fault localization. Second, CNN-FL can handle large-scale software systems and exhibits good scalability on extensive code repositories. Finally, this approach does not rely on specific feature extraction techniques but rather automatically learns the most relevant features through the network.

Li et al. proposed DEEPRL4FL [515], a fault localization approach for buggy statements/methods. DEEPRL4FL exploits the image classification and pattern recognition capability of the CNN to apply to the code coverage matrix. CNNs are capable of learning the relationships among nearby cells via a small filter and can recognize the visual characteristic features to discriminate faulty and non-faulty statements/methods. To date, DEEPRL4FL still holds the best performance in terms of the $Top\text{-}1$ metric.

Lou et al. proposed GRACE [516], a method-level FL approach based on the graph-neural network (GNN). GRACE represents aprogram by a graph, where nodes represent code elements or tests, and edges represent coverage relationships or code structures. By leveraging the power of graph representations and learning latent features, the approach enhances fault localization accuracy and helps identify faulty code locations more effectively.

Qian et al. utilized graph convolutional neural networks (GCN) to improve localization accuracy and proposed AGFL [517]. AGFL represents abstract syntax trees by adjacent matrix and program tokens by word2vec, and then combines these features to further train GCN models. AGFL applies attention and GCN to classify whether an AST node is buggy.

Qian et al. proposed GNet4FL, which is based on the GCN [518]. To improve the performance, GNet4FL collects both static features based on code structure and dynamic features based on test results. It utilizes GraphSAGE to generate node representations for source codes and performs feature fusion for entities consisting of multiple nodes, preserving the graph's topological information. The entity representations are then fed into a multi-layer perceptron for training and ranking.

Zhang et al. proposed CAN, a context-aware FL approach based on GNN [519]. CAN represents the failure context by constructing a program dependency graph, which shows how a set of statements interact with each other. Then, CAN utilizes GNNs to analyze and incorporate the context ( e.g., the dependencies among the statements) into suspiciousness evaluation.

Li et al. proposed FixLocator [520], a DLFL approach that can locate faulty statements in one or multiple methods that need to be modified accordingly in the same fix. FixLocator utilizes dual-task learning with a method-level model and a statement-level model. Similarly, Dutta et al. designed a hierarchical FL approach that uses two three-layer DNNs to first localize a function and then localize a statement [521].

Yu et al. proposed CBCFL, a context-based cluster approach that aims to alleviate the influence of coincidental correctness (CC) tests [522]. CBCFL uses the failure context, which includes statements that affect the output of failing tests, as input for cluster analysis to improve CC test identification. By changing the labels of CC tests to failing tests, CBCFL incorporates this context into fault localization techniques.

Li et al. proposed a two-phase FL approach based on GNN [523]. It extracts information from both the control flow graph and the data flow graph via GNN. The localization process is divided into two phases: (1) computing the suspiciousness score of each method and generating a ranking list, and (2) highlighting potential faulty locations inside a method using a fine-grained GNN.

Yosofvandet al. proposed to treat the FL problem as anode classification problem, where the Gumtree algorithm is used to label nodes in graphs comparing buggy and fixed code [524]. This paper uses GraphSAGE, a GNN model that handles big graphs with big neighborhoods well.

Wu et al. proposed GMBFL which improves MBFL via GNN [525]. Existing GNN-based approaches mainly focus on SBFL, while GMBFL first represents mutants and tests by a graph. The nodes of a graph are code elements, mutants, and tests. The edges are the mutation relationship between a code element and a mutant, the killed relationship between a mutant and a test, and the code structural relationship between two code elements of different hierarchies. Then GMBFL trains a Gated Graph Attention Neural Network model to learn useful features from the graph.

In addition to test-based fault localization methods, there are also approaches that localize the bugs from change sets. BugPecker [526] is the first to encode the commits and bug reports into revision graphs.

Ciborowska et al. proposed to fine-tune the BERT model for locating the buggy change set [527].

To evaluate and compare the performance of different DNN models, Zhang et al. processed a large-scale empirical study [528], which involves CNN, RNN, and multi-layer perceptron. The evaluation results show that CNNs perform the best in terms of identifying real faults.

## 11.2   Data Augmentation and Data Processing Approaches for Fault Localization

Data augmentation refers to the technique of artificially increasing the size and diversity of a dataset by applying various transformations or modifications to the original data. It plays a crucial role in improving the performance and robustness of deep learning models. In the context of fault localization, data augmentation can be used to enhance the effectiveness of deep learning-based approaches to fault localization.

Zhong and Mei proposed CLAFA [529], which employs word embedding techniques to process the names within code. Then it compares program dependency graphs from buggy and fixed code to locate buggy nodes and extracts various graph features for training a classifier.

Zhang et al. address the data imbalance problem in FL [530], which is caused by the fact that the number of failing test cases is much smaller than that of passing test cases. This paper employs test case resampling to representative localization models using deep learning, and improves the accuracy of DLFL.

Xie et al. proposed Aeneas [531], which employs a revised principal component analysis (PCA) to generate a reduced feature space, resulting in a more concise representation of the original coverage matrix. This reduction in dimensionality not only improves the efficiency of data synthesis but also addresses the issue of imbalanced data. Aeneas tackles the imbalanced data problem by generating synthesized failing test cases using a conditional variational autoencoder (CVAE) from the reduced feature space.

Hu et al. proposed Lamont, which uses revised linear discriminant analysis (LDA) to reduce the dimensionality of the original coverage matrix and leverages synthetic minority over-sampling (SMOTE) to generate the synthesized failing tests [532].

Lei et al. proposed BCL-FL [533], a data augmentation approach based on between-class learning. By leveraging the characteristics of real failed test cases, BCL-FL produces synthesized samples that closely resemble real test cases. The mixing ratio of original labels is used to assign a continuous value between 0 and 1 to the synthesized samples. This ensures a balanced input dataset for FL techniques.

Lei et al. proposed CGAN4FL [534], a data augmentation approach to address the data imbalance problem in FL. CGAN4FL employs program dependencies to create a context that exposes the root causes of failures. Subsequently, CGAN4FL harnesses a generative adversarial network (GAN) to examine this failure-inducing context and generate test cases that belong to the minority class (i.e., failing test cases). Ultimately, CGAN4FL integrates the synthesized data into the existing test cases to obtain a balanced dataset suitable for FL.

Zhang et al. proposed UNITE [535], which utilizes context information of trace data. UNITE combines three sources of information (i.e., a statement, a test case, and all test cases of a test suite), and then computes the influence relationship of the statements by program dependencies. In evaluation, UNITE significantly improves the state-of-the-art DLFL approaches.

Also, researchers try to improve DLFL from other aspects. Tian et al. proposed to use DNNs to extract deep semantic changes to construct better mutants to improve FL performance [536]. Zhang et al. proposed to synthesize failing tests to improve the performance of DLFL [537].

## 11.3   Evaluation Metrics

Evaluation metrics play a crucial role in assessing the performance of FL techniques. Similar to traditional FL approaches, DLFL studies adopt the commonly used metrics, shown as follows:

1. **Top-N**. This metric measures the number of the model in identifying the correct faulty code element within the top-N ranked elements.

2. **MAR**. MAR (Mean Average Rank) is the mean of the average rank.

3. **MFR**. MFR (Mean First Rank) is the mean of the first faulty statement's rank of all faults using a localization approach.

**Table 15** The Frequently Used Evaluation Metrics.

| Metric Name | Used Times | Reference |
|---|---|---|
| Top-N | 12 | [515] [513] [516] [517] [518] [520] [522] [526] [527] [534] [535] [525] |
| MAR(Mean Average Rank) | 8 | [515] [513] [516] [517] [518] [534] [535] [525] |
| MFR(Mean First Rank) | 8 | [515] [513] [516] [522] [534] [535] [523] [525] |
| EXAM | 6 | [509] [510] [511] [512] [514] [521] [523] |
| RImp(Relative Improvement) | 6 | [512] [514] [519] [522] [534] [535] |
| MAP(Mean Average Precision) | 2 | [526] [527] |
| MRR(Mean Reciprocal Rank) | 2 | [526] [527] |
| Hit-N(multi-defects) | 1 | [520] |

4. **EXAM**. EXAM stands for Expected Maximum Fault Localization, which measures the expected rank of the first correct fault location in a ranked list of code elements.

5. **RImp**. RImp (Relative Improvement) is to compare the total number of statements that need to be examined to find all faults using one FL approach versus the number that need to be examined by without using the FL approach.

6. **MAP**. MAP (Mean Average Rank) first computes the average precision for each fault, then calculates the mean of the average precision.

7. **MRR**. MRR (Mean Reciprocal Rank) metric calculates the mean of the reciprocal position at which the first relevant method is found.

8. **Hit-N**. Hit-N is a metric designed for multi-defects, which measures the number of bugs that the predicted set contains at least N faulty statements.

Table 15 summarizes the commonly used metrics. The most popular metrics are Top-N and EXAM, which are used eight times and six times, respectively. RImp, MFR, and MAR are used four times, which are less common. MAP and MRR are used in only two studies, while the Hit-N metric is used only once to measure the multi-defect FL approaches.

## 11.4 Datasets

The datasets used in the DLFL field are largely similar to those in the program repair and other related fields. Current approaches primarily utilize Java programs for evaluation, with a smaller portion using C language programs.

In Java, the Defects4J [538] benchmark is the most extensively used. Defects4J is a widely recognized benchmark dataset in the FL field. This dataset is well-maintained and continues to be updated. Early FL techniques uses Defects4J v1.2, primarily evaluating against six projects within it. Recent work employs Defects4J v2.0, which includes more open-source projects. Additionally, the BEARS [539] benchmark and nanoxml from the SIR [540] dataset are also employed.

In C, some C programs from the SIR dataset, as well as the ManyBugs [541] dataset, are more frequently used. Existing DLFL research has utilized programs like space from the SIR dataset, as well as python, gzip, libtiff, and others from the ManyBugs dataset.

## 11.5 Challenges and Opportunities

This section summarizes the challenges and highlights opportunities for future work in fault localization.

### 11.5.1 Challenges

Deep learning-based fault localization techniques have shown great potential in improving the accuracy and effectiveness of fault localization. However, they also face the following challenges that need to be addressed to fully exploit their capabilities.

- **The risk of overfitting**. The current existing DLFL approaches are mainly evaluated on the popular benchmark Defects4J [538], which consists of Java projects and hundreds of bugs and is used as an important dataset in debugging-related fields. Most papers use Defects4J v1.2, which only contains 395 bugs from six Java projects. Worse still, some approaches discard the Closure project and only use 224 of the bugs. This leads to the risk that the conclusions of most existing methods might be overfitting to this particular dataset or the Java programming language. Currently, there exist a few studies that target Python or JavaScript programs and we suggest evaluating the novel approaches across multiple languages and benchmarks.

- **Inadequate availability of high-quality labeled data.** The current research is limited to the Defects4J dataset due to a shortage of high-quality labeled data. This scarcity not only affects the evaluation of DLFL approaches but also impacts the training of deep learning models. In addition, existing FL data suffer from imbalance, i.e., the data from passing tests overwhelms the data from failing tests. This characteristic poses challenges for many learning-based approaches. Deep learning models require a large amount of accurately labeled data for training, which can be difficult to obtain, especially when dealing with real-world bugs across different languages.

- **Interpretability of deep learning models.** The lack of interpretability has long been a challenge for deep learning and similarly affects fault localization based on deep learning, which makes it difficult to analyze the results of fault localization, specifically the relationship between failing tests, the code elements, and the traces.

- **Occasionally low efficiency.** The current fault localization systems struggle to achieve real-time fault localization. This is partly due to their reliance on test runs to collect trace information and partly because large-scale deep-learning models could slow training and prediction.

### 11.5.2 Opportunities

Despite these challenges, deep learning-based fault localization techniques present the following promising opportunities.

- **Data augmentation.** As discussed in current challenges, existing DLFL approaches are suffering from low-quality data that are imbalanced or of limited scale. Thus data augmentation methods offer avenues for addressing the challenge.

- **Enhancing multiple FL.** Existing DLFL approaches generally assume that there is only one bug in the target project. However, a real-world buggy project often contains multiple bugs. The interaction between these bugs makes it more difficult to train and predict using traditional learning approaches, highlighting potential opportunities for improvement in this direction.

- **The integration of domain-specific knowledge.** FL relies on code syntax structures, code semantics, and execution traces, which require a deep understanding of the underlying programming languages and software engineering principles. The integration of domain-specific knowledge, such as code smells, development experience, and debugging heuristics, can significantly improve the effectiveness of FL systems. Especially, representing and learning context information of the buggy code is promising, because it enables the models to better understand the specific scenarios in which faults occur and how they relate to the text execution.

## 12   Program Repair

Program repair refers to the process of identifying and fixing software defects in programs. Program repair requires a large number of time costs and human resources from the project development team [542]. Due to the growing demand for efficient software maintenance, automatic program repair techniques have emerged as a solution [543].

Automatic program repair allows developers to automate (or nearly automate) defect detection and correction. This makes program repair efficient, reliable, and cost-effective [544]. In recent years, the development and implementation of automatic program repair techniques have been widely recognized by the software development community [545]. The success of deep learning in recent years makes it a promising approach for locating and repairing buggy programs. The family of deep learning approaches to program repair develops and applies deep learning techniques to identify software defects and generate patches [546]. Deep learning models have been widely applied to repair a wide range of program errors. We divide existing work into three categories, including compilation error repair, runtime error repair, and specific domain error repair.

## 12.1   Compilation Error Repair

A compilation error is an error that can be detected during compilation. Programs with compilation errors fail in program compilation or linking; meanwhile, programs with compilation errors cannot be directly handled by program analysis tools [547–549].

Compilation errors prevent source code from being transformed into executable machine code during compilation. These compilation errors contain issues like type errors, undefined variables, syntax errors, and other errors that violate programming language standards. Studies in the early stage have proposed error-correcting parsers to achieve program repair of faulty code [550–553]. Recently, researchers have explored the advantage of deep learning techniques for automatically fixing compilation errors. These techniques use deep neural networks to analyze extensive repair datasets and recommend precise code fixes [545].

To avoid misleading messages returned by compilers, Gupta et al. [554] trained a deep neural network, named DeepFix, to identify incorrect locations in source code and provide the corresponding repaired statements. They collected 6,971 erroneous C programs from code written by students for 93 programming tasks and found that DeepFix is able to completely repair 27% and partially repair 19% of these erroneous programs. Bhatia et al. [555] presented RLAssist, a technique that combines recurrent neural networks (RNNs) with constraint-based reasoning to fix programming assignments with syntax errors. Ahmed et al. [556] proposed an end-to-end system, called Tracer, for fixing code with multiple errors. In the same year of 2018, Santoset al. [557] proposed an approach to correcting syntax errors using n-gram and LSTM models. They evaluated the approach on the BlackBox dataset [558].

Mesbah etal. [559] proposed Deepdelta, which converts an Abstract Syntax Tree (AST) into a domain-specific language before feeding the converted tree into a Neural Machine Translation (NMT) network. Deepdelta achieves a success rate of 50% in generating correct repairs for missing symbols and mismatched method signatures. Gupta et al. [560] proposed a programming language correction framework that uses reinforcement learning to assist novice programmers with syntactic errors. This framework outperforms their previous work DeepFix [554] in 2017. Wu et al. [561] trained a deep learning model on the DeepFix dataset. The model, called graph-based grammar fix, combines token sequences and graph structures based on ASTs to predict the error position and generate correct tokens. To integrate program-feedback graphs and a self-supervised learning framework, Yasunaga et al. [562] proposed DrRepair, a program repair technique based on graph attention networks.

With the rapid development of deep learning, the effectiveness of automatic program repair has been further improved. Hajipour et al. [563] proposed a generative model for fixing compilation errors in C programs. Their model learned a distribution over potential fixes and encourages diversity over a semantic embedding space. Allamanis et al. [372] proposed to train a selector concurrently with the model that locates and fixes errors in source code. The selector was utilized to automatically generate faulty code, which is used to enrich the training set for the original model. To make the generated faulty code closely resemble real-world error scenarios, Yasunaga et al. [564] proposed BIFI, an iterative training approach for fixing syntax errors. They trained two models, including the breaker and the fixer, in an iterative manner. The breaker creates faulty code that closely resembles real-world errors while the fixer converts the faulty code into the correct version. Ahmed et al. [565] presented a lenient parser for imperfect code (i.e., the union of fragmentary code, incomplete code, and ill-formed code) and proposed an indirectly supervised approach for training the parser. To fix the parser errors in programming languages, Sakkas et al. [566] proposed a language-agnostic neurosymbolic technique in 2022. Their technique, called Seq2Parse, combined symbolic error correcting parsers and neural networks. Seq2Parse was evaluated on 1.1 million Python programs. Li et al. [567] proposed TransRepair, which utilizes a Transformer-based neural network via considering both the context of erroneous code and the feedback by compilers to fix compilation errors in C programs. Ahmed et al. [568] introduced SynShine,a three-stage approach that combines the feedback of Java compiler with models based on a Robustly Optimized BERT (Roberta) [569]. They indicated that SynShine achieves 75% of effectiveness in fixing the single-line errors of the code in the Blackbox dataset [558].

## 12.2 Runtime Error Repair

Runtime errors, also known as dynamic errors, occur when the program executes. Runtime errors result in crashes or incorrect behaviors during program execution. Automatic program repair techniques generate patches for faulty code based on test cases, crashes, references, contracts, etc [570].

Approaches to runtime error repair can help save time costs and effort in software development. These repair approaches can be roughly divided into search-based repair [571,572], constraint-based repair [573–575], template-based repair [576–578], and learning-based repair [579–581]. With the support by deep neural networks, learning-based repair approaches can generate high-quality code patches. The framework of repair approaches based on deep learning for runtime errors generally consist of five steps: fault location, data pre-processing, feature extraction, patch generation, and patch selection.

Long et al. [582] proposed a hybrid repair system, called Prophet. Prophet integrates a probabilistic model trained on the benchmark presented by Goues et al. [583] and ranks code patches to fix runtime errors. Tufano et al. [584] investigated the possibility of learning patches through the translation model based on neural networks. Their model achieves a prediction rate of 9% via training on the GitHub repositories·4) Sun et al. [585] developed a sequence-to-sequence service based on the attention techniques. White et al. [579] presented a deep learning model, DeepRepair, to produce patches that cannot be searched by redundancy-based techniques. Tufano et al. [237] explored the potential of an NMT model to create code changes made by developers during adding pull requests.

Researchers have been exploring new ways of tackling issues in program repair. Ding et al. [586] conducted a study that investigates the differences between sequence-to-sequence models and translation models for program repair. They proposed a strategy based on the empirical findings and development knowledge in patch generation. Yanget al. [587] proposed an automatic model to locate faults and generate patches. They scored the ranks between bug reports and source code based on Convolutional Neural Networks (CNNs) and auto-encoder. Then, they created patches through the Seq-GAN algorithm [588]. Lutellier et al. [589] proposed CoCoNuT, which leverages the strength of CNNs and NMTs to achieve multi-language repair for Java, C, Python, and JavaScript programs. Li et al. [21] proposed DLFix, a dual-level deep learning model designed to address the limitations of learning-based program repair. The first layer of DLFix is a tree-based RNN model that captures the context of fixed code while the second layer utilizes the output from the first layer to learn and apply code patches. The validation experiments are conducted on the datasets of Defects4J 5) [590] and Bugs.jar 6) [591]. Tian et al. [592] explored different deep learning-based approaches. Their experimental results show that embeddings from pre-trained and re-trained neural networks are beneficial to reason and generate correct patches. Dinella et al. [593] proposed Hoppity, a Javascript-targeted automatic repair model. This learning-based model focuses on graph structures of faulty code.

Tang et al. [594] proposed a grammar-based approach to syntax correct patch generation. Huang et al. [595] discussed the use of a pyramid encoder in seq2seq models to reduce computational and memory costs while achieving a similar repair rate to their non-pyramid counterparts. Their study focuses on automatic correction of logic errors. Jiang et al. [581] presented a three-stage based NMT model, called CURE. They first pre-trained a programming language model to incorporate the real-world coding style and then proposed a technique to expand the search space. They integrated subword tokenization, a technique used in natural language processing that splits words into smaller units called subword tokens, to generate precise patches. Based on the Long Short-Term Memory (LSTM) network and the bidirectional recurrent neural network, Rahman et al. [596] presented BiLSTM to identify and classify the faulty code and generate possible fixes. Chen et al. [580] proposed SequenceR, a sequence-to-sequence-based deep learning system. The model adapts a copy mechanism to deal with large-scale code instances. Berabi et al. [597] proposed TFix, a pre-trained and fine-tuned language model that generates improved patches for JavaScript programs. Tang et al. [598] proposed a Graph-to-Sequence learning model called GrasP, based on code structure. Szalontai et al. [599] focused on the uncommon parts of Python code. They constructed a neural network model to classify and generate replacement code snippets.

---

4) https://github.com/.

5) https://github.com/rjust/defects4j.

6) https://github.com/bugs-dot-jar/bugs-dot-jar.

Li etal. [600] proposed DEAR, a deep learning-based repair approach, which attempts to locate multi-bugs and generate patches to fix multiple errors in a single program. Xu et al. [601] presented M3V, an approach that integrates LSTM and GNN models for fault location and patch generation. Their approach primarily focuses on null-pointer exceptions and out-of-bounds exceptions in Java programs. Meng et al. [602] introduced Transfer, a technique that mines deep semantic information. Their model integrates multiple features to rank patches, including semantic-based features, spectrum-based features, and mutation-based features. Kim etal. [603] focused on locations of software defects. They improved the accuracy of deep learning approaches in program repair via using genetic algorithms to obtain the precise defective code. Wardat et al. [604] proposed DeepDiagnosis, which focuses on defects in deep learning applications. Using well trained models, DeepDiagnosis classifies faults into eight categories and provides feasible patches for each category of faults. Yao et al. [605] proposed Bug-Transformer, a context-based deep learning model. Bug-Transformer retains the contextual information of the faulty code and uses a transformer model for training. Yanet al. [606] proposed CREX, a transfer-learning-based technique [607] for validating C program patch correctness. Their model aims to learn the code semantic similarity to improve the accuracy validation. Chakraborty et al. [608] proposed CODIT, a tree-based deep learning model for generating code change suggestions. Ye et al. [609] proposed RewardRepair, an NMT-based model with fine-tuned loss functions. RewardRepair incorporates compilation information and test cases into the calculation of the loss functions. Ye et al. [610] also proposed ODS, a deep learning system for predicting patch correctness. They extracted code features in ASTs from patches and buggy code. Then, they employed supervised learning to determine the correctness of patches. Another technique proposed by Yeetal. [611] is SelfAPR, a self-supervised model based on employing perturbation-generated data for patch generation. Xia et al. [612] proposed AlphaRepair, a CodeBERT based model for code generation. AlphaRepair uses zero-shot learning techniques rather than training models with historical erroneous and corrected code. Kim etal. [613] investigated the efficacy of deep learning-based repair techniques for Java-to-Kotlin conversion programs. Their technique enhances the defect fixing performance by applying transfer learning techniques. Tian et al. [614] proposed BATS, a learning-based model designed to predict the correctness of patches. BATS is an unsupervised model that repairs faulty programs by detecting program behavior during failed testcases. Yuan et al. [615] proposed CIRCLE, a T5-based model for patch generation. CIRCLE employs continual learning techniques and is able to work on multiple programming languages. They also designed the Prompt feature to make it capable of understanding natural language commands. Chen et al. [616] trained an iterative model, which aims to learn from generated patches and test execution.

## 12.3 Specific Domain Error Repair

Deep learning has also been applied to domain-specific tasks related to automated program repair. The specific domain repair refers to the application of specialized knowledge or techniques from a specific domain to improve the effectiveness and efficiency of program repair.

Test repair aims at fixing errors in test cases that are unsable due to software updates. Stocco et al. [617] fixed web test cases by analyzing visual features based on an image processing approach. Panet al. [618] presented Meter, a computer vision based technique for fixing test cases in the Graphical User Interface (GUI) of mobile applications.

Build scripts are crucial components in the automatic building of software systems. Program repair for build scripts refers to the process of automatically detecting and fixing faults in build scripts [619]. Hassan et al. [620] built a benchmark of 175 build failures and the relevant remedies for Gradle·[7] They presented HireBuild as the first model of patch generation for fixing build scripts in Gradle. Based on the previous work of HireBuild, Lou et al. [621] extended the benchmark from Top-1000 GitHub projects and proposed an improved search model, which generates patches according to current test projects and external resources. Recently, Loriot et al. [622] proposed STYLER, an approach based on the LSTM neural network to generate patches for code violations against format rules.

Software vulnerability refers to security weaknesses that can compromise the integrity and availability of software systems.Ma et al. [623] developed a tool called VuRLE that automatically locates and fixes vulnerabilities in source code. Their tool generates repair templates and selectes patches based on the ASTs of source code. Harer et al. [624] proposed a technique based on Generative Adversarial Networks (GANs) that generates corrupted data and uses correct-incorrect pairs to train an NMT model. Zhou et

---

**Table 16** Evaluation Datasets for Deep Learning Based Program Repair Tools

| Year | Dataset | Language | Type | Size | URL |
|------|---------|----------|------|------|-----|
| 2014 | Defect4J | Java | Runtime Error | 835 | https://github.com/rjust/defecrrorts4j |
| 2014 | Blackbox | Java | Hybrid Error | Over 2,000,000,000 | http://www.cs.kent.ac.uk/~nccb/blackbox |
| 2016 | IntroClassJava | Java | Runtime Error | 297 | https://github.com/Spirals-Team/IntroClassJava |
| 2017 | DroixBench | Java | Hybrid Error | 24 | https://droix2017.github.io |
| 2018 | Bugs.jar | Java | Runtime Error | 1,158 | https://github.com/bugs-dot-jar/bugs-dot-jar |
| 2018 | Santos et al. | Java | Compilation Error | 1,715,313 | https://archive.org/details/sensibility-saner2018 |
| 2019 | BugSwarm | Java | Build Failure | 3,091 | https://github.com/BugSwarm/bugswarm |
| 2019 | Ponta | Java | Vulnerability | 1,068 | https://github.com/SAP/project-kb |
| 2022 | Vul4J | Java | Vulnerability | 79 | https://github.com/tuhh-softsec/vul4j |
| 2015 | ManyBugs | C | Hybrid Error | 185 | https://repairbenchmarks.cs.umass.edu |
| 2015 | IntroClass | C | Runtime Error | 998 | https://repairbenchmarks.cs.umass.edu |
| 2016 | Prutor | C | Compilation Error | 6,971 | https://www.cse.iitk.ac.in/users/karkare/prutor |
| 2017 | DBGBENCH | C | Runtime Error | 27 | https://dbgbench.github.io |
| 2017 | CodeFlaws | C | Hybrid Error | 3,902 | https://codeflaws.github.io |
| 2018 | TRACER | C | Compilation Error | 16,985 | https://github.com/umairzahmed/tracer |
| 2019 | TEGCER | C | Compilation Error | 15,579 | https://github.com/umairzahmed/tegcer |
| 2020 | Big-Vul | C/C++ | Vulnerability | 3,754 | https://github.com/ZeoVan/MSR-20-Code-vulnerability-CSV-Dataset |
| 2021 | CVEfixes | C/C++ | Vulnerability | 5,495 | https://github.com/secureIT-project/CVEfixes |
| 2021 | BugsCpp | C/C++ | Runtime Error | 215 | https://github.com/Suresoft-GLaDOS/bugscp |
| 2017 | QuixBugs | Java/Python | Runtime Error | 40 | https://github.com/jkoppel/Quixbugs |
| 2017 | HireBuild | Build Script | Build Failure | 175 | https://sites.google.com/site/buildfix2017 |
| 2019 | BugsJS | JavaScript | Runtime Error | 453 | https://bugsjs.github.io |
| 2019 | Defexts | Kotlin/Groovy/etc. | Runtime Error | 654 | https://github.com/ProdigyXable/defexts |
| 2019 | Refactory | Python | Hybrid Error | 1783 | https://github.com/githubhuyang/refactory |
| 2020 | TANDEM | Java/C/SQL/etc. | Hybrid Error | 125 | https://github.com/belene/tandem |
| 2022 | Ring | 6 Languages | Hybrid Error | 1,200 | https://github.com/microsoft/prose-benchmarks |
| 2022 | CrossVul | 40 Languages | Vulnerability | 5,131 | https://zenodo.org/record/4734050 |

al. [625] proposed SPVF, which combines ASTs, security properties, and the attention mechanism into an integrated neural network for both C/C++ and Python programs. Huang et al. [626] attempted to fix vulnerabilities by leveraging pre-trained large language models. They reported an accuracy rate of 95.47% for single-line errors and 90.06% for multiple-line errors. Chen et al. [627] hypothesized that there could be a correlation between program repairing and vulnerability fixing. They proposed VRepair, a transfer learning model designed to solve security vulnerabilities in C programs with limited data. Due to the increasing number of reported vulnerabilities, Chi et al. [628] developed SeqTrans, an NMT-based tool that automatically fixed vulnerabilities. Their approach involves learning from historical patches and contextual features of source code.

### 12.4 Datasets

The diverse range of programming languages leads to the creation of various types of datasets for program repair. In light of existing research, there are numerous available datasets that are specifically tailored for the application of automatic program repair tools.

The evaluation datasets are listed in Table 16. We divide the datasets into three categories according to the programming languages: in Java, in C/C++, and in other programming languages. We briefly introduce typical datasets as follows. Prutor [629] is a tutorial system, which helps students solve programming problems. Prutor is used in the introductory programming course in IIT Kanpur. Thus, Prutor collects many pieces of C code, including the buggy code and the correct code.

Blackbox [558] is a project since 2013. It collects data from the BlueJ IDE,[8] a tutorial environment for JAVA learners. After five years of collection, the Blackbox dataset has amassed over two terabytes of data [630]. In 2018, Santos et al. [557] refined the Blackbox dataset to evaluate their Sensibility approach. They selected 1,715,312 program pairs of previous and current versions from the Blackbox dataset, including 57.39% pairs with one syntax error and 14.48% with two syntax errors.

Defects4J [631] is one of the most widely-used benchmark in program repair [589,609,632,633]. The latest version of Defects4J is a collection of 835 reproducible bugs from 17 open-source Java projects. Each bug in Defects4J corresponds to a set of test cases that can trigger the bugs. GrowingBugs [634] is

8) BlueJ IDE, https://www.bluej.org/

highly similar to Defects4J in that bug-irrelevant changes in bug-fixing commits have been excluded from the patches. The current version of GrowingBugs contains 1,008 real-world bugs collected from open-source applications. The only difference between GrowingBugs and Defects4J is that the latter excludes bug-irrelevant changes from bug-fixing commits manually, whereas the former does it automatically by BugBuilder [635,636].

Vul4J [637] is a dataset of reproducible Java vulnerabilities. All Vulnerabilities in Vul4J correspond to human patches and Proof-of-Vulnerability (POV) test cases. CrossVul [638] contains vulnerabilities over 40 programming languages. Each file is corresponding to an ID of Common Vulnerability Exposures (CVEs) and its source repository. This dataset also contains commit messages, which may serve as human-written patches.

## 12.5 Challenges and Opportunities

We present challenges and opportunities in deep learning for program repair in this section.

### 12.5.1 Challenges

There are several challenges in deep learning for program repair. These challenges reveal that there is still a long way to go in applying deep learning techniques to automatic program repair.

**Training data**. Deep learning requires a considerable number of data to train learnable models. In program repair, labeled data of buggy code and patches are limited. Obtaining high-quality erroneous and patched code is a severe challenge due to the limited datasets of program repair. Another challenge is to determine how to use buggy code and patches in model training. A deep learning model can involve the location of buggy code, its specific type, its context, its syntax, and semantics. The mapping between buggy code and patched code is a key step in training models in program repair. To date, there is no theoretical analysis for such a challenge.

**Model interpretability**. The lack of interpretability for deep learning models makes it difficult to ensure the correctness of the generated patches. The original goal of program repair is to assist the real-world developers. Thus, it may be difficult to persuade developers to use deep learning-based program repair in real-world development.

**Self-validation of data**. Models of deep learning are built on training data. If the training data contains errors, these errors may propagate to the generated patches. This hurts the effectiveness of automatic program repair. Developers are unable to confirm the reliability of patches generated by deep learning.

### 12.5.2 Opportunities

Despite the challenges, we identify the following opportunities in the field of deep learning for program repair.

**Data collection and generation**. Automatic data collection and generation can benefit program repair. Researchers can train deep learning models to handle a wide range of programming scenarios based on high-quality data. Additionally, data generation methods such as program synthesis may be able to enrich the existing data.

**Hybrid approaches**. Hybrid approaches that combine deep learning with other existing techniques, like symbolic reasoning, template-based repair, and rule-based search, have shown promising results in program repair. These hybrid approaches can improve the effectiveness and efficiency of current program repair tools by integrating the strength of multiple techniques.

**Model optimization**. Researchers are able to explore ways to optimize deep learning models for program repair. Such optimization contains model re-sizing knowledge distillation, neural architecture search, and so on. Model optimization can help researchers transfer available knowledge from a large and complex model to new-coming or unknown scenarios.

In conclusion, the opportunities in deep learning for program repair are vast and exciting. Continuous research and development in this field may lead to more advanced and effective approaches for program repair.

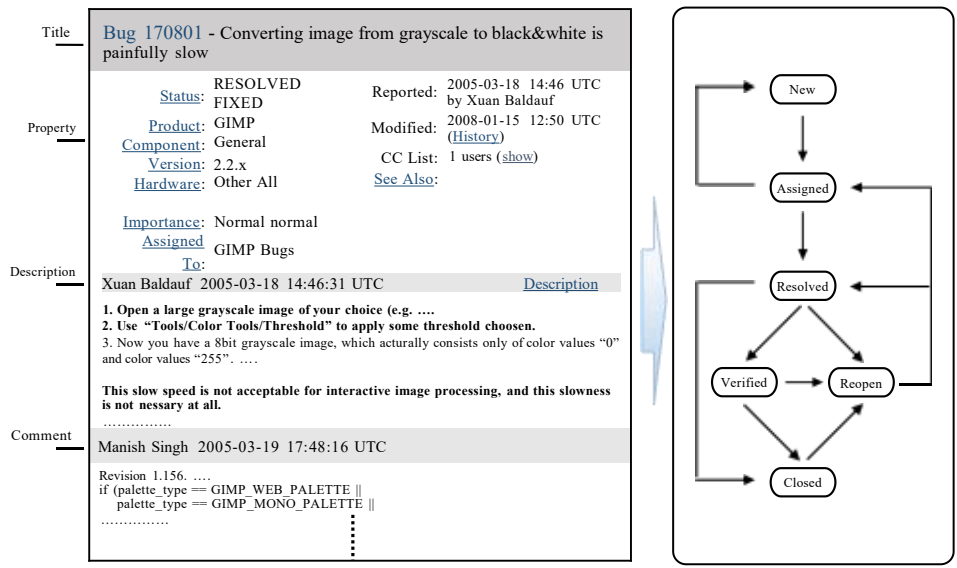

**Figure 6** An example of bug report and its lifecycle.

## 13 Bug Report Management

Given the intricate nature of software systems, bugs are unavoidable. A previous study shows that a collection of 606 software failures reported in 2017 has affected approximately 3.7 billion users and caused financial losses of $1.7 trillion. Therefore, efficiently fixing bugs becomes a fundamental step for software projects [639].

To fix bugs, most software projects use bug reports to manage bugs. A bug report is used to record specific details of the bug such as the title, the descriptions (e.g., reproducible steps and stack traces), the property fields, and the comments, which assist developers in identifying and rectifying buggy code [640]. Figure 6 is an example of a bug report from the Gnome project with bug report ID 170801[9]. In this example, Xuan Baldauf issued a bug report titled "converting image from grayscale to black&white is painfully slow", and provided detailed information on how to reproduce the bug in the description field (e.g., "Open a large … at all"). Additionally, the reporter specified the property of the bug, such as the product and component containing the bug. After the submission, other participants contributed comments in the comment field. The bug report and its associated comments are typically utilized by software developers to facilitate subsequent software activities.

The general activities to manage bug reports for bug fixing involve six steps [641] (as shown in Figure 6). Upon the submission of anew bug report, its initial status is labeled as new. Subsequently, the bug report undergoes manual scrutiny to verify its validity and prevent duplication. Once the bug is confirmed, the bug report is assigned to a developer. The developer then proceeds to fix the bug and sets its status to resolved after fixing the bug. For resolved bugs, additional developers conduct code review to verify the bug resolution. If the bug resolution is verified, developers can close this bug; otherwise, they label the bug report as reopen, which means that the bug is not successfully fixed. In a typical bug report management (BRM) cycle, these processes are mainly carried out manually by software developers.

However, with the exponential growth in the number of bug reports submitted to many large-scale software projects, the size and complexity of bug report repositories increase significantly. It becomes a laborious task to manually manage bug reports, which consumes a significant amount of time for developers [642]. For instance, the Eclipse project received a total of 552,334 bug reports in recent 20 years (from January 2003 to December 2022), averaging around 76 new bug reports daily. Additionally, given the varied reporting experience of bug reporters [640], not all bug reports provide sufficient information to assist developers with bug fixing. Developers have to spend tremendous time understanding and managing these bug reports.

To tackle this challenge, many studies propose to utilize text mining and machine learning (ML) [639, 643,644] to automate BRM. These studies employ classical ML techniques such as Naïve Bayes, Random

**Table 17** BRM tasks addressed by deep learning techniques.

| # | Tasks | Description | Total |
|---|---|---|---|
| 1 | Bug report refinement | Generate a high-quality bug report by enriching/modifying an existing one. | 1 |
| 2 | Duplicate bug detection | Detect duplicate or similar bug reports in bug repositories. | 8 |
| 3 | Bug assignment | Recommend the most appropriate developers to fix the bug. | 5 |
| 4 | Bug severity/priority prediction | Predict the severity/priority of a bug report before fixing. | 2 |
| 5 | Bug fixing time prediction | Predict how long it will take to fix the bug. | 1 |
| 6 | Bug report summarization | Summarize a bug report into a much shorter form. | 6 |
| 7 | Bug localization | Locate relevant source code files or methods that possibly contain the bug based on the bug report. | 16 |
| 8 | Bug-Commit linking | Link bug reports with bug fixing commits or bug inducing commits. | 3 |

Forest, Support Vector Machine, and k-nearest neighbors to automatically detect duplicate bug reports, triage bug reports, and identify reopened bug reports.

However, the effectiveness of these classical ML techniques is limited [527,645]. Therefore, recent studies have utilized deep learning (DL) techniques to enhance the automation of BRM. DL uses its powerful feature engineering capability to deeply analyze bug reports. The exponential growth number of bug reports also becomes an important source to effectively train DL models for different classification and regression tasks. For example, Li et al. [646] introduced an autoencoder architecture that extracts multiple features from bug reports to generate bug report summary. Fang et al. [647] proposed the use of Recurrent Neural Network (RNN) to capture sequential information in bug reports and source code for bug localization. These studies provide compelling evidence of the effectiveness of DL in improving automated BRM. In this section, we survey the BRM tasks improved by DL techniques and discuss the challenges and opportunities in this area.

Table 17 summarizes the main BRM tasks addressed by DL techniques, including the task name, the task description, and the number of related papers for each task. In total, researchers have applied DL techniques for eight tasks in BRM.

## 13.1 Bug report refinement

When a bug report is submitted, bug report refinement aims to refine the bug report and improve its quality for better understanding. For example, we can enrich the bug report with more information collected from similar bug reports or reformulate the content of the bug report. DL can be used to improve the refinement process. Zhou et al. [648] reformulated a bug report as a query representation by leveraging multi-level embeddings through Convolutional Neural Networks (CNNs) with the self-attention mechanism. The reformulation is used to help developers understand the bug report and retrieve similar bug reports in the repository.

## 13.2 Duplicate bug detection

Duplicate bug report detection aims to identify whether a given bug report is a duplicate of an existing one. The detection not only eliminates redundant data but also provides additional insights into the reported issues. To automate this task, different types of deep neural networks have been used to train duplicate bug report detection models, such as simple DNN, CNN, RNN, and LSTM [649–651]. For example, Deshmukh et al. [652] used CNN and LSTM to retrieve duplicate bug reports, achieving a precision value of 90% and a recall value of 80%.

The main goal of these DL techniques is to represent the content of bug reports for distinguishing them. Evaluation demonstrates that DL techniques such as DNN-based 2D embedding and BERT show higher performance in detecting duplicate bug reports than traditional language models (e.g., n-gram based model) [653–655]. Despite the promising performance, a recent study [645] analyzed the effectiveness of state-of-the-art duplicate bug report detection approaches in a realistic setting using industrial datasets. The results of the study shows that a simple technique already adopted in practice can achieve comparable results as a recently proposed research tool.

## 13.3 Bug assignment

A submitted bug report should be assigned to a developer for resolution. This assignment process is known as bug assignment (or triaging), which can be automated to reduce the manual effort. An existing empirical study shows that DL techniques outperform the other traditional ML techniques for this task [656]. For bug assignment, DL models are employed to extract word sequences, semantic and syntactic features from bug report textual contents [657]. The features contain discriminative information that indicates who should fix the bug.

Lee et al. [658] proposed an automatic bug triager using CNN and word embedding. The results demonstrate benefits in both industrial and open source projects. Mani et al. [659] proposed a novel bug report representation algorithm using an attention-based deep bidirectional recurrent neural network (DBRNN-A), which addresses the problem that existing bag-of-words models fail to consider syntactical and sequential word information in unstructured texts. The empirical results show that DBRNN-A provides higher rank-10 average accuracy. Liu et al. [660] proposed BT-RL model, which uses deep reinforcement learning for bug triaging. It utilizes deep multi-semantic feature fusion for high-quality feature representation, and an online dynamic matching model employing reinforcement learning to recommend developers for bug reports.

## 13.4　Bug severity/priority prediction

The severity/priority of bug reports indicates the importance of fixing a bug, which is often manually decided by developers. Bug severity/priority prediction aims to automatically assign the severity/priority property of a bug report based on the knowledge of historical bug reports. Han et al. [661] used word embeddings and a one-layer shallow CNN to automatically capture discriminative word and sentence features in bug report descriptions for predicting the severity. Gomes et al [662] conducted a survey on bug severity/priority prediction, which also confirms the effectiveness of DL techniques on this task.

## 13.5　Bug fixing time prediction

This task predicts the time to be taken to fix a bug, which helps software team better allocate the work of developers. Previously, many popular ML methods such as KNN and Decision trees have been used to predict the fixing time of bug reports. In recent years, DL has also been employed. For example, Noyori et al. [663] adopted a CNN and gradient-based visualization approach for extracting bug report features related to bug fixing time from comments of bug reports. These features can significantly improve the effectiveness of bug fixing time prediction models.

## 13.6　Bug report summarization

The common practice for software developers to fix newly reported bugs is to read the bug report and similar historical bug reports. Statistics show that nearly 600 sentences have to be read on average if a developer refers to only 10 historical bug reports during bug fixing [646]. Bug report summarization automatically identifies important sentences in a bug report or directly generates a short, high-level summary of the bug report [664]. Regarding DL techniques, Huang et al. [28] proposed a CNN-based approach to analyze the intention of each sentence in bug report for the ease of reading the content. Li et al. [646] and Liu et al. [665] proposed auto-encoder networks with different structures for informative sentence extraction in bug reports. In recent studies, DL based language models are also employed to automatically generate titles of bug reports [666,667].

## 13.7　Bug localization

Bug localization is the main BRM task facilitated by DL techniques, where 16 related papers are found. For this task, DL techniques are used to bridge the gap between the natural language in bug reports and the source code [668,669]. For example, DL networks such as CNNs can be used to extract semantic correlations between bug reports and source files. Then, these features are merged and passed to a classification layer to compute the relevancy scores between bug reports and source files [670]. The relevancy scores can be combined with the scores computed from other ML or information retrieval techniques [671–673] (such as learning to rank [674]) to better associate source code with a bug report.

In recent years, advanced DL techniques such as BERT [527], adversarial transfer learning [675], and attention mechanism [676] have been employed to improve the localization effectiveness. Using these powerful DL techniques, a bug report can be located to the source code on different levels (e.g., file level [677], component level [678], and method level [526]). These techniques also enable both within-project bug localization and cross-project bug localization [675,679].

**Table 18**  Datasets used for BRM tasks.

| # | Tasks | <1000 | 1000–10000 | 10001–30000 | 30001-100000 | >100000 |
|---|---|---|---|---|---|---|
| 1 | Bug report refinement | * | * | * | * | 1 |
| 2 | Duplicate bug detection | * | * | * | * | 5 |
| 3 | Bug assignment | * | * | * | 1 | 2 |
| 4 | Bug severity/priority prediction | * | * | * | 1 | * |
| 5 | Bug fixing time prediction | * | * | * | 1 | * |
| 6 | Bug report summarization | 2 | * | * | 1 | 1 |
| 7 | Bug localization | 1 | 3 | 10 | * | 1 |
| 8 | Bug-Commit linking | 1 | * | 2 | * | * |

## 13.8  Bug-Commit linking

Bug-commit linking associates bug reports with bug fixing commits or bug inducing commits. Consequently, developers can better understand which commit fixes the bug and why/how/when the bug is introduced [639]. A variant of this task is to link bug reports with reviews or comments in software engineering forums (e.g., APP reviews) [680]. For this task, DL techniques (e.g., word embedding and RNN) learn the semantic representation of natural language descriptions and code in bug reports and commits, as well as the semantic correlation between bug reports and commits [681,682].

## 13.9  Datasets

The type of datasets used for BRM tasks is determined by the nature of each BRM task. For the majority of BRM tasks, the main datasets are the bug reports (also called issue reports). Such BRM tasks include bug report refinement, duplicate bug detection, bug assignment, bug severity/priority prediction, bug fixing time prediction, and bug report summarization. Regarding bug localization and bug-commit linking, datasets require bug reports, source code, and commit messages for analysis. Existing studies evaluate their DL techniques using datasets from different software repositories, such as Open Office, Eclipse, Mozilla, and Net Beans [652,653,683,684]. Since many software projects nowadays are managed by distributed collaboration platforms such as GitHub, bug reports from GitHub are also important dataset sources for BRM tasks, such as TensorFlow, Docker, Bootstrap, and VS Code [28].

Table 18 shows the sizes of datasets used for BRM tasks. The number in each cell represents the number of datasets with the certain size used for a BRM task. All BRM tasks have datasets with more than 10,000 BRM-related items (e.g., bug reports). For five out of eight BRM tasks, they have constructed large datasets with more than 100,000 BRM-related items. The large datasets for BRM tasks not only improve the reliability of the evaluation, but also facilitate the training of DL techniques.

## 13.10  Challenges and opportunity

Based on the preceding analysis of deep learning-based BRM, we present the challenges and opportunities for future research.

### 13.10.1  Challenges

- Computation cost. Computation cost of deep learning is one of the major concerns in BRM. Deep learning works well for many tasks but generally costs a large amount of computation resources. Sometimes the training time lasts for hundreds of hours of CPU/GPU time, especially for complex network architectures. The less computation cost is important for using DL in real BRM scenario in industry, because the long computation time leads to power consumption and heat dissipation issues, which increase the total financial cost of software companies [685].

- Training datasets. The size of a training set is important for DNNs. Biswas et al. [686] found that there are sometimes no performance increases for domain-specific training of DL, due to the small-scale training set. Nizamani et al. [687] also observed a trend of performance improvement for deep learning when the training set size is increased. For deep learning applications in BRM, small training sets often lead performance decline and over-fitting.

- Interpretability. The interpretability is important in BRM. Results provided by machine learning algorithms are sometimes difficult to be understood. DL techniques are even worse due to the complex network architectures. Therefor, more interpretable DL techniques are important to help developers get trustworthy prediction results.

### 13.10.2   Opportunities

- DL acceleration in the BRM context. SE studies accelerate the network training with optimization strategies in AI, e.g., batch gradient descent and RMSprop. For a given BRM task, some studies select faster neural networks as substitutes for the slow ones. For example, CBOW is a faster model than Skip-gram in Word2Vec. Li et al. [688] compared the two models and used CBOW for their task, as the two models achieve similar performance. Meanwhile, existing studies also reduce the computation cost by using the distributed model training [689] and dynamic GPU memory manager [690].

- BRM data enrichment. The challenge of training datasets can be alleviated as the exponential growth in the number of bug reports. BRM studies also adopt DL techniques such as fine-tuning [654] and transfer learning [675] to address this problem. In addition, we can automatically generate (relatively low-quality) artificial data to enlarge the training set. Typically, data generation is treated in a case-by-case manner. For instance, Morgan et al. [691] used APP screenshots and the labeled GUI-component names in screenshots to train DNNs for GUI-component classification. They synthesized APP screenshots by placing GUI-components of specified types on a single screen with randomized sizes and values. They also performed color perturbation on the images to further enlarge the training set.

- Interpretable DL techniques for BRM tasks. Existing studies try to interpret the prediction results of DL by visualization, including t-SNE and heat map visualizations. The t-SNE (t-distributed Stochastic Neighbor Embedding) technique projects high-dimensional vectors into two-dimensional spaces. It is useful to understand the embedding (vectors) generated by DNNs. With t-SNE, we can understand the semantically related APIs and SE terms [692] calculated by deep learning. A heat map is usually used to visualize network parameters. By visualizing parameters of a layer, we can understand which part of information on which the network focuses more. A heat map assumes the more important (in deeper color) a region is, the more weight the network assigns to features in that area. SE studies use heat maps to visualize the important part in SE images for classification and the attention of RNN [693].

## 14   Developer Collaboration

Software development usually relies on highly collaborative efforts among developers and is widely known as a type of social-technical activity. Hence effective collaboration among software developers is one of the most important factors that greatly benefit productive software development. Brooks highlights the collaboration cost in software development in his famous book, The Mythical Man-Month [694]. On one hand, for large-scale software projects inside an organization, hundreds or thousands of developers could be involved, in which developers may not know each other well, and it is a great challenge to establish effective collaboration among developers. On the other hand, open source and crowdsourcing projects are becoming increasingly prevalent software development paradigms, and millions of developers are loosely organized in Internet-based development platforms, where both the development tasks and developers are characterized by variety and scale, and how to support collaborative development in such open environments is another big challenge. While developer collaboration may involve many aspects, development tasks are the pivotal points. Thus, the core issues for developer collaboration include: (1) understanding developers, and (2) assigning a task to one or multiple proper developers. In this regard, thanks to the availability of the large volume of software development data, machine learning (especially deep learning) has been employed to analyze the data and provide intelligent support for facilitating developer collaboration. In particular, we survey developer collaboration from the following three angles: developer expertise profiling, intelligent task assignment, and development team forming.

## 14.1 Developer Expertise Profiling

Developer expertise profiling mainly aims at realizing certain forms of representation of the skills that a developer has mastered, which is the basis for collaborative development. Profiling developer expertise has received much attention in the software engineering community [695–699]. The typical approach is to mine developers' past experience data to measure their expertise with machine learning and data mining techniques.

We first review the research efforts with traditional machine learning techniques. For instance, [695] presents an approach, Expertise Browser, to measure expertise of developers with the data in change management systems, and [698] studies how developers learn their expertise by quantifying the development fluency. [699–701] evaluate developers by graph-based algorithms. [702] present a conceptual theory of software development expertise for programming mainly by a large-scale survey of real-world software developers. [703] evaluate developers' contributions by development values consisting of the effect of code reuse and the impact on development. Expertise profiling is also used for modeling developers in open source software community. For example, Venkataramani al. [704] captured the expertise of developers by mining their activities from the open source code repositories. Saxena et al. [705] annotated GitHub code with tags in Stack Overflow and then created a detailed technology skill profile of a developer based on code repository contributions of the developer. Considering single-community data could be insufficient for accurately characterizing developers, several techniques have been proposed to connect users in different software communities [706–712]. In [709], the authors conducted an empirical study of user interests across GitHub and Stack Overflow and they found that developers do share common interests in the two communities. Huang et al. [710] proposed CPDScorer to model the programming ability of developers across CQA sites and OSS communities. They first analyzed the answers posted in CQA and the projects submitted in OSS to score developer expertise in the two communities, respectively. They then computed the final expertise by summing up the two scores. However, they did not consider the interactions among developers, which have implications for evaluating the expertise of developers. Furthermore, most approaches to developer expertise profiling ignore the fact that developer expertise evolves over time due to learning or forgetting. To fill this gap, Yan et al. [711] and Song et al. [713] proposed heterogeneous information network-based approaches to profiling developer expertise with GitHub and Stack Overflow data, where there are four types of nodes including developers, skills, questions, and projects, and nine types of edges in the network. As a result, the problem of profiling developers is formulated as estimating the distance of developer nodes and skill nodes. That work combines the historical contributions of developers, the dynamics of expertise due to forgetting, and collaborations among developers, which is particularly featured by incorporating the collaboration relationships into the estimation of developers' expertise. Montandon et al. [712] employed data from social coding platforms (i.e., Stack Overflow and GitHub), built three different machine-learning models to identify the technical roles of open source developers such as backend, frontend, full-stack, etc, and they simply leveraged the data from Stack Overflow to build a ground truth for evaluating the performance of their approach.

Apart from traditional machine learning, deep learning techniques are also introduced in profiling developer expertise as vectors. The vectors can be utilized for predicting or recommending downstream tasks to developers. Dey et al. [714] proposed a deep neural network based approach to generating vectorized representation of software developers based on the APIs they have used. Specifically, the authors collected the open source data with WoC (World of Code) [715], extracted the mappings of projects, developers and the APIs, and then used Doc2Vec [236] to obtain the vectors representing APIs, developers and projects. Similarly, Dakhel et al. [716] also used Doc2Vec to generate vector embeddings of developers, but they incorporated more data including the textual data of repositories and issues beyond APIs. Javeedetal. [717] proposed to use LSTM and convolutional neural networks to train deep learning models to classify whether a developer is an expert or a novice according to six attributes of source code, including security, reliability, complexity, lines of code, maintainability and duplication.

## 14.2 Intelligent Task Assignment

Given a software development task, one important issue is how to find appropriate developers to handle the task, which is also known as development task assignment. In the following, we summarize how machine learning and deep learning are used to support development task assignments in various scenarios including crowdsourcing software development, open source development, bug triage etc.

## 14.2.1  Crowdsourcing developer recommendation

Crowdsourcing software development usually adopts the open-call mode to solicit workers (i.e., developers) for various tasks published online by requesters. Developers hope to find appropriate tasks by considering the task factors such as reward, difficulty, and needed efforts while requesters need to find capable developers to complete the tasks with guaranteed quality. As a result, recommending suitable developers to a task, or vice versa, becomes an important research topic in crowdsourcing software development.

As prior studies often overlook the skill improvement of developers over time, Wang et al. [718] proposed a new technique for crowdsourcing developer recommendations. After conducting an empirical study of 74 developers on Topcoder and re-calculating developers' scores, they found that skill improvement of developers fits well with the negative exponential learning curve. Based on the learning curve, a skill prediction technique is designed and a skill improvement-aware framework for recommending developers is proposed. Zhang et al. [719] proposed a meta-learning-based policy model to address the challenge of identifying developers who are most likely to win a given task in crowdsourcing software development. This model firstly filters out developers unlikely to participate or submit to a given task, then recommends the top k developers with the highest possibility of winning. Yu et al. [720] proposed a new deep model, which is a cross-domain developer recommendation algorithm using feature matching based on collaborative filtering, for T-shaped expert finding. Their recommendation model leverages the data from software crowdsourcing platforms (i.e., ZhuBaJie and Joint Force), solves the problem of data sparsity, and finally improves the recommendation performance to some extent. Wang et al. [721] presented PTRec, a context-aware task recommendation technique, capturing in-process progress-oriented information and crowdworkers' traits through a testing context model. Using random forest, it dynamically recommends tasks aligned with worker skills and interests. The evaluation shows its excellence in precision and recall, saving efforts and enhancing bug detection. Furthermore, Wang et al. [722] presented iRec2.0, which integrates dynamic testing context modeling, learning-based ranking, and multi-objective optimization for crowdworker recommendations in crowdtesting. It aims to detect bugs earlier, shorten non-yielding windows, and alleviate recommendation unfairness, demonstrating the potential to improve cost-effectiveness.

## 14.2.2  Reviewer recommendation

Code review is one of the important tasks for ensuring code quality, which relies on professional developers, known as reviewers, to identify defects by reading source code. Thus finding appropriate reviewers is a core issue for achieving effective code reviews. To address this issue, researchers have conducted extensive studies on recommending reviewers for code review tasks, especially in the context of open source development.

Ying et al. [723] proposed a reviewer recommendation approach (EARec) for a given pull request, considering developer expertise and authority simultaneously. Jiang et al. [724] provided an approach to recommending developers to comment a PullRequest in social-coding platforms like GitHub. Zhang et al. [725] presented CORAL, a reviewer recommendation technique using a socio-technical graph and a graph convolutional neural network. Trained on 332 Microsoft repositories, CORAL identifies qualified reviewers missed by traditional systems, excelling in large projects, while traditional systems perform better in smaller ones. Rebai et al. [726] framed code reviewer recommendation as a multi-objective search problem, balancing expertise, availability, and collaboration history. Validation on 9 open-source projects confirms its superiority over existing approaches. Zanjan et al. [727] introduced cHRev, an approach for automatic reviewer recommendation based on historical contributions of reviewers in their previous reviews. Due to leveraging specific previous review information, cHRev outperforms existing approaches on three open-source systems and a Microsoft code base. Hannebauer et al. [728] empirically compared six modification expertise-based algorithms and two reviews expertise-based algorithms on four FLOSS projects. The study concludes that review expertise-based algorithms, particularly the Weighted Review Count (WRC), are more effective. Rong et al. [729] introduced HGRec, a recommendation system utilizing hypergraph techniques to model complex relationships involving multiple reviewers per pull-request in code review. Evaluated on 12 OSS projects, HGRec demonstrates superior accuracy, emphasizing the potential of hypergraphs in this field. Kovalenko et al. [730] evaluated a reviewer recommendation system in a company setting, covering over 21,000 code reviews. Despite the relevance of recommendations, it identifies no evidence of influence on user choices, highlighting the need for more user-centric design and evaluation in reviewer recommendation tools. Ahasanuzzaman et al. [731] proposed KUREC, a code reviewer recommender utilizing Java programming language knowledge units (KUs) to generate developer expertise profiles and select reviewers. Evaluated against baselines, KUREC is found to be equally effective but more stable. Besides, combining KUREC with baselines further enhances performance. Gon¸calves et al. [732] identified 27 competencies vital for code review through expert validation and ranked them using a survey of 105 reviewers. The findings reveal that technical skills are essential and commonly mastered, but improvements are needed in clear communication and constructive feedback.

### 14.2.3 Other tasks

Besides the aforementioned tasks, there are also a series of other development tasks that benefit from machine learning and deep learning technologies,e.g. bug assignment (Section 13.3), question answering tasks in online Q&A sites, and various development tasks for open source contributors.

Huanget al. [733] proposed an approach to recommending appropriate answerers for questions posted to Q&A sites. Specifically, they leveraged graph attention networks to represent the interactive information between candidate answerers, and an LDA topic model to capture the text information. It was verified by experiments that the approach outperforms the state-of-the-art techniques of that time. Jin et al. [734] proposed CODER, a graph-based code recommendation framework for OSS developers, that models user-code and user-project interactions via a heterogeneous graph to predict developers' future contributions. CODER has shown superior performance in various experimental settings, including intra-project and cross-project recommendations. Xiao et al. [735] introduced RecGFI, an approach for recommending "Good First Issues" to newcomers in open-source projects. Utilizing features from content, background, and dynamics. Employing an XGBoost classifier, RecGFI achieves up to 0.853 AUC in evaluation, demonstrating superiority over alternative models.

Santos [736] proposed an automatic open issue labeling strategy to assist OSS contributors in selecting suitable tasks and helping OSS communities attract and retain more contributors. The technique uses API-domain tags to label issues and relies on qualitative studies to formulate recommendation strategies and quantitative investigations to analyze the relevance between API-domain labels and contributors. The results show that the predicted labels have an average precision of 75.5%, demonstrating the superiority of the technique. Costa et al. [737] presented TIPMerge, an approach for recommending participants for collaborative merge sessions within large development projects with multiple branches. TIPMerge builds a ranked list of developers appropriate to collaborate by considering their changes in previous history, branches and dependencies, and recommends developers with complementary knowledge. The approach demonstrates a mean normalized improvement of 49.5% for joint knowledge coverage compared to selecting the top developers.

## 14.3 Development Team Formation

Complex development tasks often ask for a team of developers. Thus how to find a cohort of developers who can collaboratively handle a complex task is an important issue in software development.

In order to address the complexity of finding collaborators with shared interests in large open-source software, Constantinoetal. [738] proposed a visual and interactive web application tool named CoopFinder. They further presented and evaluated two collaborator recommendation strategies based on co-changed files [739]. The strategies utilize TF-IDF scheme to estimate the importance of files modified by developers and measure developers' similarity using the Cosine metric. Through an extensive survey of 102 real-world developers, the strategies show up to an 81% acceptance rate, enhancing collaboration efficiency among developers. Surian et al. [740] introduced a technique to find compatible collaboration among developers. They first created a collaboration network using information of developers and projects from Sourceforge.Net, then recommended collaborators for developers based on their programming skills and past projects through a random walk with restart procedures. Canfora et al. [741] introduced Yoda, a technique for recommending mentors to newcomers in software projects. By mining data from mailing lists and versioning systems, developers with experience meanwhile actively interacting with newcomers are selected as their mentors. Evaluation on five open-source projects and surveys with developers shows the potential usefulness of Yoda in supporting newcomers in a team and indicates that top committers are not always the best mentors. Ye et al. [742] introduced a personalized teammate recommendation approach for crowdsourcing developers. Through an empirical study on Kaggle, three factors influencing developers' teammate preferences are identified and a linear programming-based technique is proposed to compute developers' teammate preferences. Finally, a recommendation approach with an approximation algorithm to maximize collaboration willingness is designed.

## 14.4 Datasets

Here, we present some commonly used datasets in the three main kinds of research efforts toward intelligent developer collaboration. For developer expertise profiling, data primarily originate from collaborative software development platforms such as GitHub, encompassing version control data from various open-source projects with diverse developer information [743] [744] [716]. For intelligent task assignment, datasets are sourced from crowdsourcing platforms like Topcoder and Baidu CrowdTest, where historical data contain developer information and the corresponding task categories [718] [721]. In the case of development team formation, datasets are predominantly obtained from platforms like SourceForge and Kaggle, showcasing richer collaboration patterns among developers [740] [742]. These datasets provide abundant information for studying developer behavior, intelligent task assignment, and team formation.

- **Developer expertise profiling.** The World of Code (WoC) dataset [743] is a versioned and expansive repository of version control data from Free/Libre and Open Source Software (FLOSS) projects using Git. The dataset, collected in March 2020, contains 7.9 billion blobs, 2 billion commits, 8.3 billion trees, 17.3 million tags, 123 million distinct repositories, and 42 million unique author IDs. WoC supports various research tasks, including developer expertise profiling. Using the WoC dataset, Fry et al. [744] proposed a technique to identify all author IDs belonging to a single developer in the entire dataset, revealing aliases. Using machine learning, Fryetal. processed around 38 million author IDs, identifying 14.8 million with aliases linked to 5.4 million developers. This dataset enhanced models of developer behavior at the global open-source software ecosystem level, facilitating rapid resolution of new author IDs. Meanwhile, Dakhel et al. [716] collected a dataset to determine the domain expertise of developers using information from GitHub. Their data collection process involved three main types of information: repositories that developers have contributed to, issue-resolving history, and API calls involved in a commit. This dataset contained information about 1,272 developers with expertise labels in five job roles. The dataset consisted of textual information from 58,000 repositories, issue-resolving history from 60,000 issues, and API calls from 21 million commits across different GitHub repositories. The dataset aimed to provide comprehensive insights into developers' expertise by considering their contributions to repositories, issue resolution history, and API usage in commits across diverse projects on GitHub.

- **Intelligent task assignment.** Crowdsourcing platforms provide the feasibility of data collection for intelligent task assignments. For instance, Topcoder is a competition-based crowdsourcing software development platform. Topcoder offers various types of tasks, such as "Test Suites," "Assembly," and "Bug Hunt," each representing a category of challenges. Challenges are instances of task types, and developers choose whether to participate in these challenges. In this respect, Wang et al. [718] collected a dataset from Topcoder. Their dataset involved 32,565 challenges, 7,620 developers, and 59,230 submissions spanning from 2006 to 2016. The dataset focused on 100 developers with over 100 submissions, containing the evolution of the skills of each developer. After filtering out submissions with a final score of 0 (indicating unfinished submissions), there were 74 developers left in the dataset. There has also been a lot of work on dataset building using other crowdsourcing platforms. For example, Wanget al. [721] collected a dataset from Baidu CrowdTest. The dataset involved 2,404 crowdworkers and comprised 80,200 submitted reports. For each testing task, comprehensive information was gathered, including task-related details and all submitted test reports with associated information, such as submitter and device.The dataset serves as a valuable resource for analyzing crowdtesting dynamics and outcomes.

- **Development team formation.** Work related to development team formation is often obtained from open-source software development platforms with collaborative and social nature, such as SourceForge and Kaggle. For example, Surian et al. [740] collected a dataset by analyzing the SourceForge database, which contained information about projects, project categories, and programming languages. There were 209,009 developers associated with 151,776 projects. The dataset included details such as 354 project categories and information about the usage of 90 different programming languages within the projects. Meanwhile, Yeetal. [742] collected a dataset by crawling data from Kaggle, including details from 275 competitions and 74,354 developers. The data spanned from April 2010 to January 2018 and encompassed 191,300 submissions. Additionally, developers' social data from Kaggle's communities were crawled, enhancing the dataset with information about interactions and discussions among developers.

## 14.5 Challenges and Opportunities

In general, software is becoming more complex as the major enabling force for the infrastructure of various information systems. Consequently, more developers participate in software projects, and software development is increasingly exhibiting a social-technical characteristic, which requires more effective collaboration among developers. Thus, more research efforts are needed to produce new theories, techniques, and tools to further improve collaborative development. To that end, we summarize the challenges and opportunities in this research area.

### 14.5.1 Challenges

The use of deep learning to improve developer collaboration faces the following research challenges in terms of collaborative tasks, software development data, and evaluation benchmarks.

- **Complex collaboration among multiple developers.** Existing research mainly considers the collaboration between two developers. In other words, one developer posts a task requirement, and another developer is required to fulfill the task. However, at higher levels (e.g., from a project's perspective), we must consider the collaboration among a group of developers, where global collaboration effects and constraints should be of greater concern.

- **Continuously growing data size and heterogeneity.** On the one hand, developers generate more data, which are often distributed across various platforms, including personal IDEs, proprietary enterprise environments, open-source platforms (e.g., GitHub), and public forums (e.g., Stack Overflow). On the other hand, software development data are essentially heterogeneous, involving natural language data, source code, AI models, and even graphical data. Dealing with highly dispersed, heterogeneous, and large-scale development data is a great challenge for using deep learning to achieve more effective and efficient collaborative development.

- **Lack of benchmarks for effective evaluation.** As the effectiveness of developer collaboration usually cannot be observed in a short time, evaluating a newly proposed technique is difficult. Therefore, having benchmark datasets on one or multiple collaborative development tasks is highly desirable.

### 14.5.2 Opportunities

Although deep learning has shown promising results in improving developer collaboration, there are still abundant research opportunities for advancing this direction.

- Application of deep learning to a wider spectrum of collaborative development tasks. As collaborative activities are pervasive in software development processes, deep learning can be introduced to deal with more tasks other than those in existing studies.

- Incorporating advanced deep learning technologies. Although deep learning has developed dramatically for nearly two decades, new technologies are still being put forward continually, such as graph neural networks and large language models. Applying these technologies to handle large-scale heterogeneous development data embraces more opportunities for intelligent collaborative development.

- Novel collaborative development activities enabled by deep learning. Large language models like ChatGPT are becoming more powerful in handling development tasks, such as code generation and program repair. It is possible to see novel collaborative development activities among human developers and deep learning-based AI models.

## 15 Conclusion

In this paper, we present the first task-oriented survey on deep learning-based software engineering. Our survey focuses on twelve of the important tasks in software development and maintenance: requirements engineering, code generation, code search, code summarization, software refactoring, code clone detection, software defect prediction, bug finding, fault localization, program repair, bug report management, and developer collaboration. For each of the selected tasks, we summarize the mostrecent advances concerning the application of deep learning for the given task, as well as relevant challenges and future opportunities. On the basis of the surveyed deep learning-based subareas of software engineering, we make the following observations:

- **Widespread applications and impressive results.** Deep learning techniques have been widely applied across various subareas of software engineering and have achieved impressive results. First, deep learning techniques typically improve the performance of most tasks. For example, deep learning-based code clone detection always achieves higher recall and better precision than the best classical approaches. An existing empirical study also shows that deep learning techniques outperform other traditional machine learning techniques for bug assignments. Moreover, in requirements engineering, most studies report precision and recall exceeding 80%, and the $F_1$ score is often above 75%. Second, the powerful feature engineering capability of deep learning enables the capture of semantic information. Typically, some tasks (e.g., software defect prediction and software refactoring) involve complex feature engineering. Deep learning helps release researchers and practitioners from tedious feature engineering. Third, deep learning can further contribute to the automation of software engineering processes. Traditional techniques may rely on complex preprocessing and/or postprocessing techniques to automate the whole process. Deep learning can turn the whole process in an end-to-end manner. For example, deep learning techniques can directly deal with bug reports and help automate the process of bug report management.

- **Challenges in high-quality training data acquisition.** The primary challenge faced by most subareas of software engineering is obtaining high-quality training data. First, the sizes of datasets are often limited. For example, publicly accessible data for requirements typically are small in volume and lack some details, making it difficult to train effective deep learning models. Second, the quality of datasets is often unguaranteed. For example, in code generation, it is unclear whether datasets contain vulnerable code snippets that may result in unsafe codes. Third, some existing datasets are the results of specially designed data production activities and may not well align with real-world scenarios. Although this can lead to strong model performance during evaluation, the performance may not be effectively transferred into practical use.

- **Interpretability challenge.** The interpretability of deep learning models is also a common challenge across various subareas of software engineering. First, the lack of interpretability for deep learning models makes ensuring correctness difficult. For example, ensuring whether generated patches are correct is difficult. Second, deep learning models are known for their "black-box" nature, making it challenging for developers to understand their rationale. Therefore, it is hard for developers to work with deep learning models.

- **Generalization across languages.** Developing deep learning models that generalize well across different languages is a considerable challenge. Different programming languages have their own syntax and semantics, making it difficult to develop a universal model that performs well across diverse languages. Therefore, the current practice is to deal with each programming language individually. For example, for different programming languages, researchers often need to train different models for the same code summarization task. Furthermore, to incorporate additional code structure information in deep learning models, researchers must parse code snippets with their corresponding parsers, thus resulting in significant differences for different programming languages.

Our survey demonstrates that deep learning-based software engineering has achieved significant advances recently and has the potential for further improvement. However, some critical challenges should be resolved before deep-learning based software engineering can reach its maximal potential. We believe that future research should focus on resolving these challenges.

**Acknowledgements** We thank the following persons for their prior contributions to the manuscript preparation (in alphabetical order): Yuze Guo (Beihang University), Ruiqi Hong (Beihang University), Mingwei Liu (Fudan University), Xiaofan Liu (Wuhan University), Di Wu (Beihang University), Hongjun Yang (Beihang University), Yanming Yang (Zhejiang University), Binquan Zhang (Beihang University), and Zhuang Zhao (Wuhan University).